\documentclass[prd,aps,a4paper,eqsecnum,nofootinbib,floatfix,twocolumn]{revtex4} 

\newif\ifusesec
\usesectrue  
   
\usepackage{graphicx} 
\usepackage{amsmath,amsfonts,amssymb}

\newcommand{\be}{\begin{equation}}
\newcommand{\ee}{\end{equation}}
\newcommand{\beq}{\begin{equation}}
\newcommand{\eeq}{\end{equation}}
\newcommand{\bea}{\begin{eqnarray}}
\newcommand{\eea}{\end{eqnarray}}
\newcommand{\eq}{\eqref}

\newcommand{\g}{{\gamma}}

\newcommand{\pinf}{p_{\infty}}
\newcommand{\cI}{{\cal I}}

\begin{document}

\title{Radiative contributions to gravitational scattering}

\author{Donato Bini$^{1,2}$, Thibault Damour$^3$, Andrea Geralico$^1$}
  \affiliation{
$^1$Istituto per le Applicazioni del Calcolo ``M. Picone,'' CNR, I-00185 Rome, Italy\\
$^2$INFN, Sezione di Roma Tre, I-00146 Rome, Italy\\
$^3$Institut des Hautes Etudes Scientifiques, 91440 Bures-sur-Yvette, France\\
}

\date{\today}

\begin{abstract}
The linear-order effects of radiation-reaction on the classical scattering of two point masses, in General
Relativity, are derived by a variation-of-constants method. Explicit expressions for the
radiation-reaction contributions to the changes of 4-momentum during scattering
 are given to linear order in the
radiative losses of energy, linear-momentum and angular momentum. The
polynomial dependence on the masses of the 4-momentum changes is shown to
lead to non-trivial  identities relating the various radiative losses.
At order $G^3$ our results
lead to a streamlined classical derivation of results recently derived within a quantum approach.
At order $G^4$ we compute  the needed radiative losses to next-to-next-to-leading-order
in the post-Newtonian expansion, thereby reaching the absolute fourth and a half post-Newtonian level of
accuracy in the 4-momentum changes. We also provide  explicit expressions, at the  absolute sixth post-Newtonian accuracy, for the
radiation-graviton contribution to {\it conservative} $O(G^4)$ scattering.
At orders $G^5$ and $G^6$ we derive explicit theoretical expressions for the last
two hitherto undetermined parameters describing the fifth-post-Newtonian dynamics.
Our results at the fifth-post-Newtonian level confirm results of [Nucl. Phys. B \textbf{965}, 115352 (2021)]
but exhibit some disagreements with results of [Phys. Rev. D \textbf{101}, 064033 (2020)].
\end{abstract}

\maketitle

\section{Introduction}

A new angle of attack on the time-honored two-body problem in General Relativity has been recently
the focus of active research, namely the study of the (classical or quantum) scattering
of two, gravitationally interacting, massive bodies. Several complementary avenues 
towards this problem have been pursued, using various approaches:
post-Minkowskian (PM), e.g., \cite{Damour:2016gwp,Damour:2017zjx,Vines:2017hyw,Bini:2018ywr,Bjerrum-Bohr:2019kec,Kalin:2019inp,Kalin:2020mvi,Mougiakakos:2021ckm}, post-Newtonian (PN), e.g., \cite{Bini:2017wfr}, Effective One Body (EOB), e.g., \cite{Damour:2016gwp,Bini:2017xzy}, Effective Field Theory (EFT), 
e.g., \cite{Cheung:2018wkq,Mogull:2020sak},
Tutti Frutti (TF) \cite{Bini:2019nra}, and  scattering amplitudes (including generalized unitarity, the double copy, eikonal resummation and advanced multiloop integration methods), e.g., \cite{Bern:2019nnu,Bern:2019crd,KoemansCollado:2019ggb,DiVecchia:2019kta,Cheung:2020gyp,Cristofoli:2020uzm,Bern:2021dqo,Bjerrum-Bohr:2021din}.

A new twist in this story has recently gained prominence: the issue of radiative corrections to scattering.
Indeed, a paradoxical discrepancy between the high-energy limit of the $O(G^3)$ scattering result of  
Bern et al. \cite{Bern:2019nnu} and the massless scattering result of Amati, Ciafaloni and Veneziano 
\cite{Amati:1990xe} was highlighted in Ref. \cite{Damour:2019lcq}. This discrepancy was confirmed
in recalculations of massless scattering \cite{Bern:2020gjj,DiVecchia:2020ymx}.
It is only quite recently that the
resolution of this paradox was understood to be rooted in radiative contributions to scattering
\cite{DiVecchia:2020ymx,Damour:2020tta,DiVecchia:2021ndb, DiVecchia:2021bdo,Herrmann:2021tct}.

In Ref. \cite{Damour:2020tta}, one of us derived the (leading-order) $O(G^3)$ radiation-reaction
contribution to the scattering angle of two (classical) bodies in General Relativity. This result was
confirmed by scattering-amplitude computations, \cite{DiVecchia:2021bdo,Herrmann:2021tct,Bjerrum-Bohr:2021din}, using various techniques (eikonal resummation or
an observable-based formalism \cite{Kosower:2018adc}). In addition, Ref. \cite{Herrmann:2021tct}
completed the scattering-angle result of \cite{Damour:2020tta} by deriving 
(from the  observable-based formalism 
of Ref. \cite{Kosower:2018adc}) the full expressions for the 4-momentum changes,
$\Delta p_{a \mu} \equiv p_{a \mu}^+- p_{a \mu}^-$, $a=1,2$, of the two scattering bodies.

The first aim of the present paper is to give a purely classical derivation of the linear-order radiation-reaction
contributions to $\Delta p_{a \mu}$ by using a (first-order) variation-of-constant approach.
[Our method generalizes the one introduced in Ref. \cite{Bini:2012ji} by including {\it recoil effects}.]
The application of our general result to the $G^3$ level will be shown to reproduce the result
obtained in Ref. \cite{Herrmann:2021tct} via a quantum computation. 
In addition, we shall apply our results to the $G^4$ level, thereby completing the recent $G^4$-level
result of Ref. \cite{Bern:2021dqo} by showing how to compute $O(G^4)$ linear-order radiation-reaction contributions
to scattering. Our limitation to linear-order in radiation-reaction implies that our $O(G^4)$ {\it dissipative} results will be complete only up to the $O(1/c^9)$  PN order included. [The $O(G^4)$ {\it potential-graviton} dynamics derived in Ref. \cite{Bern:2021dqo}
has been partially checked in \cite{Blumlein:2021txj}, and recently fully rederived within a
classical approach in \cite{Dlapa:2021npj}.]
Our theoretical result for the latter $O(G^4)$ radiation-reaction contribution notably involves
 the $O(G^3)$ radiative loss of angular momentum, which has not yet been computed in PM
 gravity. We have, however, computed it (as well as the other needed PM radiative losses)
 with 2PN fractional accuracy. We also show how the mass polynomiality of the radiation-reacted
 momentum changes $\Delta p_{a \mu}$ yield remarkable a priori constraints on several
 PM radiative losses.
 
Besides providing expressions for the $O(G^4)$ radiation-reaction contributions
to scattering, we shall also complete the {\it potential-graviton} contribution
to $O(G^4)$ conservative scattering, by explicitly extracting from the current TF results another 
$G^4$-level radiative contribution to scattering, namely the conservative, {\it radiation-graviton} 
contribution linked (in the EFT approach) to nonlocal (soft) graviton exchanges. 
Finally, we shall also go beyond the $O(G^4)$ level by combining information from the 
5PN EFT results of Refs. \cite{Foffa:2019eeb,Blumlein:2020pyo,Foffa:2021pkg}
and the 5PN TF results of Refs. \cite{Bini:2019nra,Bini:2020nsb,Bini:2020hmy}. This will allow
us to derive explicit theoretical expressions for the two hitherto undetermined $O(G^5)$ and
$O(G^6)$ parameters entering the 5PN dynamics (as described through the TF approach).
Our results confirm and extend results of  Bl\"umlein et al. \cite{Blumlein:2020pyo}, but
exhibit several disagreements with results of Foffa and Sturani  \cite{Foffa:2019eeb,Foffa:2021pkg}.
We point out that part of these disagreements might be rooted in subtleties linked to contributions
that are nonlinear in radiation-reaction.

\section{Decomposing the various radiative contributions to the impulse} 

In the present work we focus on the total change 
\be \label{decomp0}
\Delta p_{a \mu} \equiv p_{a \mu}^+- p_{a \mu}^-,
\ee
 between the infinite past
and the infinite future, of the (classical) 4-momentum $p_{a \mu} = m_a u_{a \mu}$ of a particle experiencing a 
gravitational two-body scattering. Here, the subscript $a=1,2$ labels the particle (of mass $m_a$),  $u_{a \mu}^{-}$ 
denotes its incoming 4-velocity, and $u_{a \mu}^{+}$ its outcoming 4-velocity. These
 4-velocities are measured in the asymptotic incoming and outgoing Minkowski spacetimes.
Following the terminology used in 
Ref. \cite{Kosower:2018adc}, the momentum change $\Delta p_{a \mu}$ will often be called the ``impulse'' of particle $a$.

In the following,  we decompose $\Delta p_{a \mu}$ as the sum of two contributions, namely
\be \label{decomp1}
\Delta p_{a \mu}(u_1^{-}, u_2^{-}, b)= \Delta p_{a \mu}^{\rm cons}(u_1^{-}, u_2^{-}, b)+ \Delta p_{a \mu}^{\rm rr, tot}(u_1^{-}, u_2^{-}, b)\,.
\ee
Here, $  \Delta p_{a \mu}^{\rm cons}(u_1^{-}, u_2^{-}, b)$ denotes the 
{\it conservative} contribution to the impulse, obtained by neglecting all radiative losses, and expressed
as a function of the incoming 4-velocities and of the (vectorial) impact parameter. As in our previous works, the conservative impulse $\Delta p_{a \mu}^{\rm  cons}(u_1^{-}, u_2^{-}, b)$
is defined as the impulse that follows from the Fokker-Wheeler-Feynman-type time-symmetric gravitational
interaction of two masses (using a half-retarded-half-advanced Green's function at each PM order). The complementary
{\it radiation reaction} contribution, $\Delta p_{a \mu}^{\rm rr, tot}(u_1^{-}, u_2^{-}, b)$,  is linked to the radiative losses occurring when the two bodies interact via {\it retarded} gravitational interactions. It will be obtained, by a linear-response computation, as the sum of two separate radiative effects, namely
\be  \label{decomp2}
\Delta p_{a \mu}^{\rm rr, tot}(u_1^{-}, u_2^{-}, b)= \Delta p_{a \mu}^{\rm rr, rel}+ \Delta p_{a \mu}^{\rm rr, rec} + O({\cal F}_{\rm rr}^2)\,.
\ee
Here, the first contribution, $ \Delta p_{a \mu}^{\rm rr, rel}$, is linked to radiative effects acting on
the {\it relative motion} of the binary system, while the second one,
$\Delta p_{a \mu}^{\rm rr, rec}$, is linked
to the over-all {\it recoil}\footnote{By {\it recoil} of the two-body system, we mean the loss of total mechanical linear momentum of
the system via radiation of linear momentum in the form of gravitational waves.} of the two-body system.
We have added an error term, $ O({\cal F}_{\rm rr}^2)$ in Eq. \eq{decomp2}, as a reminder that our derivation
of $\Delta p_{a \mu}^{\rm rr, rel}$ and $ \Delta p_{a \mu}^{\rm rr, rec}$ is valid only to first
order in the radiation-reaction force ${\cal F}_{\rm rr}$. We will also include a discussion of some
of the effects that are quadratic in ${\cal F}_{\rm rr}$ (which start at order $G^4/c^{10}$).

The decomposition \eq{decomp2} does not exhaust all the {\it radiation-related} contributions to the impulse. Indeed, when
following the EFT approach to binary dynamics, the
conservative impulse $\Delta p_{a \mu}^{\rm  cons}$ is, itself, naturally decomposed 
into two contributions:
\be  \label{decomp3}
\Delta p_{a \mu}^{\rm  cons}(u_1^{-}, u_2^{-}, b)= \Delta p_{a \mu}^{\rm cons, pot}+ \Delta p_{a \mu}^{\rm cons, rad} \,.
\ee
Here $ \Delta p_{a \mu}^{\rm cons, pot}$ denotes (when using the method of regions) the part of the conservative
impulse that is due to the mediation of {\it potential graviton modes}, while  $ \Delta p_{a \mu}^{\rm cons, rad}$ denotes
 the part of the conservative impulse that is due to the mediation of {\it radiation graviton modes}.
 In other words, there are three different impulse effects linked to soft (radiative-like) graviton modes:
 a conservative effect $\Delta p_{a \mu}^{\rm cons, rad}$  linked to the exchange of {\it time-symmetric} radiation gravitons,
 and two separate dissipative effects linked to {\it time-antisymmetric} radiation-reaction forces acting on the system:
  $ \Delta p_{a \mu}^{\rm rr, rel}$ and   $ \Delta p_{a \mu}^{\rm rr, rec}$.

 Finally, let us also note that, from the technical point of view, we will  PM-expand  each impulse (and also, 
 each partial contribution to the impulse), say
\be
\Delta p_{a \mu}= \sum_{n\geq 1}\Delta p_{a \mu}^{G^n}\,,
\ee
each PM contribution $\Delta p_{a \mu}^{G^n}$  being, eventually, further expanded in a PN series:
\be
\Delta p_{a \mu}^{G^n} \sim \frac{G^n}{c^m} \left(1 + \frac1{c^2} + \cdots  \right)\,.
\ee
Let us also mention that it will be often convenient to decompose  the impulse of each particle,
along an appropriate  basis of 4-vectors.
 A first possible basis is
${\hat b}_\mu$, $ u_{1 \mu}^- $ and $ u_{2 \mu}^- $. 
This decomposition involves three scalar coefficients (for each $a=1,2$), $c_b^a, c_{u_1}^a, c_{u_2}^a$ ($a=1,2$), namely
\be \label{decompbu1u2}
\Delta p_{a \mu}(u_1^{-}, u_2^{-}, b)= c_b^a {\hat b}_\mu + c_{u_1}^a u_{1 \mu}^- +   c_{u_2}^a u_{2 \mu}^-\,,
\ee
where ${\hat b}^\mu \equiv b^\mu/b$ is the unit (spacelike) 4-vector in the direction of 
the vectorial impact parameter\footnote{We recall that $b^\mu = b \, \hat {b}^\mu$
is orthogonal to both incoming 4-velocities $ u_{1 \mu}^-$, $ u_{2 \mu}^-$,
and is taken in the direction going from body 2 towards body 1.}  $b^\mu$. 
The decomposition of any impulse $\Delta p_{a \mu}$ along the vectors ${\hat b}_\mu$, $ u_{1 \mu}^- $ and $ u_{2 \mu}^- $
is a linear operation which commutes with any other linear decomposition of $\Delta p_{a \mu}$.
This means in particular that the decomposition \eq{decomp1} induces a
corresponding linear decomposition of the corresponding coefficients in the basis \eq{decompbu1u2},
say (with a label $X= b, u_1, u_2$)
\bea
\label{sum_contrib}
c_X^a &=& c_X^{a\, \rm cons} + c_X^{a\, \rm rr, rel} +  c_X^{a\, \rm rr, rec}+ O({\cal F}_{\rm rr}^2)\,. 
\eea

\section{Application of the method of variation of constants to the relative motion, and the recoil, of two-body systems} \label{sec3}

In the present work, we  assume that the PM-expanded, physical, {\it retarded}~\footnote{I.e., obtained by
using a retarded Green function at each step of the PM iteration.}
equations of motion of a binary system can be described
by adding to the (Fokker-Wheeler-Feynman time-symmetric) Poincar\'e-invariant conservative dynamics 
additional radiation-reaction forces ${\cal F}^\mu_{a}$
which have (at least to first-order) a {\it time-antisymmetric} character, and which cause losses of the total mechanical Noetherian
quantities $P_{\rm system}^\mu, J_{\rm system}^{\mu \nu}$ of the conservative dynamics of the two-body system
that balance the corresponding radiated quantities $P_{\rm rad}^\mu, J_{\rm rad}^{\mu \nu}$.
In the case of the  linear momentum (for which PM-expanded interaction terms are better controlled in all Lorentz frames),
we have also the property that $P_{\rm system}^\mu$ reduces to the usual total mechanical linear momentum 
$p_1^\mu + p_2^\mu = m_1 u_1^\mu + m_2 u_2^\mu$ in the asymptotic incoming ($u_{1,2}\to u_{1,2}^-$), and outgoing ($u_{1,2}\to u_{1,2}^+$), states.

The balance law in this case states that the additional radiation-reaction forces ${\cal F}^\mu_{a}$ imply that
\be \label{pbalance1}
P_{\rm system}^{\mu  -}=P_{\rm system}^{\mu  +} + P_{\rm rad}^\mu\,,
\ee
i.e., explicitly,
\be \label{pbalance2}
p_1^{\mu  -}+ p_2^{\mu -}=p_1^{\mu  +}+ p_2^{\mu +} + P_{\rm rad}^\mu\,.
\ee
The early studies (at the 2PM and 3PM levels) of the PM-expanded
two-body dynamics in the 1980's \cite{Westpfahl:1979gu,Bel:1981be,Damour:1982wm} have explicitly checked,   at the leading order in radiation reaction,
the validity of the decomposition of the PM-expanded
two-body dynamics in conservative plus radiation-reaction effects (balancing the radiative losses)
\cite{Damour:1981bh,Damour:1982wm,Damour:1983tz}.
[These early PM-based works were motivated by unsatisfactory aspects of earlier PN-based
studies of radiation damping  \cite{Chandrasekhar1970,Burke:1970wx}.]
Other approaches (notably Hamiltonian-based studies \cite{Jaranowski:1996nv}, 
and PN-based ones \cite{Nissanke:2004er}) have confirmed the validity of
balance laws between mechanical properties (energy, linear momentum, angular momentum)
of the radiation-reacted system  and
corresponding fluxes radiated as gravitational waves at the next-to-leading-order (NLO) in 
PN-expanded radiation reaction.
Recently, a quantum-based computation \cite{Herrmann:2021tct} 
(using the formalism of Ref. \cite{Kosower:2018adc}) has explicitly checked the validity of the balance 
Eq. \eq{pbalance2} at the 3PM order.

Formulating the time-retarded dynamics in terms of 
 additional radiation-reaction forces ${\cal F}^\mu_{a}$ acting on the time-symmetric
conservative dynamics allows one to apply the general method (due to Lagrange) of variation of constants. 
In a PN context, this was done for ellipticlike bound motions in Refs \cite{Damour:1983tz,Damour:2004bz},
and for hyperboliclike scattering motions in \cite{Bini:2012ji}. We have generalized these treatments
in two ways: (i) by working within a PM context; and (ii) by including the effects linked to the overall recoil
of the binary system. [The previous treatments applied the method of varying constants only to the 
{\it relative motion} of the two bodies,
considered in the c.m. system, using an approximation where the total linear momentum of the 2-body system
was conserved.]
In our PM context,  it will be crucial to take into account the non-conservation of the 
the total linear momentum of the 2-body system, i.e. the non-zero value of $ P_{\rm rad}^\mu$ in Eqs. \eq{pbalance1}, \eq{pbalance2}. Indeed, $ P_{\rm rad}^\mu= O(G^3)$, so that a complete PM
description of radiation-reaction effects
beyond the 2PM [$O(G^2)$] level must take into account recoil effects. [In a PN context, recoil
is viewed as being $O(\frac{G^3}{c^7})$, i.e., of 3.5PN order, while the radiative energy loss is
 $O(\frac{G^3}{c^5})$, i.e., of 2.5PN order.]

\subsection{Separating the relativistic two-body dynamics in relative, and center-of-mass, dynamics}

In order to compute the  effect of the overall recoil of the binary system,
 \be \label{pbalance3}
\Delta P_{\rm system}^{\mu }= P_{\rm system}^{\mu  +} - P_{\rm system}^{\mu  -}=- P_{\rm rad}^\mu\,,
\ee
 on the impulse  $\Delta p_{a \mu}$ of each particle,
one can make use of  results from the literature on the separation of the relativistic dynamics 
 of two-body systems in relative dynamics and center-of-mass (c.m.) dynamics (see, e.g., \cite{Alba:2006hs} for an introduction).
 In particular, Sch\"afer and collaborators \cite{Rothe:2010jj,Georg:2015afa} have given a very concrete, and directly relevant, application of this separation
 to the case of gravitationally interacting binary systems. More precisely, Refs. \cite{Rothe:2010jj,Georg:2015afa} showed
 how to perturbatively construct a {\it canonical} transformation between standard (Arnowitt-Deser-Misner-type) two-body
 phase-space variables ${\bf x}_1, {\bf x}_2, {\bf p}_1, {\bf p}_2$ and new phase-space variables ${\bf r}, {\bf p}, {\bf R}, {\bf P}$
 (where ${\bf r}, {\bf p}$ describe the relative dynamics, while ${\bf R}, {\bf P}$ describe the dynamics of the c.m.)
 such that the square of the total Hamiltonian,
 \bea \label{Htot1}
  H_{\rm tot}({\bf x}_1-{\bf x}_2, {\bf p}_1, {\bf p}_2) &=&  \sqrt{m_1^2 c^4 + c^2 {\bf p}_1^2}+ \sqrt{m_2^2 c^4 + c^2 {\bf p}_2^2} \nonumber\\
&&  + V_{\rm int}({\bf x}_1- {\bf x}_2, {\bf p}_1, {\bf p}_2)\,,
  \eea
 describing the {\it conservative} dynamics of a gravitationally interacting binary system
 takes, when reexpressed in terms of the new variables, a form neatly showing the separation between the
 relative dynamics and the c.m. one, namely
 \be \label{Htot2}
 H^2_{\rm tot}({\bf r}, {\bf p}, {\bf R}, {\bf P})=  H^2_{\rm rel}({\bf r}, {\bf p})+  c^2 {\bf P}^2\,.
 \ee
 Here the function $H_{\rm rel}({\bf r}, {\bf p})$ is the usual c.m.-reduced relative Hamiltonian, namely
 \be
 H_{\rm rel}({\bf r}, {\bf p}) \equiv  
 \left[H_{\rm tot}({\bf x}_1- {\bf x}_2, {\bf p}_1, {\bf p}_2)\right] ^{{\bf x}_1- {\bf x}_2 \to {\bf r}}_{{\bf p}_1 \to {\bf p}, {\bf p}_2 \to -{\bf p}} \, .
 \ee
 The canonical transformation between  ${\bf x}_1, {\bf x}_2, {\bf p}_1, {\bf p}_2$ and  ${\bf r}, {\bf p}, {\bf R}, {\bf P}$ is complicated
 and can only be constructed perturbatively, either in a PN-expansion starting from its usual Newtonian analog \cite{Rothe:2010jj,Georg:2015afa},
 or in a combined expansion in powers of  ${\bf P}= {\bf p}_1+ {\bf p}_2$, and in powers of $G$.  Some quantities, however,
 have  simple (Newtonian-looking) expressions. Namely, the  total linear momentum is given by
 \be \label{Pvsp1p2}
  {\bf P}= {\bf p}_1+ {\bf p}_2\,,
 \ee
and the total angular momentum (for spinless particles) is given by
 \be
  {\bf J}= {\bf x}_1 \times {\bf p}_1+{\bf x}_2 \times  {\bf p}_2= {\bf r} \times {\bf p}+{\bf R} \times  {\bf P}\,.
 \ee
 In addition, using the definition of the c.m. position variable, ${\bf R}$ \cite{Rothe:2010jj,Georg:2015afa}, 
 namely (denoting ${\mathcal M} c^2 \equiv \sqrt{H_{\rm tot}^2- c^2 {\mathbf P}^2}$)
 \be
 {\bf R} = \frac{ c^2 {\bf G}}{H_{\rm tot}} +\frac{1}{{\mathcal M}(H_{\rm tot}+{\mathcal M}c^2)}\left[{\mathbf J}-\left(\frac{ c^2 {\bf G}}{H_{\rm tot}}\times {\mathbf P}\right)  \right]\times {\mathbf P}\,,
 \ee
 where $G^i= t P^i+ K^i$, with $K^i \equiv J^{i0}$, we have checked that the  square of the relative (3-dimensional) angular momentum
 ${\mathbf J}^{\rm rel}={\bf r} \times {\bf p}$ is simply equal to the square of the  (4-dimensional) Pauli-Lubanski spin 4-vector,
  \be
 S_\mu= \frac1{2 {\mathcal M} c}\eta_{\alpha \beta \gamma \mu} P^\alpha J^{\beta \gamma}\,,
 \ee
 where $\eta_{\alpha \beta \gamma \mu}$ (with $\eta_{0123} = +1$) denotes the Levi-Civita tensor and 
$J^{\beta \gamma}=(J^{i0},J^{ij})=(K^i,\epsilon^{ijk}J^k)$.
Explicitly, we have 
 \be \label{JrelvsPL}
 ({\bf r} \times {\bf p})^2 = S_\mu S^\mu\,.
 \ee
 The other expressions relating relative variables to the original phase-space variables ${\bf x}_1, {\bf x}_2, {\bf p}_1, {\bf p}_2$ 
 are complicated, and contain interaction terms. E.g., one has
 \bea 
\label{pvsp1p2}
 {\bf p}&=& \frac{E_2({\bf p}_2)}{E_1({\bf p}_1)+E_2({\bf p}_2)} {\bf p}_1- \frac{E_1({\bf p}_1)}{E_1({\bf p}_1)+E_2({\bf p}_2)} {\bf p}_2\nonumber\\
&& + O( {\bf p}_1+ {\bf p}_2) + O\left(\frac{G}{c^2} \right)\,,
 \eea
 where we used the shorthand notation
 \be
 E_a({\bf p}) \equiv  \sqrt{m_a^2 c^4 + c^2 {\bf p}^2}\,.
 \ee
 The above results on the relativistic separation between relative and c.m. dynamics hold in the conservative case.
However, the existence, in the conservative case, of such a relativistic decomposition between relative motion and c.m. motion 
offers a useful framework for applying the method of varying constants
in the non-conservative case where the two-body system
loses energy, momentum and angular momentum in the form of gravitational radiation. 
To do so, it is convenient to work within the
{\it incoming} c.m. frame of the system, with time axis
\be
U^{ \mu -} \equiv \frac{ p_1^{ \mu -} + p_2^{ \mu -}}{ |p_1^{ -}+  p_2^{ -} |}\,.
\ee
In addition, when needed, we will use the sharper c.m. condition that the incoming value of the time-space components $J_{0i}^-$
of the incoming relativistic angular momentum vanish: $J_{0i}^-=0$, or
\be
\left( P^\mu J_{\mu \nu} \right)^-=0 \,.
\ee
Working in this frame, and working to first order in radiation-reaction effects, we can 
separately consider (before adding them together, as indicated in Eq. \eq{decomp2}), 
the effects of radiative losses in 
energy, linear momentum and angular momentum.

 \subsection{Effect of radiative losses on the relative dynamics}

If we first neglect the linear-momentum loss (i.e. the overall recoil of the two-body system), we can set $ {\bf P}$ to zero
in Eq. \eq{Htot2}, and consider the effect of radiative losses
 on the relative dynamics, i.e., on the dynamics described by the variables ${\bf r}, {\bf p}$
in the Hamiltonian \eq{Htot2}. Let us give here a new derivation of  the effect of radiative losses
on the relative scattering angle.

Radiation damping can be described by adding to the {\it relative} Hamilton equations
a {\it relative} radiation reaction force ${\mathcal F}^{\rm rr}$ in the evolution equation for ${\bf p}$ (as done in the EOB formalism \cite{Buonanno:2000ef}),
say
\bea \label{releqs}
\dot {\bf r} &=& + \frac{\partial H_{\rm tot}}{\partial {\bf p}}, \nonumber\\
\dot {\bf p} &=& - \frac{\partial H_{\rm tot}}{\partial {\bf r}} + {\mathcal F}^{\rm rr}. 
\eea
After the neglect of $ {\bf P} \approx {\bf P}^-=0 $ (in the incoming c.m. frame), $H_{\rm tot}({\bf r}, {\bf p}, {\bf R}, {\bf P})$
reduces to $ H_{\rm rel}({\bf r}, {\bf p})$,  as follows from Eq. \eqref{Htot2}. Writing the first set of Hamilton equations \eq{releqs} in polar coordinates  $r, \phi$ (in the
conserved\footnote{In the present spinless case, ${\mathcal F}^{\rm rr}$ lies in the initial ${\mathbf r}$-${\mathbf p}$ plane.} plane of motion), say
\be
{\bf r}= r  (\cos \phi \, {\bf e}_x + \sin \phi \, {\bf e}_y)\,,
\ee
yield equations for $\dot r$ and $\dot \phi$ in terms of $r, p_r$ and $p_\phi$, where $p_\phi$ is equal to the magnitude (and the
$z$ component) of the relative angular momentum ${\bf r} \times {\bf p}$. Eliminating the time by
considering the ratio $\frac{\dot \phi}{\dot r}= \frac{d \phi}{d r}$
and solving the  energy equation $E_{\rm rel}=  H_{\rm rel}(r, p_r, p_\phi)$ in $p_r$ yields an equation for the trajectory
of the form
\be \label{phivsr}
 \frac{d \phi}{d r}= \Phi[r, E_{\rm rel}, p_\phi]= \Phi[r, E_{\rm rel}, J_{\rm rel}],
\ee
where we denoted by $ J_{\rm rel}$ the magnitude of the relative angular momentum ${\mathbf  J}_{\rm rel}= {\bf r} \times {\bf p}$.
A crucial feature of  Eq. \eq{phivsr} is that the quantities  $ E_{\rm rel}, J_{\rm rel}$ entering the right-hand side (rhs) are {\it not}
conserved in presence of the (relative) radiation-reaction force ${\mathcal F}^{\rm rr}$ in Eqs. \eq{releqs}, but that they
adiabatically evolve during the scattering to interpolate between their incoming values $ E_{\rm rel}^-, J_{\rm rel}^-$
and their outgoing ones $ E_{\rm rel}^+, J_{\rm rel}^+$. In other words, Eq. \eq{phivsr} is a technically precise way of proving that
the relative variables ${\bf r}, {\bf p}$ feel the external radiative losses only through the
adiabatic variation of the two ``constants of motion" they depend upon in
the conservative case:  the relative energy $ E_{\rm rel}= H_{\rm rel}({\bf r}, {\bf p})$ and the relative angular momentum 
$J_{\rm rel}^2= \left( {\bf r} \times {\bf p} \right)^2$, both quantities being considered in the (incoming) c.m. frame.
 At this stage, one should also remember that Eq. \eq{JrelvsPL} shows that $J_{\rm rel}^2$ is equal to the square of the
 Pauli-Lubanski 4-vector $S_\mu$.
 
The varying-constant trajectory equation \eq{phivsr} yields a simple proof of the  result obtained in Ref. \cite{Bini:2012ji}
for the effect of radiation reaction on the (relative) scattering angle. Indeed, if we expand the 
right-hand side (rhs) of \eq{phivsr} to
first order in the variation of $ E_{\rm rel}$ and $J_{\rm rel}$, either around their initial values, $ E_{\rm rel}^-, J_{\rm rel}^-$,
or, more conveniently around their values at the moment of closest approach, say
\bea
E_{\rm rel}(r) &=& E_{\rm rel}(r_{\rm min})+ \delta E(r), \nonumber\\
J_{\rm rel}(r) &=& J_{\rm rel}(r_{\rm min})+ \delta J(r) \,,
\eea
we get (to first order in the radiative variations $\delta E(r)$ and $\delta J(r)$) a trajectory equation of the form
\bea \label{phivsr2}
 \frac{d \phi}{d r}&=&  \Phi[r, E_{\rm rel}(r_{\rm min}), J_{\rm rel}(r_{\rm min})]\nonumber\\
&&+ \frac{\partial \Phi}{\partial E_{\rm rel}}\delta E(r)+ \frac{\partial \Phi}{\partial J_{\rm rel}}\delta J(r)\,.
\eea
Here, we formally expressed the time evolution in terms of the (relative) radial distance $r$, with the usual understanding that 
$r$ decreases ($p_r<0$) during the first (approaching) half of the scattering, while it increases ($p_r>0$) during the
second (receding) half. [As the (single-valued) function $\phi(t)$ is continuously increasing with time,
the (multi-valued) function $\phi(r)$ should decrease (with, correspondingly, $ \Phi[r] <0$) during the approach and 
increase (with, correspondingly, $ \Phi[r] >0$) during the outgoing motion.]
The advantage of having expressed Eq. \eq{phivsr2} in terms of quantities 
referring to the closest-approach is that we can use the fact that  the conservative motion 
is {\it time-symmetric} with respect to the closest approach,  
namely the solution $\phi^{\rm cons}(r, E,J)$  of the conservative scattering equation,
\be \label{phivsrcons}
 \frac{d \phi^{\rm cons}(r,E,J)}{d r}=  \Phi[r, E, J]\,,
\ee
defines a curve in the $(r, \phi)$ plane which is symmetric with respect to the line  $\phi= \phi_{\rm min}$,
corresponding to the point of closest approach.

 The main physical observable extracted from the latter curve
is the  total conservative scattering angle defined as
\be
\chi^{\rm cons}(E,J) \equiv \phi^{\rm cons +} - \phi^{\rm cons -} - \pi ,
\ee
where $ \phi^{\rm cons -}(E,J)$ and  $\phi^{\rm cons +}(E,J)$ denote the incoming and outgoing asymptotic
values of the solution $\phi^{\rm cons}(r,E,J)$ of Eq. \eq{phivsrcons}.

The time symmetry of the conservative dynamics has two consequences when solving the (first-order) radiation-reacted
scattering equation \eq{phivsr2}:  the first term on the rhs, namely $\Phi[r, E_{\rm rel}(r_{\rm min}), J_{\rm rel}(r_{\rm min})]$, yields a time-symmetric solution; while the two terms on the second line
yield  a correction to the scattering angle which, when multiplied by $dr= \dot r dt$ is {\it time-antisymmetric}\footnote{The
latter time-antisymmetry (around the closest approach) of the integrand giving the radiation-reaction contribution to the scattering angle
 is a direct consequence of the assumed global time-antisymmetry of the radiation-reaction force.}
with respect to the moment of closest approach. As a consequence, the integrated effect of the two terms 
$\frac{\partial \Phi}{\partial E_{\rm rel}}\delta E(r)+ \frac{\partial \Phi}{\partial J_{\rm rel}}\delta J(r)$ in Eq. \eq{phivsr2}
is zero. The final result is then that, to first order in radiation reaction, the relative scattering angle is given by
\be
\chi^{\rm rel}= \chi^{\rm cons}(E_{\rm rel}(r_{\rm min}), J_{\rm rel}(r_{\rm min}))\,.
\ee
Using again the time symmetry between the two halves of the scattering, one can further replace the  midpoint values 
$E_{\rm rel}(r_{\rm min}), J_{\rm rel}(r_{\rm min})$ of the slowly varying constants by the averages of the
corresponding  slowly varying constants: $E_{\rm rel}(r_{\rm min})\approx \frac12 (E_{\rm rel}^- + E_{\rm rel}^+)$,
and $J_{\rm rel}(r_{\rm min})\approx \frac12 (J_{\rm rel}^- + J_{\rm rel}^+)$. Another form is obtained by Taylor expanding
around the incoming values, with the (first-order-accurate) result
\be \label{chirelwithrr}
\chi^{\rm rel}= \chi^{\rm cons}(E_{\rm rel}^-, J_{\rm rel}^-) + \delta^{\rm rr} \chi^{\rm rel}\,,
\ee
where
 \be \label{deltarrchirel}
 \delta^{\rm rr} \chi^{\rm rel}= \frac12 \frac{\partial \chi^{\rm cons} }{\partial E_{\rm rel}^-} \delta^{\rm rr} E_{\rm rel} + \frac12 \frac{\partial \chi^{\rm cons} }{\partial J_{\rm rel}^-} \delta^{\rm rr} J_{\rm rel}\,.
 \ee
 Here
 \bea \label{ErrJrr}
 \delta^{\rm rr} E_{\rm rel} &=& E_{\rm rel}^+ - E_{\rm rel}^-= - E_{\rm rad}\,,\nonumber\\
 \delta^{\rm rr} J_{\rm rel} &=& J_{\rm rel}^+ - J_{\rm rel}^- = - J_{\rm rad}\,,
 \eea
 denote the  radiation-reaction-related changes in $E_{\rm rel}$ and $J_{\rm rel}$, i.e. (using 
 $E$ and $J$ balance) minus the radiative losses of $E$ and $J$ (computed in the incoming
 c.m. system). 
 
\subsection{Effects of radiative losses on the c.m. dynamics}
 
The total linear momentum $ P^\mu=  p_1^{ \mu  }+   p_2^{ \mu  }$ of the two-body system 
will change, because of radiative losses, from its incoming value, $ P^{\mu -}$, in the infinite past, to a
different outgoing value, $ P^{\mu +}$, in the infinite future. Let us compute now the  contribution $\Delta p_{a \mu}^{\rm rr, rec}$
to the impulse caused by the overall {\it recoil}, i.e. the spatial part, $ \delta^{\rm rec} P^{\mu}$ (wrt the incoming c.m. frame) of the momentum loss
of the binary system defined by the following orthogonal split,
\be
\Delta P^{\mu} =P^{\mu +} -  P^{\mu -} = \delta^{\rm rr} E_{\rm rel} U^ -{}^{\mu}+ \delta^{\rm rec} P^{\mu}\,.
\ee
Here, $\delta^{\rm rr} E_{\rm rel}=- U_\mu^- \Delta P^{\mu}$ is the radiation-reaction-induced 
energy change (measured
in the incoming c.m. frame), while $ \delta^{\rm rec} P^{\mu}$ is orthogonal to $ U^-{}^{\mu }$.
When working in the incoming c.m. frame, so that the incoming 3-momentum ${\bf P}^-=0$,
the 4-vector $\delta^{\rm rec} P^{\mu}$  is purely spatial, and simply equal to the outgoing 3-momentum
\be
\label{Prr}
{\bf P}^+= \delta^{\rm rec} {\bf P}= - {\bf P}^{\rm rad}.
\ee

The computation of $\Delta p_{a \mu}^{\rm rr, rec}$ can again be estimated by using the method of varying constants.
Here, the constant that we vary is the total 3-momentum, ${\bf P}$, which varies (in the incoming c.m. frame)
between its initial value ${\bf P}^-=0$ and its final value  ${\bf P}^+$. 

In the two (incoming and outgoing) asymptotic regions the interactions terms in Eq. \eq{pvsp1p2} vanish.
Combining then Eq. \eq{Pvsp1p2} and Eq. \eq{pvsp1p2} allows one to relate the asymptotic values of
${\bf p}_1$ and ${\bf p}_2$ to the asymptotic values of ${\bf p}$ and ${\bf P}$. 
When working in the incoming c.m. frame (i.e. when using ${\bf P}^-=0$), this yields
\bea
&&{\bf p}_1^{-} = {\bf p}^{-}\,, \nonumber\\
&&{\bf p}_2^{-} = - {\bf p}^{-}\,, \nonumber\\
&&{\bf p}_1^{+} = {\bf p}^{+} + \frac{E_1^{+}}{(E_1+E_2){^+}} {\bf P}^{+} + O({\bf P}_{+}^2)\,,\nonumber\\
&&{\bf p}_2^{+} = -{\bf p}^{+} + \frac{E_2{^+}}{(E_1+E_2){^+}} {\bf P}^{+} + O({\bf P}_+^2)\,.\qquad
\eea
The corresponding asymptotic kinetic energies are
\bea
&& E_1^{-} \equiv  \sqrt{m_1^2  + ({\bf p}_1^-)^2}\,,\nonumber\\
&& E_2^{-} \equiv  \sqrt{m_2^2  + ({\bf p}_2^-)^2}\,,\nonumber\\
&& E_1^{+} \equiv  \sqrt{m_1^2  + ({\bf p}_1^+)^2}\,,\nonumber\\
&& E_2^{+} \equiv  \sqrt{m_2^2  + ({\bf p}_2^+)^2}\,.
\eea
By expanding the latter relations to first order in ${\bf P}^+$, one finds that  the radiation-reacted impulses
\be
\Delta p_a^{\mu }= p_a^{\mu +}- p_a^{\mu -}= (E_a^+ - E_a^-, {\bf p}_a^+- {\bf p}_a^-)\,,
\ee
read
\bea \label{dpwithrecoil}
\Delta p_1^{0} &=& \sqrt{m_1^2  + ({\bf p}^+)^2}-  \sqrt{m_1^2  + ({\bf p}^-)^2}+ \frac{{\bf p}^{+} \cdot {\bf P}^{+}}{(E_1+E_2){^+}}\,, \nonumber\\
\Delta {\bf p}_1&=& {\bf p}^{+} - {\bf p}^{-} + \frac{E_1^{+}}{(E_1+E_2){^+}} {\bf P}^{+}\,, \nonumber\\
\Delta p_2^{0} &=& \sqrt{m_2^2  + ({\bf p}^+)^2}-\sqrt{m_2^2  + ({\bf p}^-)^2}- \frac{{\bf p}^{+} \cdot {\bf P}^{+}}{(E_1+E_2){^+}}\,, \nonumber\\
\Delta {\bf p}_2&=& -({\bf p}^{+} - {\bf p}^{-}) + \frac{E_2^{+}}{(E_1+E_2){^+}} {\bf P}^{+}\,,
\eea
where the temporal component refers to the incoming c.m. time axis $e_0 =U^-$.  
 
 In these expressions the terms which do not involve ${\bf P}^{+}$ exactly correspond to the
 impulses that would be derived by neglecting the recoil and only
 considering the effect of radiative losses on the relative motion. More precisely, they would correspond
 (when separating out the relative radiation-reaction effects from the conservative contribution)
 to the sum $\Delta p_{a \mu}^{\rm cons} + \Delta p_{a \mu}^{\rm rr, rel}$ in Eq. \eq{decomp1},
 when inserting Eq. \eq{decomp2} and neglecting the last, recoil contribution.
 In other words, the terms involving ${\bf P}^{+}$  in Eqs. \eq{dpwithrecoil} describe the looked-for
 recoil contributions $\Delta p_a^{\mu \rm rec}$ to the impulses.
 
 We conclude that the explicit expressions of the recoil contributions (viewed in the
 incoming c.m. frame) read
 \bea \label{dprec}
 \Delta p_a^{0 \, \rm rec} &=& + \frac{{\bf p}_a^{+} \cdot {\bf P}^{+}}{E{^+}}, \nonumber\\
  \Delta {\bf p}_a^{ \rm rec} &=&+ \frac{E_a^{+}}{E{^+}} {\bf P}^{+}\,,
 \eea
 where  $E{^+} = (E_1+E_2){^+}$ denotes the total outgoing c.m. energy.
 Note that the recoil contributions
  can be simply interpreted as coming from
 the linearized effect of a small boost of velocity vector equal to the outgoing c.m. velocity
 (in the ingoing c.m. frame). Indeed, let us consider the recoil velocity 
 \be
 {\bf V}= \frac{{\bf P}^{+}}{E{^+}} \approx \frac{{\bf P}^{+}}{E{^-}}\,,
 \ee
 where $E{^+} = (E_1+E_2){^+}$ denotes the total outgoing c.m. energy, which can be replaced
 (as indicated in the second equation) by the incoming one when neglecting terms bilinear in
 the radiative losses $\Delta {\bf P}$ and $\Delta E= E{^+} - E{^-}$. The usual Lorentz-transformation
 formula for a 4-momentum vector, under an infinitesimal boost, reads
 \bea
 \delta^{{\bf V}} E_a &=& {\bf p}_a \cdot {\bf V}, \nonumber\\
 \delta^{{\bf V}} {\bf p}_a &=& E_a {\bf V}\,.
 \eea
 The recoil contributions Eq. \eq{dprec} are indeed obtained by applying this transformation formula to the outgoing momenta $p_a^{\mu +}$.
 
\subsection{Final results for the radiation-reacted impulses}

Let us summarize our results for the radiation-reacted impulses. 
To get fully explicit results we need to choose a vectorial basis in the plane of motion (still working in the incoming c.m. frame).
As a first basis we can use the two orthogonal (spatial) unit vectors $\hat {\bf b}$ and ${\bf n}_-$,
where $\hat {\bf b}= \frac{{\bf b}}{b}$ is along the
incoming vectorial impact parameter ${\bf b}$, and where the unit vector ${\bf n}_-$ lies 
along the direction of  ${\bf p}_1^{-}$, namely
\bea
\label{p12_min}  
{\bf p}_1^{-} &=&p^-  {\bf n}_-  \equiv  P_{\rm c.m.}^- {\bf n}_- \,,\nonumber\\
{\bf p}_2^{-} &=&-p^-  {\bf n}_-  \equiv  - P_{\rm c.m.}^- {\bf n}_-\,.
\eea
Here $p^- =P_{\rm c.m.}^-$ is linked to the incoming c.m. energy by
\bea
P_{\rm c.m.}^-&=&m_1m_2 \frac{\sqrt{(u_1^-\cdot u_2^-)^2-1}}{E_{\rm c.m.}^-}= m_1m_2 \frac{\pinf}{E_{\rm c.m.}^-}
\,,\nonumber\\
\eea
where $\pinf = \sqrt{\g^2-1} $, $\g \equiv -u_1^-\cdot u_2^-$, and
\bea
E_{\rm c.m.}^-&=&(E_1+E_2)^-= M h(\g, \nu) \equiv M \sqrt{1 + 2 \nu (\g-1)}
\,.\nonumber\\
\eea 
We note that ${\bf n}_-$ can  be expressed as a linear combination of
the incoming four velocities of the two bodies, 
\beq
\label{n_meno}
{\mathbf n}_- = \frac{m_1m_2}{P_{\rm c.m.}^- E_{\rm c.m.}^-}\left(\frac{E_2^-}{m_2}u_1^--\frac{E_1^-}{m_1}u_2^-  \right)\,,
\eeq
as follows, for example,  by subtracting  the relations
\bea
p_1^-&=& m_1u_1^-=E_1^-U +p^-{\mathbf n}_-\,,\nonumber\\
p_2^-&=&m_2u_2^-=E_2^-U -p^-{\mathbf n}_-\,.
\eea

With respect to this basis the outgoing  3-momentum ${\bf p}^+$ reads
\be \label{bfp+}
{\bf p}^+= p^+ \left(\cos \chi^{\rm rel} \, {\bf n}_ - -  \sin \chi^{\rm rel}  \, \hat {\bf b} \right)\,.
\ee
In Eq. \eq{bfp+}, one must insert 
both the radiation-reacted value of the
magnitude $ p^+$ of the outgoing relative momentum ${\bf p}^+$, and the radiation-reacted value
of the (relative-motion) scattering angle, $\chi^{\rm rel}$. In the latter radiation-reacted scattering angle
$\chi^{\rm rel}= \chi^{\rm cons} + \delta^{\rm rr} \chi^{\rm rel}$, Eq. \eq{chirelwithrr}, the radiation-damping contribution is given by Eq. \eq{deltarrchirel}.
Concerning the radiation-reacted value of the magnitude $ p^+$ of the outgoing relative momentum
it is determined by writing that the total  outgoing relative energy 
$E_{\rm rel}^+=\sqrt{m_1^2  + ({\bf p}^{+})^2}+ \sqrt{m_2^2  + ({\bf p}^{+})^2}$
is equal to $E_{\rm rel}^- + \delta^{\rm rr} E_{\rm rel}= E_{\rm rel}^- - E_{\rm rad}$,
where $E_{\rm rel}^- =  \sqrt{m_1^2  + ({\bf p}^{-})^2}+ \sqrt{m_2^2  + ({\bf p}^{-})^2}$.
Computing $\Delta p \equiv p^+- p^-$, to first order, from the latter condition yields
\bea
\label{mod_p_plus_min_mod_p_min}
 \Delta p  &\equiv& |{\bf p}^{+}| - |{\bf p}^{-}|= \frac{E_{ 1\, \rm rel}^{-}E_{ 2\, \rm rel}^{-}}{|{\bf p}^{-}|E_{\rm rel}^- }  \delta^{\rm rr} E_{\rm c.m.}\nonumber\\
&&+O((\delta^{\rm rr} E_{\rm c.m.})^2) \,.
\eea

Inserting all those results in the above expressions for the impulses, Eqs. \eqref{dpwithrecoil}, finally yields explicit expressions
for the impulses as the sum of three contributions: a conservative part, a relative-motion radiation-reaction
part (computed from $E_{\rm rad}$ and $J_{\rm rad}$, Eq. \eqref{deltarrchirel}) and a recoil radiation-reaction part
(computed from $\delta^{\rm rec}P^\mu$, Eq.  \eqref{dprec}), namely
\be \label{decompconsrr}
\Delta p_{a \mu}= \Delta p_{a \mu}^{\rm cons}(E_{\rm rel}^-, J_{\rm rel}^-)+ \Delta p_{a \mu}^{\rm rr, rel}+ \Delta p_{a \mu}^{\rm rr,  rec}\,.
\ee
Following Eqs. \eq{decompbu1u2} and \eq{sum_contrib}, the decomposition
of  $\Delta p_{a \mu}$ along the 4-vectors ${\hat b}^\mu \equiv b^\mu/b$, $ u_{1 \mu}^-$, 
and $ u_{2 \mu}^-$ yields six scalar coefficients $c_b^a, c_{u_1}^a, c_{u_2}^a$ ($a=1,2$).
Each scalar coefficient is readily computed as\footnote{
Spatial vectors, differently from spacetime vectors, are usually denoted by boldface symbols.
Sometimes it is convenient to represent a spatial vector with the same symbol as a spacetime one  (but not in boldface).}
\bea
\label{gen_expr_ci}
 c_b^{a, X}  &=&  \hat { b}  \cdot \Delta p_a^X \,,\nonumber\\
 c_{u_1}^{a, X} &=& \frac{u_1^- - \g \, u_2^-}{\g^2-1} \cdot \Delta p_a^X \,, \nonumber\\
 c_{u_2}^{a, X} &=& \frac{u_2^- - \g \, u_1^-}{\g^2-1} \cdot \Delta p_a^X \,, 
\eea 
with $X=$ cons, rr rel, rr rec, the dot product is the Minkowski one, and we recall that $\g \equiv - u_1^-  \cdot u_2^-$.

All these coefficients, conservative and radiation-reaction,  are listed in Table \ref{tab:impulse_coeffs1}.
In this table $P^+_b$,  and $P^+_n$ denote the components of ${\bf P}^+$ along the 
$({\bf b}, {\bf n}_-)$ basis  in the plane of motion, i.e.,
\be
{\bf P}^+= P^+_b \hat {\bf b}+ P^+_n {\bf n}_-\,.
\ee


\begin{table*}  
\caption{\label{tab:impulse_coeffs1} Conservative, radiation-reaction-relative, and radiation-reaction-recoil contributions to the impulse coefficients $c_b^1$, $c_{u_1}^{1}$, $c_{u_2}^{1}$ and $\widehat c_{u_1}^{1}$. 
}
\begin{ruledtabular}
\begin{tabular}{lll}
      & $c_b^{1\rm cons}$ & $- P_{\rm c.m.}^- \sin \chi^{\rm cons}$\\
 cons & $c_{u_1}^{1\rm cons}$ & $(m_1 \g + m_2) \frac{m_1 m_2}{E_{\rm c.m.}^2}( \cos \chi^{\rm cons}-1)$\\
      & $ c_{u_2}^{1\rm cons}$ & $- (m_2 \g + m_1) \frac{m_1 m_2}{E_{\rm c.m.}^2}( \cos \chi^{\rm cons}-1)$\\
			& $\widehat c_{u_1}^{1\rm cons}$ & $-\frac{(P_{\rm c.m.}^-)^2}{m_1}( \cos \chi^{\rm cons}-1)$\\
\hline
      & $c_b^{1\rm rr\,  rel}$ &$-\frac{E_1^-E_2^-}{P_{\rm c.m.}^- E_{\rm c.m.}^-}\sin \chi^{\rm cons} \delta^{\rm rr}E_{\rm rel}-P_{\rm c.m.}^- \cos \chi^{\rm cons}\delta^{\rm rr}\chi^{\rm rel}$ \\
 rr\,, rel & $c_{u_1}^{1\rm rr\,  rel}$ & $ \frac{m_1E_2^-}{(E_{\rm c.m.}^-)^2}\left(\frac{E_1^-E_2^-}{(P_{\rm c.m.}^-)^2}\cos \chi^{\rm cons}+1\right) \delta^{\rm rr}E_{\rm rel}- \frac{m_1E_2^-}{E_{\rm c.m.}^-}\sin \chi^{\rm cons}\delta^{\rm rr}\chi^{\rm rel}$ \\
      & $ c_{u_2}^{1\rm rr\,  rel}$ & $-\frac{m_2E_2^-}{(E_{\rm c.m.}^-)^2}\left(\frac{(E_1^-)^2 }{(P_{\rm c.m.}^-)^2}\cos \chi^{\rm cons}-1\right)\delta^{\rm rr}E_{\rm rel}+\frac{m_2E_1^-}{E_{\rm c.m.}^-}\sin \chi^{\rm cons}\delta^{\rm rr}\chi^{\rm rel}$\\
			& $\widehat c_{u_1}^{1\rm rr\,  rel}$ & $-\frac{E_1^-E_2^-}{m_1E_{\rm c.m.}^-} (\cos \chi^{\rm cons}-1) \delta^{\rm rr}E_{\rm rel}+ \frac{(P_{\rm c.m.}^-)^2}{m_1}\sin \chi^{\rm cons}\delta^{\rm rr}\chi^{\rm rel}$ \\

\hline
      & $c_b^{1\rm rr\,  rec}$ & $\frac{E_1^-}{E_{\rm c.m.}^-} P^+_b$ \\
			&  & $=-\frac{E_1^-}{E_{\rm c.m.}^-} \sin \frac{\chi^{\rm cons}}{2}  P^+_y$ \\
 rr\,, rec & $c_{u_1}^{1\rm rr\,  rec}$ & $\frac{m_1}{(E_{\rm c.m.}^-)^3}\left[
-m_1m_2\sqrt{\gamma^2-1} \sin \chi^{\rm cons}P^+_b+\left[\gamma (E_{\rm c.m.}^-)^2+m_1m_2 \left(1+(\gamma^2-1)\cos \chi^{\rm cons}\right) \right]\frac{P^+_n}{\sqrt{\gamma^2-1}}
\right]$\\
	&  & $=\frac{m_1\gamma}{E_{\rm c.m.}^-\sqrt{\gamma^2-1}} \cos \frac{\chi^{\rm cons}}{2}  P^+_y$ \\
      & $ c_{u_2}^{1\rm rr\,  rec}$ & $\frac{m_1 }{E_{\rm c.m.}^-\sqrt{\gamma^2-1}}\cos \frac{\chi^{\rm cons}}{2}\left( \sin \frac{\chi^{\rm cons}}{2} P^+_b -\cos \frac{\chi^{\rm cons}}{2} P^+_n\right)$ \\
				&  & $=-\frac{m_1}{E_{\rm c.m.}^-\sqrt{\gamma^2-1}} \cos \frac{\chi^{\rm cons}}{2}  P^+_y$ \\
			& $\widehat c_{u_1}^{1\rm rr\,  rec}$ & 0\\
 \end{tabular}
\end{ruledtabular}
\end{table*}

When inspecting the  expressions of $ c_{u_1}^{1, \rm rr, rel}$ and $ c_{u_2}^{1, \rm rr, rec}$,
 one notices the presence of denominators involving powers of $P_{\rm c.m.}^-$ or of $\pinf^2=\g^2-1$. 
 These factors imply some loss of PN accuracy. It is useful to
minimize the appearance of such small denominators by computing the impulse
coefficients along a slightly different basis. For instance,  we can decompose (when $a=1$),
$\Delta p_1^X$ along the new basis
\beq \label{newbasis}
 \hat { b}\,,\qquad  u_1^- \,,\qquad u_{2\perp 1} \equiv u_2^-- \g u_1^-\,,
\eeq
where $u_2^-$ is replaced by ${ {u}}_{2\perp 1}=u_2^-- \g u_1^-=\Pi_{u_1}(u_2)$, with $\Pi_{u}(v) \equiv v+ (v \cdot u) u$ denoting the projection of the vector $v$ orthogonally to the (unit, timelike) vector $u$.
[This basis is orthogonal, but the third vector is not a unit vector.] Mutatis mutandis, $\Delta p_2^X$ is conveniently
decomposed along $ \hat { b},  u_2^- $ and $u_{1\perp 2}=u_1^-- \g u_2^-=\Pi_{u_2}(u_1)$.
The corresponding expansion coefficients differ from the previous ones only for the second one.
For instance, we have  
\be
\label{Delta_p1_con_u2perp1}
\Delta p_1^X=c_b^{1,X} \hat { b} +\widehat c_{u_1}^{1,X} u_1^{-} +  c_{u_2}^{1,X} u_{2\perp 1}\,,
\ee
where $c_b^{1,X} $ and $ c_{u_2}^{1,X}$ are the same as before, and where
\bea
\widehat c_{u_1}^{1,X}&=& -u_1^- \cdot \Delta p_1^X=c_{u_1}^{1,X}+\gamma c_{u_2}^{1,X}\,.
\eea

In addition, the time-symmetry of the conservative hyperbolic motion around closest approach
implies that the radiated  linear momentum (evaluated in a time-symmetric way\footnote{\label{foot_bisec}When working to first-order in radiation reaction one can consistently evaluate the gravitational wave (GW) radiation absorbed and then emitted in a  time-symmetric way by the conservative dynamics, with equal amounts of incoming radiation recorded on past null infinity and of outgoing radiation recorded on future null infinity. See Appendices \ref{App_tail_E}, \ref{App_tail_J}  and \ref{App_tail_P} below for further discussion.} along the conservative
hyperbolic motion) is aligned along the special direction ${\mathbf e}_y$  in the plane of motion,
defined as the bisector between the incoming velocity and the outgoing one (taken in the conservative dynamics). 
The unit vector  ${\mathbf e}_y$ is part of a basis   ${\mathbf e}_x$,  ${\mathbf e}_y$
in the plane of motion defined as
\bea
{\mathbf e}_x&=& \cos \frac{\chi_{\rm cons}}{2}\hat {\mathbf b} +\sin \frac{\chi_{\rm cons}}{2}{\mathbf n}_-\,,\nonumber\\
{\mathbf e}_y&=& -\sin \frac{\chi_{\rm cons}}{2}\hat {\mathbf b} +\cos \frac{\chi_{\rm cons}}{2}{\mathbf n}_-\,,
\eea
with inverse relations
\bea
\hat {\mathbf b}&=& \cos \frac{\chi_{\rm cons}}{2}{\mathbf e}_x -\sin \frac{\chi_{\rm cons}}{2}{\mathbf e}_y\,,\nonumber\\
{\mathbf n}_-&=& \sin \frac{\chi_{\rm cons}}{2}{\mathbf e}_x +\cos \frac{\chi_{\rm cons}}{2}{\mathbf e}_y\,.
\eea

The collinearity of the recoil with the ${\mathbf e}_y$ direction, i.e., the fact that one can write
 ${\bf P}^+= P^+_y {\mathbf e}_y$ yields  the links
 \bea \label{Pylinks}
  P^+_b &=& - \sin \frac{\chi_{\rm cons}}{2}  P^+_y\,, \nonumber\\
   P^+_n &=& + \cos \frac{\chi_{\rm cons}}{2}  P^+_y\,.
 \eea
Inserting the links  \eq{Pylinks} in the expressions of $ c_{u_1}^{1, \rm rr, rec}$ and 
$ c_{u_2}^{1, \rm rr, rec}$, and using the new basis Eq. \eq{newbasis}, leads to simplified expressions
for the radiation-reaction impulse coefficients (see again Table \ref{tab:impulse_coeffs1}).

\subsection{Mass polynomiality of the impulses}

In the following sections we are going to use the present knowledge on radiative losses
to explicitly compute the impulse coefficients at successive  PM orders. PN related information about these coefficients are postponed to Appendix \ref{PNAppendix}. 
It will be especially convenient to express the  impulse coefficients
 in terms of the impact parameter $b$, of the
 relative Lorentz factor $\g$,  and of the two masses, $m_1, m_2$. Indeed, the argument given 
 just below Eq. (2.9) of Ref. \cite{Damour:2019lcq} is valid in the general non-conservative case
 considered here and shows that the 4-vectorial impulses 
  $ \Delta p^\mu_a$, when decomposed on the basis $b^\mu$, $u_{ 1 -}^\mu$, $u_{ 2 -}^\mu$, have the general structure (say for $a=1$)
 \be \label{deltapmu2}
\Delta p_{1 \mu}= - 2G m_1 m_2 \frac{2 \g^2-1}{\sqrt{\g^2-1}} \frac{b_\mu}{b^2} +  \sum_{n\geq 2} \Delta p_{1 \mu}^{n \rm PM}\,.
\ee 
 Here each term $ \Delta p_{1 \mu}^{n \rm PM}$
is a combination of the three vectors $b^\mu/b$, $u_{1 -}^{\mu}$ and 
$u_{2 -}^{\mu}$, with coefficients that are, at each order in $G$,  {\it homogeneous polynomials} in $m_1$ and $m_2$, containing the product $m_1 m_2$ as an overall factor. In other words, it has the structure
\bea \label{deltapmugennPM}
 \Delta p_{1 \mu}^{n \rm PM} &\sim& \frac{G m_1 m_2}{b^n} \left[  (Gm_1)^{n-1} \right.\nonumber\\
&+&\left.  (Gm_1)^{n-2} G m_2+ \cdots + (Gm_2)^{n-1}\right]\,,\qquad\quad
\eea  
 where each term is a combination of the three vectors $b^\mu/b$, $u_{1 -}^{\mu}$ and 
$u_{2 -}^{\mu}$, with coefficients that are functions of $\g$.

This implies that the impulse coefficients $c_{b}^{a\, \rm nPM}, c_{u_1}^{a\, \rm nPM},
 c_{u_2}^{a\, \rm nPM}$ are (when expressed in terms of $b$ and $\g$)
 polynomials in $m_1$ and $m_2$ of the form 
 $G m_1 m_2 \left[  (Gm_1)^{n-1} + (Gm_1)^{n-2} G m_2+ \cdots + (Gm_2)^{n-1}\right].$
 
 The polynomiality in the masses of the impulses was shown to imply, {\it in the conservative case}, a 
 corresponding polynomiality in $\nu =m_1m_2/(m_1+m_2)^2$ of the scattering angle \cite{Damour:2019lcq}, which played an important role in determining the structure of the 5PN and 6PN
 Hamiltonians \cite{Bini:2019nra,Bini:2020nsb,Bini:2020hmy,Bini:2020rzn}. 
 We shall see below that the simple polynomiality rule satisfied by the conservative scattering angle
 is violated when considering the radiation-reacted {\it relative} scattering angle.  
 However, we shall explicitly check that the more general property of
 mass polynomiality is restored when adding the effect of recoil, i.e. considering the
 total, radiation-reacted impulses
 \be \label{decompconsrr2}
\Delta p_{a \mu}(u_1^{-}, u_2^{-}, b)= \Delta p_{a \mu}^{\rm cons}(u_1^{-}, u_2^{-}, b)+ \Delta p_{a \mu}^{\rm rr, tot}(u_1^{-}, u_2^{-}, b)\,,
\ee
with
 \be  \label{decomprr2}
\Delta p_{a \mu}^{\rm rr, tot}(u_1^{-}, u_2^{-}, b)= \Delta p_{a \mu}^{\rm rr, rel}+ \Delta p_{a \mu}^{\rm rr, rec} + O({\cal F}_{\rm rr}^2)\,.
\ee
As a consequence, the expansion coefficients $c_X^{a\, \rm rr, tot}$ of the 
total radiation reaction contributions (including relative and recoil effects) to the impulses,
\beq
c_X^{a\, \rm rr, tot}=c_X^{a\, \rm rr, rel} +  c_X^{a\, \rm rr, rec}\,,\quad X=b,u_1,u_2\,,
\eeq
must satisfy (as we shall check in the cases where they are known) the  mass-polynomiality of the impulse coefficients. 

\section{Scattering at orders $O(G)$ and $O(G^2)$}

 For completeness, let us recall that the scattering at orders  $O(G)$ and $O(G^2)$ is 
 conservative\footnote{Though the radiation-reaction force starts at order $O(G^2)$ and
 causes a $O(G^2)$ angular momentum loss \cite{Damour:1981bh,Damour:2020tta}, the
 total linear momentum of the system is conserved at order $O(G^2)$ \cite{Westpfahl:1985}.
 }. Therefore, the knowledge of the scattering angle suffices to determine
 all the impulse coefficients. Inserting in the general expressions listed in Table \ref{tab:impulse_coeffs1} the
 2PM-accurate scattering angle~\cite{Westpfahl:1985} (given in Appendix \ref{PMAppendix} for convenience)
yields PM-exact values for these impulse coefficients. 
As explained above, it is useful to express the impulse coefficients 
 in terms of the impact parameter $b$, of the
 relative Lorentz factor $\g$ (or equivalently of $\pinf = \sqrt{\g^2-1}$),  and of the two masses, $m_1, m_2$. This will allow us
 to exhibit their polynomial structure in the masses.
  
 As exhibited in Eq. \eq{deltapmu2}, the only non-zero coefficients at 1PM are the $c_b^a$ ones, namely
 \be
 c_b^{1, \rm 1PM}= \frac{G m_1 m_2}{b} \left(-\frac{2}{p_\infty} -4 p_\infty\right)\,,
 \ee
 and $ c_b^{2, \rm 1PM} = -c_b^{1, \rm 1PM}$. We have then
\bea
\Delta p_1^{1, \rm 1PM}=-\frac{G m_1 m_2}{b} \frac{2}{p_\infty}\left(1 +2 p_\infty^2\right)\hat b\,.\qquad
\eea
 
 At the 2PM order, we have
 \be
  c_b^{1, \rm 2PM}=  \frac{G m_1 m_2}{b}  \frac{G M}{b} \pi \left(-\frac{3}{p_\infty} -\frac{15}{4}p_\infty\right)\,,
 \ee
 and
 \bea
\label{hat_cis}
c_{u_1}^{1\, \rm 2PM} &=& -\frac{G m_1 m_2}{b}  \frac{2 G (m_1\gamma +m_2)}{b} \left(\frac{1}{p_\infty^2}+2 \right)^2\,,\nonumber\\
c_{u_2}^{1\, \rm 2PM} &=& +\frac{G m_1 m_2}{b}  \frac{2 G (m_2\gamma +m_1)}{b} \left(\frac{1}{p_\infty^2}+2 \right)^2\,,  \nonumber\\
\widehat c_{u_1}^{1, \rm 2PM}&=& + \frac{G m_1 m_2}{b}  \frac{2 G m_2}{b}   p_\infty^2 \left( \frac{1}{p_\infty^2}+2 \right)^2\,.
\eea
 The polynomiality in the masses of these coefficients is clearly exhibited in these expressions.
The complete expression for $\Delta p_1^{1, \rm 2PM}$ then reads

\bea
\Delta p_1^{1, \rm 2PM}&=&-\frac{G^2m_1m_2}{b^2}\left[
(m_1+m_2)\frac{3\pi}{4p_\infty}(4+5p_\infty^2)\hat b\right.\nonumber\\
&+&
2\frac{(1+2p_\infty^2)^2}{p_\infty^4}(m_2 (u_1-\g u_2)
\nonumber\\
&&\left.- m_1 (u_2-\g u_1))
\right]\,.
\nonumber\\
\eea

\section{Classical scattering at order $O(G^3)$}

The scattering at order $O(G^3)$ has been the topic of several recent works. First, the
conservative part of the 3PM scattering has been computed in Refs.  \cite{Bern:2019nnu,Bern:2019crd,Kalin:2020fhe}.
Second, the radiation-reaction contribution to the relative scattering angle has been computed
both in supergravity \cite{DiVecchia:2020ymx}, and in General Relativity \cite{Damour:2020tta,DiVecchia:2021ndb}.
Ref. \cite{Herrmann:2021tct} has completed the determination of the 3PM scattering
by computing the full impulses $\Delta p_a^\mu$ from the $O(G^3)$ quantum scattering amplitude, using the
approach of Ref. \cite{Kosower:2018adc}. The result of Ref.  \cite{Herrmann:2021tct} was recently
confirmed in Ref. \cite{Bjerrum-Bohr:2021din}.
Let us show here how our purely classical approach
to radiation-reacted scattering leads to a quite simple rederivation of the quantum-derived
result of \cite{Herrmann:2021tct}.

It has been known for a long time \cite{Blanchet:1987wq} that nonlocal-in-time effects, related to radiation-graviton (a.k.a. soft-graviton) exchange, start to arise at
the $\frac{G^4}{c^8}$ level, i.e. at the 4PM and 4PN level. As a consequence, the  potential-graviton contribution to conservative
scattering (which has been derived in Refs. \cite{Bern:2019nnu,Kalin:2020fhe}) fully describes  the {\it conservative} part of the scattering
at the third PM order, $O(G^3)$. This yields a 3PM conservative impulse 
contributing to the conservative part of the scattering angle.
To determine the additional $O(G^3)$ {\it radiation-reaction} contributions to scattering,
one can start from the general expressions of the impulse coefficients listed in Table \ref{tab:impulse_coeffs1}
above, and use the facts that the radiation-reaction angular momentum loss (entering Eq. \eq{deltarrchirel})
starts at order $O(G^2)$ \cite{Damour:1981bh}, while the radiation-reaction linear momentum loss starts at order  $O(G^3)$
\cite{Kovacs:1977uw,Kovacs:1978eu}. This shows  that there exists, at $O(G^3)$, both  {\it relative-motion} radiation-reaction
contributions (coming either from the additional contribution \eq{deltarrchirel} to the relative scattering angle, or from
energy-loss effects), and {\it recoil} contributions. These radiative contributions enter differently the various components 
 $c_b^{a}, c_{u_1}^{a}, c_{u_2}^{a}$ of the decomposition of $\Delta p_{a \mu}^{\rm rr}$ along the $b, u_1^-, u_2^-$ basis.
 
 Let us start by discussing the $b$ component $c_b^{1, \rm 3PM}$. It reads
 \be
 c_b^{1 \, \rm tot, 3PM}= c_b^{1 \,\rm cons \, 3PM}+ c_b^{1 \,\rm rr \, 3PM}\,,
 \ee
 where the conservative contribution is
\bea
c_b^{1 \,\rm cons \, 3PM}&=&- \frac{\mu \sqrt{\gamma^2-1}}{h}  \frac{[2\chi_3^{\rm cons}-\frac43 (\chi_1^{\rm cons})^3]}{j^3}\nonumber\\
&=& - \frac{\mu  h^2}{(\gamma^2-1)}\left[2\chi_3^{\rm cons}-\frac43 (\chi_1^{\rm cons})^3\right] \left(\frac{G M }{b }\right)^3\,,\nonumber\\
\eea
with $\chi^{\rm cons}$ expanded as
\beq
\label{chi_cons_PM_exp}
\frac12 \chi^{\rm cons}= \sum_{n\geq 1}\frac{  \chi_n^{\rm cons}(\g,\nu)}{j^n} \,,
\eeq
with coefficients $\chi_1^{\rm cons}$,  $\chi_2^{\rm cons}$  and  $\chi_3^{\rm cons}$ given in Eq. \eqref{chi3PMcons},
and where the radiation-reaction contribution only comes from the  effect $ \delta^{\rm rr} \chi^{\rm rel}$ on the relative scattering angle. Moreover, at the $G^3$ order $ \delta^{\rm rr} \chi^{\rm rel}$ is entirely
determined by the angular-momentum-loss effect in Eq. \eq{deltarrchirel}. Namely 
\beq
[\delta^{\rm rr} \chi^{\rm rel}]^{\rm 3PM}=\frac12 \chi_1^{\rm  cons\, 1PM}\left[\frac{J^{\rm rad}}{J}\right]^{\rm 2PM}\,.
\eeq
Here,
\be
 \chi_1^{\rm  cons\, 1PM}=
\frac{2 (2 \g^2-1)}{(\g^2-1)}\frac{GMh}{ b}
= \frac{2 (2 \g^2-1)}{\sqrt{\g^2-1} }\,\frac{1}{j}\,,
\ee
while the 2PM-accurate fractional angular-momentum loss was found in Ref. \cite{Damour:2020tta} to be
\be \label{JradbyJ}
\left[\frac{J^{\rm rad}}{J}\right]^{\rm 2PM}= \frac{2 (2 \g^2-1)}{\sqrt{\g^2-1}} \frac{G^2 m_1 m_2 }{b^2} {\cI}(v) \,,
\ee
where, denoting $ v \equiv  \sqrt{1- \frac1{\g^2}}$,
\be
\label{cal_I_di_v_def}
{\cI}(v) = -\frac{16}{3}+ \frac{2}{v^2}+ \frac{2(3 v^2-1)}{v^3} {\rm arctanh}(v)\,.
\ee
Inserting these results in the general expressions listed in Table \ref{tab:impulse_coeffs1} yields the explicit
result
\be
c_b^{1 \,\rm rr \, rel \, 3PM} =- \frac{G^3  m_1^2 m_2^2 }{ b^3}    \frac{2 (2 \g^2-1)^2}{(\g^2-1)}\,    {\cI}(v)\,.
\ee
The beginning of the PN expansion of $c_b^{1 \,\rm rr \, rel \, 3PM}$ reads
\be
c_b^{1 \,\rm rr \, rel\, 3PM} =- \frac{G^3  m_1^2 m_2^2 }{ b^3}  \left( \frac{16}{5} +\frac{80}{7} p_\infty^2+\frac{512}{63} p_\infty^4 + \cdots \right) \,,
\ee

Concerning  $ c_{u_1}^{1}$, $c_{u_2}^{1}$ the  conservative part is given by
\bea
c_{u_1}^{1 \,\rm cons \, 3PM}&=& -\mu \left(\frac{GM}{b}\right)^3 \frac{4(m_1\gamma+m_2)h \chi_2^{\rm cons}\chi_1^{\rm cons}}{M (\gamma^2-1)^{3/2}}\,, \nonumber\\
c_{u_2}^{1 \,\rm cons \, 3PM}&=& \mu \left(\frac{GM}{b}\right)^3 \frac{4(m_2\gamma+m_1)h \chi_2^{\rm cons}\chi_1^{\rm cons}}{M (\gamma^2-1)^{3/2}} \,.\nonumber\\
\eea
Contrary to  $ c_{b}^{a}$, which received radiation-reaction contributions only from 
$ \delta^{\rm rr} \chi_{\rm rel}$, the two other coefficients, $ c_{u_1}^{a}, c_{u_2}^{a}$, of the decomposition of $\Delta p_{a \mu}^{\rm rr}$   receive contributions both from  the relative-motion
impulse, $\Delta p_{a \mu}^{\rm rr, rel}$, and from the recoil impulse,
$\Delta p_{a \mu}^{\rm rr, rec}$. Expressing them in terms of the c.m.-frame radiated\footnote{Remember
that the radiated quantities are opposite to the corresponding changes in the two-body quantities used
in Table \ref{tab:impulse_coeffs1} above: e.g., $\delta^{\rm rr}E_{\rm rel}\big|^{\rm 3PM}=-E_{\rm c.m.}^{\rm rad  \, 3PM}$.}
energy or momentum, we have 
 \bea
\label{cu1_u2_3PM}
 &&c_{u_1}^{1 \,\rm rr \, rel \, 3PM}= -\frac{(m_1\gamma+ m_2)\gamma }{(\gamma^2-1) E_{\rm c.m.}}E_{\rm c.m.}^{\rm rad  \, 3PM}\,,\nonumber\\
\nonumber\\
&& c_{u_2}^{1 \,\rm rr \, rel \, 3PM}= -\frac{1}{\gamma}c_{u_1}^{1\, \rm rr \, rel\, 3PM}\,,
 \eea
and
 \bea
\label{eq:cu1_3PM}
c_{u_1}^{1 \,\rm rr \, rec \, 3PM}&=& -  \frac{m_1\gamma}{\sqrt{\gamma^2-1}E_{\rm c.m.}} 
{P}_{n \, \rm c.m.}^{\rm rad \, 3PM} \,,\nonumber\\
\nonumber\\
c_{u_2}^{1 \,\rm rr \, rec \, 3PM}&=& -\frac{1}{\gamma}c_{u_1}^{1\, \rm rr \, rec\, 3PM}\,.
 \eea

As was mentioned in \cite{Bini:2020hmy} (See Eq. (6.19)), the symmetry of the gravitational-radiation emission in the center-of-velocity
frame proven by Kovacs and Thorne
\cite{Kovacs:1977uw} implies the following 4-vectorial structure for the radiated 4-momentum 
\bea
P^{\mu}_{ \rm rad \, 3PM}&=& \pi\frac{G^3 m_1^2 m_2^2}{b^3} \frac{\widehat {\cal E}(\g)}{\g+1} (u_{1}^-{}^\mu+ u_{2}^-{}^\mu)\,.\\
&&\nonumber 
\eea

This structure in turn implies  simple links\footnote{At higher PM orders we shall derive analogs of these links
by exploiting the mass-polynomiality of the impulses.}  
between the energy radiated in the rest frame of one of the particles
(namely $\pi\frac{G^3 m_1^2 m_2^2}{b^3} \widehat {\cal E}(\g)$), the energy radiated
in the c.m. frame, and the linear momentum radiated in the c.m. frame:
\bea
\label{E3_and_P3}
E_{\rm c.m.}^{\rm rad  \, 3PM}&=&\frac{\pi G^3 m_1^2 m_2^2}{b^3}\frac{\widehat {\cal E}(\g)}{h}\,,\nonumber\\
{\mathbf P}_{\rm c.m.}^{\rm rad  \, 3PM} &=&E_{\rm c.m.}^{\rm rad  \, 3PM}\sqrt{\frac{\gamma-1}{\gamma+1}}
\frac{m_2-m_1}{M}{\mathbf n}_-\,.
\eea
Here we used the  link \eq{n_meno}  between  ${\mathbf n}_-$, $u_1^-$ and $u_2^-$.
Let us also note that we have ${\bf e}_y = {\bf n}_- + O(G)$ so that
${P}_{y \, \rm c.m.}^{\rm rad \, 3PM}= {P}_{n \, \rm c.m.}^{\rm rad \, 3PM} $
(at the 3PM approximation).

Inserting Eqs. \eqref{E3_and_P3} into Eqs. \eqref{cu1_u2_3PM} above  leads then  to
\bea
c_{u_1}^{1 \,\rm rr \, rel \, 3PM}&=& -\pi \left(\frac{GM}{b}\right)^3\frac{\nu^2}{h^2}
\widehat {\mathcal E}(\gamma)
\frac{(m_1\gamma+ m_2)\gamma }{(\gamma^2-1)}\,,\nonumber\\
c_{u_1}^{1\, \rm rr \, rec\, 3PM} &=& \pi \left(\frac{GM}{b}\right)^3 \frac{\nu^2}{h^2} \widehat {\cal E}(\gamma)\frac{(m_1-m_2)m_1\gamma }{M (\gamma+1)}\,, \nonumber\\
c_{u_2}^{1 \,\rm rr \, rel \, 3PM} &=& -\frac{1}{\gamma}c_{u_1}^{1\, \rm rr \, rel\, 3PM}\,,\nonumber\\
 c_{u_2}^{1 \,\rm rr \, rec \, 3PM}   &=&-\frac{1}{\gamma}c_{u_1}^{1\, \rm rr \, rec\, 3PM}\,.
\eea
 A crucial point here is that the separate expressions of $c_{u_1}^{1\, \rm rr \, rel\, 3PM} $,
$c_{u_1}^{1\, \rm rr \, rec\, 3PM} $, $c_{u_2}^{1\, \rm rr \, rel\, 3PM} $,
$c_{u_2}^{1\, \rm rr \, rec\, 3PM} $, are {\it not polynomials in the two masses} (both because
of the factor $\nu^2=m_1^2 m_2^2/(m_1+ m_2)^4$,
and of the presence of
$h^2=1+ 2 \nu (\g-1)$ in the denominators.
However, one finds that the total radiation-reacted values $c_{u_1}^{1\, \rm rr \, tot\, 3PM}=c_{u_1}^{1\, \rm rr \, rel\, 3PM}+c_{u_1}^{1\, \rm rr \, rel\, 3PM}$
obtained by adding the relative-motion and the recoil
contributions, are polynomial in the masses (and actually simply proportional to  $ m_1^2 m_2^2$). 
This property follows from the (non evident) identity
\bea \label{identity}
&&m_1\gamma+ m_2- (\g-1) \frac{(m_1-m_2)m_1 }{M }  \equiv  M h^2 \nonumber\\
&&\qquad =M\left[1 + 2 \nu (\g-1)\right]\,.
\eea
 Indeed,
\bea
\label{3PM_impulse_coeffs}
c_{u_1}^{1\, \rm rr \, tot\, 3PM} &=&  -\pi \left(\frac{GM}{b}\right)^3\frac{\nu^2 \gamma \widehat {\mathcal E}(\gamma)}{h^2 (\gamma+1)}
\left[
\frac{(m_1\gamma+ m_2) }{(\gamma-1)}\right.\nonumber\\
&&\left.
-\frac{(m_1-m_2)m_1 }{M }\right]  \nonumber\\
&=& -\pi \frac{G^3 m_1^2 m_2^2}{b^3}  \frac{ \gamma }{ (\gamma^2-1)}
\widehat {\mathcal E}(\gamma)\,,\nonumber\\
c_{u_2}^{1\, \rm rr \, tot\, 3PM} &=& -\frac{1}{\gamma}c_{u_1}^{1\, \rm rr \, tot\, 3PM}\nonumber\\
&=& + \pi \frac{G^3 m_1^2 m_2^2}{b^3}  \frac{ 1}{ (\gamma^2-1)}
\widehat {\mathcal E}(\gamma)\,.
\eea
This mass-polynomiality is a nice check of our derivation above of the two types of radiation-reaction effects.

Summarizing so far, our simple, linear-response computation of the radiation-reaction contributions to the
impulse yield, when considered at the 3PM level, a result of the form
\bea \label{3PMrrimpulse}
\Delta p_{1}^{\rm rr, 3PM}&=& \frac{G^3 m_1^2 m_2^2}{b^3} \left[ -  \frac{2 (2 \g^2-1)^2}{(\g^2-1)}\,    {\cI}(v)\hat b \right. \nonumber\\
&&\left. + \pi  \frac{\widehat {\mathcal E}(\gamma)}{\g+1} (u_2^- - \g u_1^- )\right]
\,.
\eea
This result agrees with Ref. \cite{Herrmann:2021tct}.
To make the results Eqs. \eq{3PM_impulse_coeffs}, \eq{3PMrrimpulse}, explicit, we need to insert the explicit expression
of the radiated-energy function $\widehat  {\mathcal E}(\gamma)$. 
Kovacs and Thorne \cite{Kovacs:1977uw} have given explicit expressions for the time-domain radiation pattern at $O(G^3)$
(see also Ref. \cite{Mougiakakos:2021ckm} for explicit expressions of the frequency-domain radiation pattern at $O(G^3)$).
We did not succeed in computing the exact total radiated energy $\widehat {\cal E}(\g)$  by integrating
over time and angles the radiation pattern of Ref. \cite{Kovacs:1977uw}, but we could compute 
its PN expansion  up to order $v^{15}$ included, namely:
\bea
\label{hat_E_def_exp}
\widehat {\cal E}(\g)&=&\frac{37}{15} v +\frac{2393}{840}v^3+\frac{61703}{10080}v^5+\frac{3131839}{354816}v^7\nonumber\\
&+& 
\frac{513183289}{46126080}v^9 +
\frac{60697345}{4612608}v^{11} +
\frac{588430385}{39207168} v^{13} \nonumber\\
&+& 
\frac{12755740946147}{762814660608}v^{15}+
O(v^{17})\,.
\eea 

Recently Refs. \cite{Bern:2021dqo,Herrmann:2021tct,DiVecchia:2021ndb} have succeeded in obtaining
the exact expression of the the 3PM radiated (energy or) 4-momentum. It reads
\bea
\label{hat_E_def}
\widehat {\cal E}(\g)&=&f_1(\gamma)+f_2(\gamma)\ln\left(\frac{\gamma+1}{2}\right)\nonumber\\
&+&f_3(\gamma) {\rm arccosh}(\gamma)  \,, \qquad
\eea
where 
\bea
f_1(\gamma)&=&\frac{1}{48(\gamma^2-1)^{3/2}} (1151-3336\gamma+3148\gamma^2\nonumber\\
&&-912\gamma^3+339\gamma^4-552\gamma^5+210\gamma^6)\,, \nonumber\\
f_2(\gamma)&=& -\frac{(76\gamma -150\gamma^2+60\gamma^3+35\gamma^4-5)}{8\sqrt{\gamma^2-1}}\,,\nonumber\\
f_3(\gamma)&=& \frac{\gamma (-3+2\gamma^2) (11-30\gamma^2+35\gamma^4)}{16(\gamma^2-1)^2}\,.
\eea
Its high-energy limit, $\gamma \to \infty$, is $\widehat {\cal E}(\g) \approx \widehat C \g^3$,
with $\widehat C = \frac{35}{8}(1+2\ln(2)) \approx 10.4400$, i.e., ${\cal E}(\g) \approx  C \g^3$,
with $C = \pi \widehat C \approx 32.7983$.

The fact that $\widehat {\cal E}(\g)$ grows faster than $\gamma$ in the high-energy limit, i.e., that the energy loss at fixed impact parameter can exceed the energy of the system, shows that the domain of validity of PM gravity is limited. For discussions of the  domain of physical validity of the PM expansion see e.g. Eqs. (6.46)-(6.50) in Ref. \cite{Damour:2019lcq} (and references therein).

\section{Radiation-graviton  contribution to the conservative $O(G^4)$ scattering angle}

Within the EFT approach \cite{Goldberger:2004jt}, the conservative scattering angle $\chi^{\rm cons, 4PM}$
(which encodes the full $O(G^4)$  conservative impulse,  $ \Delta p_{\mu a}^{\rm cons, 4PM}$)
can be decomposed into two contributions:
a {\it potential-graviton} contribution, and a {\it radiation-graviton} one:
\be
\chi^{G^4 \rm cons}= \chi^{G^4 \rm cons,  pot}+ \chi^{G^4 \rm cons, radgrav}.
\ee
These two conservative contributions derive from corresponding (subtracted) radial actions, say 
\be \label{chi4pot}
\chi^{G^4, \rm cons, pot} = - \frac{ \partial I_{r, 4}^{\rm pot}(J) }{\partial J}\,,
\ee
\be \label{chi4radgrav}
\chi^{G^4, \rm cons, radgrav} = - \frac{ \partial I_{r, 4}^{\rm radgrav}(J) }{\partial J}\,.
\ee
Here $ I_{r, 4}^{\rm pot}(J)$ is linked to the exchange of potential gravitons
(with momenta $k^0 \sim v |{\bf k}|$, $|{\bf k}| \sim \frac1b$), while $I_{r, 4}^{\rm radgrav}(J)$
is linked to the
{\it time-symmetric} exchange of soft gravitons propagating on large spatiotemporal distances
($k^0 \sim  |{\bf k}| \ll \frac1b $).

Recently, Bern et al. \cite{Bern:2021dqo} derived the {\it potential-graviton} contribution 
$I_{r, 4}^{\rm pot}(J)$  to the radial action (see also \cite{Dlapa:2021npj}). 
The latter quantity is  IR-divergent, and was written
in Ref. \cite{Bern:2021dqo} in the form
\bea \label{Ir4pot}
I_{r, 4}^{\rm pot}(J, \g, m_1, m_2)&=& -f_{I_1}f_{I_\epsilon}\left[{\mathcal M}_4^p(\g)\right. \nonumber\\
&+&\left. \nu \left(\frac{{\mathcal M}_4^t(\g)}{\epsilon} +{\mathcal M}_4^{\rm f-t}(\g) \right)  \right]\,,\qquad
\eea
Here $\epsilon \equiv -\frac{D-4}{2}$, the prefactors are defined as\footnote{\label{foot2}The notations
here are
$
\mu_0=\frac{1}{\ell_0}\,, 
\tilde \mu^2=4\pi e^{-\gamma}\mu_0^2\,, \tilde \mu^2e^{2\gamma}=4\pi e^{\gamma}\mu_0^2\,, 4\pi e^{\gamma}=\bar q\,,$
with 
$ 
r_0=\ell_0 \frac{e^{-\gamma/2}}{2\sqrt{\pi}}=\frac{\ell_0}{\sqrt{4\pi e^{\gamma}}}\equiv \frac{\ell_0}{\sqrt{\bar q}}\,.
$
}
\bea
f_{I_1}&=&\frac{\pi}{8}\frac{G^4M^7\nu^2 p^2}{E J^3}\,,\nonumber\\
f_{I_\epsilon}&=&\left(\frac{4\tilde \mu^2 e^{2\gamma}J^2}{p^2} \right)^{-2\epsilon} \,,
\eea
while the (amplitude-related) summands are
\be
{\mathcal M}_4^p(\g)=-\frac{35}{8} \frac{(1-18\g^2 + 33\g^4)}{(\g^2-1)}\,,
\ee
\be
{\mathcal M}_4^{\rm t}(\g)=4 p_\infty \widehat {\mathcal E}(\gamma)\,,
\ee
and
\be
 {\mathcal M}_4^{\rm f-t}(\g) \equiv {\mathcal M}_4^{\rm f}(\g) -14 {\mathcal M}_4^{\rm t}(\g) \,.
 \ee
Here, the term ${\mathcal M}_4^p(\g)$ encodes the test-particle (Schwarzschild background) dynamics,
the IR-divergent, tail-related term ${\mathcal M}_4^{\rm t}(\g)$ involves 
the $O(G^3)$ radiated energy loss defined in Eq. \eqref{hat_E_def}, while
 the term ${\mathcal M}_4^{\rm f}(\g)$, entering $ {\mathcal M}_4^{\rm f-t}(\g)$, 
   is a complicated, transcendental function of $\g$ which encodes the finite piece of the 
   potential-graviton amplitude. It is
 given in the last Eq. (6) of Ref. \cite{Bern:2021dqo} (with the change of notation $\sigma \to \g$).
 For concreteness, the  PN expansions of the various building blocks of the 4PM radial action can be
 found in Appendix \ref{PNAppendix}.
 
The aim of the present section is to complete the result of  Ref. \cite{Bern:2021dqo} by giving the
explicit expression (at the 6PN accuracy) of the complementary radiation-graviton contributions
to the scattering angle, $\chi^{G^4, \rm cons, radgrav}$ or equivalently, 
to the radial action $ I_{r, 4}^{\rm radgrav}(J)$. For clarity, we shall express our results
in terms of a radial action $ I_{r, 4}^{\rm radgrav}(J)$ written in a form closely connected with
the form of the potential radial action used in Ref. \cite{Bern:2021dqo}. Namely, we shall write
the complementary radiation-graviton contribution to the radial action in the form
\bea \label{Ir4rad}
I_{r, 4}^{\rm radgrav}&=& -f_{I_1}f_{I_\epsilon}\nu \left(-\frac{{\mathcal M}_4^t(\g)}{\epsilon} +{\mathcal M}_4^{\rm rad}(\g) \right) \,.
\eea
Here, we have conventionally separated out the same prefactors used in the potential-graviton contribution \eq{Ir4pot}, and the
opposite of the IR-divergent contribution $+\nu \frac{{\mathcal M}_4^t(\g)}{\epsilon}$ present in Eq. \eq{Ir4pot}.
Indeed, past works on tail effects \cite{Foffa:2011np,Galley:2015kus,Porto:2017dgs}
have shown that the IR divergence exhibited in the near-zone (or potential-graviton) dynamics
is cancelled by a (UV) divergence present in the multipolar expansion of the wave-zone (or radiation-graviton) dynamics.

The recently developed TF formalism has allowed one to compute the $O(G^4)$ conservative, non-local (radiation-graviton) dynamics
up to the 6PN accuracy included. In this formalism, the dynamics is decomposed into local-in-time and non-local-in-time parts.
Though the latter decomposition is closely linked to the decomposition in potential-graviton and radiation-graviton dynamics,
one cannot simply identify the time-symmetric local-in-time dynamics of Refs. \cite{Bini:2019nra,Bini:2020nsb} to the EFT time-symmetric potential-graviton 
dynamics. However, one can identify the {\it total}, TF-derived conservative scattering angle to the {\it total}, conservative scattering
angle that would result from completing the potential-graviton contribution \eq{chi4pot}, by the (still to be computed)  
$O(G^4)$ radiation-graviton contribution \eq{chi4radgrav}. In particular, each term in the PM expansion of the (gauge-invariant)
total scattering angle should coincide. In other words, we can write
\bea
\label{chi_pot_rad}
\chi^{G^4, \rm TF \, tot}&=& \chi^{G^4, \rm TF \, loc}+ \chi^{G^4, \rm TF \, nonloc}\nonumber\\
&=& \chi^{G^4, \rm pot} + \chi^{G^4, \rm radgrav} \,.
\eea
When adding Eq. \eqref{Ir4pot} and Eq. \eqref{Ir4rad} the $\frac1{\epsilon}$ divergence cancels so that we can directly work in the $\epsilon \to 0$ limit
(and neglect the factor $f_{I_\epsilon} \to 1$). This yields the equation
\bea
\chi^{G^4, \rm TF \, tot}&=& \frac{ \partial f_{I_1} }{\partial J} \left[ {\mathcal M}_4^p(\g)\right.\nonumber\\
&+&\left.
\nu \left({\mathcal M}_4^{\rm f-t}(\g)+ {\mathcal M}_4^{\rm radgrav}(\g) \right) \right]\,.\quad
\eea
In the TF formalism it is usual to represent the PM-expansion of the scattering angle in the form \eq{chi_cons_PM_exp}, so that
the 4PM contribution is denoted 
\bea
\chi^{G^4, \rm TF \, tot}&=&\frac{2 \chi_4}{j^4}= \frac{2 (\chi_4^{\rm loc} + \chi_4^{\rm nonloc})}{j^4}\,.
\eea
In addition, it is convenient to introduce the corresponding energy-rescaled scattering angle 
\be
\tilde \chi_4 \equiv h^3 \chi_4 = h^3 (\chi_4^{\rm loc} + \chi_4^{\rm nonloc})\,,
\ee
where $h = \frac{E_{\rm c.m.}}{M} = \sqrt{1+2\nu(\gamma-1)}$.

This yields the equation
\beq
\tilde \chi_4 =-\frac{3}{16  } \pi  p_\infty^2 \left[{\mathcal M}_4^p+\nu \left({\mathcal M}_4^{\rm f-t}  +{\mathcal M}_4^{\rm radgrav} \right)\right]\,.
\eeq
As the $\nu \to 0$ limit of $\tilde \chi_4 $ is equal to the Schwarzschild (test-mass) result, equivalent to
\beq
\chi_4^{\rm Schw}=-\frac{3}{16  } \pi  p_\infty^2  {\mathcal M}_4^p\,,
\eeq
we get the following simple equation to determine the radiation-graviton contribution ${\mathcal M}_4^{\rm rad}$:
\beq
\label{tildechi4radgrav}
\tilde \chi_4-\chi_4^{\rm Schw}=-\frac{3}{16  } \pi  p_\infty^2  \nu \left({\mathcal M}_4^{\rm f-t}(\g)  +{\mathcal M}_4^{\rm radgrav}(\g) \right)\,. 
\eeq
[The linearity of $\tilde \chi_4-\chi_4^{\rm Schw}$ in the symmetric mass ratio $\nu$ is a consequence of a 
general rule pointed out in Ref. \cite{Damour:2019lcq}.]

Using now our previous results (namely Eq. (8.2) in Ref. \cite{Bini:2020nsb} for  $\tilde \chi_4^{\rm loc}$  and  Eqs. (6.11),  
(7.22) and (7.28) in Ref.  \cite{Bini:2020hmy} for $\tilde \chi_4^{\rm nonloc}$), we deduce that the TF-computed 6PN-accurate
value of $\tilde \chi_4-\chi_4^{\rm Schw}$ is equal to
\be
\label{tildechi4TF}
\tilde \chi_4-\chi_4^{\rm Schw}= \pi \nu \widehat A(p_\infty)- \widehat {\mathcal E}(p_\infty)\, \ln \left(\frac{p_\infty}{2}\right)\,,
\ee
where
\bea
\widehat A(p_\infty)&=& -\frac{15}{4} 
+\left(\frac{123}{256}\pi^2-\frac{557}{16}\right) p_\infty^2\nonumber\\
&+&\left(\frac{33601}{16384} \pi^2-\frac{6113}{96} \right) p_\infty^4\nonumber\\
&+&\left(\frac{93031}{32768} \pi^2-\frac{615581}{19200} \right) p_\infty^6\nonumber\\
&+&\left(\frac{29201523}{33554432} \pi^2-\frac{5824797}{627200} \right) p_\infty^8\,.\qquad
\eea

Here we used the relation, known from previous work, between the coefficient of  the logarithmic term $\ln \left(\frac{p_\infty}{2}\right)$
and the energy radiated in gravitational waves  (see, e.g., Ref. \cite{Bini:2020hmy}).
The 6PN-accurate expression \eqref{tildechi4TF} applies to the conservative  (in the sense Fokker-Wheeler-Feynman time-symmetric sense)  $O(G^4)$ scattering angle. If we were considering instead the non-conservative, retarded scattering angle  one should add to 
\eqref{tildechi4TF} radiation-reaction-related contributions. As we shall see, terms quadratic in radiation-reaction will start contributing at the 5PN order, namely at  $O(G^4/c^{10})$. We gave here the expression \eqref{tildechi4TF} for the purpose of comparison with  forthcoming  $O(G^4)$ conservative computations, under the assumption that the meaning of conservative is the same as the (Fokker-Wheeler-Feynman) one used here. 

Identifying Eq. \eq{tildechi4radgrav} with Eq. \eq{tildechi4TF} finally leads to the following structure for the yet-uncomputed radiation-graviton
contribution to the conservative 4PM radial action
\bea
{\mathcal M}_4^{\rm radgrav} &=& {\mathcal M}_4^{\rm radgrav, finite}
+4 {\mathcal M}_4^t  \ln \left(\frac{p_\infty}{2}\right)\nonumber\\
&=& {\mathcal M}_4^{\rm radgrav, finite}
+ 16 p_\infty \widehat {\mathcal E}\, \ln \left(\frac{p_\infty}{2}\right)\,,\qquad\quad
\eea
where the 6PN-accurate value of the non-logarithmic contribution is
\beq
\label{Mrad_finite}
{\mathcal M}_4^{\rm radgrav, finite}=\frac{12044}{75} p_\infty^2+ \frac{212077}{3675} p_\infty^4+ \frac{115917979}{793800} p_\infty^6\,.
\eeq
This result offers a benchmark to any future computation of the radiation-graviton contribution to conservative dynamics.

Our result \eq{Mrad_finite} represents only the beginning of the PN expansion of the
conservative part of the radiation-related contributions to the $O(G^4)$ scattering. 
The dissipative $O(G^4)$ contribution will be discussed below.

\section{Mass-polynomiality of the $O(G^4)$ radiation-reaction contribution to the impulses and its consequences}

In the present section we shall explicitly show how the mass polynomiality of the 
4-vectorial impulses $ \Delta p^\mu_a$  (considered as  functions of  $u_{ 1 -}^\mu$, $u_{ 2 -}^\mu$, and $b^\mu = b {\hat b}^\mu$) allows one to derive strong restrictions on the various theoretical building blocks entering our general radiation-reaction formulas derived in Section  \ref{sec3}. For definiteness, we shall focus 
on the 4PM level, but it will be clear that our arguments extend to higher PM levels.

Let us recall the structure of the impulses that will constitute the starting point of our reasonings:
the PM expansion of the impulse (say for $a=1$) is of the form
 \be \label{deltapmugen}
\Delta p_{1 \mu}= - 2G m_1 m_2 \frac{2 \g^2-1}{\sqrt{\g^2-1}} \frac{b_\mu}{b^2} + \sum_{n\geq 2} \Delta p_{1 \mu}^{n \rm PM}\,,
\ee 
where the $n$-PM ($O(G^n)$) contribution $\Delta p_{1 \mu}^{n \rm PM}$ must be $\propto \frac{G^n}{b^n}$, and be the product of $m_1 m_2$ by a homogeneous polynomial of degree $n-1$ in the masses. 
See Eq. \eq{deltapmugennPM}.

The structure \eq{deltapmugennPM} applies both to the total (radiation-reacted) 
impulse and to its conservative contribution. [Therefore, it also separately applies
to the radiation-reaction contribution to the impulse.]
Its consequences for the conservative scattering angle were discussed in Ref. \cite{Damour:2019lcq}.
For the conservative part of the $O(G^4)$ (and $O(\frac1{b^4})$) coefficients $c_{b, G^4}^{1 \rm cons}$, $c_{u_1,G^4}^{1 \rm cons}$ and $c_{u_2,G^4}^{1 \rm cons}$
\bea
c_{b,G^4}^{1 \rm cons}&=& \frac{G^4 m_1 m_2 (m_1+m_2)^3}{b^4} \frac{4h^2 \tilde \chi_2^{\rm cons}(\chi_1^{\rm cons}) ^2- \tilde \chi_4^{\rm cons}  }{(\gamma^2-1)^{3/2}}\,,\nonumber\\
c_{u_1,G^4}^{1 \rm cons}
&=&-\frac{G^4 m_1 m_2 (m_1+m_2)^2}{b^4} \frac{2(m_1\gamma+m_2)}{3(\gamma^2-1)^2} \nonumber\\
&& \times\left(-h^2(\chi_1^{\rm cons})^4+3(\tilde \chi_2^{\rm cons})^2+6\chi_1^{\rm cons}\tilde\chi_3^{\rm cons}\right)\,,\nonumber\\
c_{u_2,G^4}^{1 \rm cons}
&=&\frac{G^4 m_1 m_2 (m_1+m_2)^2}{b^4}\frac{2 (m_2\gamma+m_1)}{3(\gamma^2-1)^2} \nonumber\\
&& \times\left(-h^2(\chi_1^{\rm cons})^4+3(\tilde\chi_2^{\rm cons})^2+6\chi_1^{\rm cons}\tilde\chi_3^{\rm cons}\right) \,.\nonumber\\
\eea
Here we have expressed them in terms of the $h$-rescaled scattering coefficients
$\tilde \chi_n \equiv h^{n-1} \chi_n$ (where $h=\sqrt{1+ 2 \nu(\g-1)}$) which are polynomials in $\nu$ of order $[\frac{n-1}{2}]$.
It is then easy to check (remembering that $\nu=\frac{m_1m_2}{(m_1+m_2)^2}$)
that $c_{b,G^4}^{1 \rm cons}$, $c_{u_1,G^4}^{1 \rm cons}$ and 
$c_{u_2,G^4}^{1 \rm cons}$ are polynomials in the masses.
 
Let us now discuss the consequences of  such a structure for the radiation-reaction contribution
to the impulse.

We have seen above that, at the $O(G^3)$ level,  the mass dependence of the total radiation-reaction 
contribution to the impulse was $\Delta p_a^{\rm rr} \sim G^3 m_1^2 m_2^2/ b^3$. 
The presence of a factor $m_1^2 m_2^2$ in the LO contribution to $\Delta p_a^{\rm rr}$ is linked
to the fact that radiation-reaction effects balance radiated fluxes that are quadratic in the emitted
waveform, and that the (time-dependent part of the) emitted waveform contains a factor $m_1 m_2$.
As a consequence, next-to-leading-order, and higher, contributions to $\Delta p_a^{\rm rr} $ must
all contain at least a factor $m_1^2 m_2^2$. 
Polynomiality in the masses then shows that the radiation-reaction 
contribution to the impulses  (and therefore to their coefficients $c_b^a, c_{u_1}^a, c_{u_2}^a$)
must have the structure
\bea
\Delta p_a^{\rm rr} &\sim& \frac{G^3 m_1^2 m_2^2}{b^3} \left[1+ \frac{G m_1}{b} +   \frac{G m_2}{b} \right.\nonumber\\
&+&\left.
   \frac{G^2 m_1^2}{b^2}+  \frac{G^2 m_1 m_2}{b^2}+ \frac{G^2 m_2^2}{b^2}+ \cdots \right]\,,
\eea
with (unwritten) coefficients depending only on $\g$.
We now show how to use this mass-polynomiality, in conjunction with our general formulas above for the
radiation-reaction effects, to restrict the mass dependence of the various radiative losses.

Previous studies have found convenient to express the PM expansion of the conservative scattering angle as
a power expansion in the dimensionless variable $\frac1j= \frac{G m_1 m_2}{J}$, namely,
\beq \label{chiexpj}
\chi^{\rm cons} = \sum_{n=1}^\infty \frac{2\chi^{\rm cons}_n }{j^n}\,.
\eeq
Remembering that (see Appendix \ref{PMAppendix})
\be \label{1byj}
\frac1j= \frac{G M h}{ b \pinf},
\ee
any expansion in powers of $\frac1j$ corresponds to an expansion in powers of $\frac{G}{b}$. 
However, the transcription between the two expansions involve powers of $M h= M \sqrt{1+ 2\nu(\g-1)}$.
Keeping in mind the presence of such (non polynomial in masses) factors,
we can similarly parametrize the PM expansions of the radiative losses by the coefficients of their
 power expansion in $\frac1j$, say
\bea
\label{various_expansions}
-\frac{ \delta^{\rm rr} J}{J_{\rm c.m.}}=\frac{ J^{\rm rad}}{J_{\rm c.m.}} &=& + \nu^1  \sum_{n=2}^\infty \frac{{ J}_{n}}{j^n} \,, \nonumber\\
-\frac{ \delta^{\rm rr} E}{M}=\frac{E^{\rm rad}}{M} &=&  + \nu^2   \sum_{n=3}^\infty \frac{ E_{n}}{j^n} \,, \nonumber\\
-\frac{ \delta^{\rm rr} P_y}{M}=\frac{P_y^{\rm rad}}{M}&=& + \frac{m_2-m_1}{M} \nu^2 \sum_{n=3}^\infty \frac{P_{n}}{j^n}\,.
\eea
Here we have adimensionalized the left-hand sides, and pulled out some powers of $\nu$ on the 
right-hand sides, to ensure that the expansion coefficients $J_{n}$, $E_n$, $P_n$ are
dimensionless, and that their LO PN contribution is $\nu$-independent. [ 
$J_{\rm c.m.} = b \mu \pinf/h=GM^2\nu j$, in the denominator of the first equation
contains a factor $\nu$ so that $J^{\rm rad}$ actually contains the same factor $\nu^2$ 
as the other losses.] Note  the sign difference between the (negative) mechanical losses of the 
two-body system (denoted with $\delta^{\rm rr} $) and the (positive) radiated quantities 
($ J^{\rm rad}>0$, $ E^{\rm rad}>0$ and also, with our convention,  $\frac{P_y^{\rm rad}}{m_2-m_1}>0$,
see below).

The quantities $ J^{\rm rad}$ and $ E^{\rm rad}$ are symmetric under the exchange of the two bodies.
Therefore the expansion coefficients $J_{n}$ and $E_n$ are functions of $\g$ and of the
symmetric mass-ratio $\nu$. The same is true for the expansion coefficients $P_n$ (obtained after 
factoring the antisymmetric mass ratio $\frac{m_2-m_1}{M}$). It was shown in Ref. \cite{Bini:2020hmy} 
 that the
$\nu$ dependence of $E_n$ is restricted by the property
\be \label{Ennu}
h^{n+1} E_n(\g, \nu) = P^\g_{[\frac{n-2}{2}]}(\nu)\,,
\ee
where $ P^\g_N(\nu)$ denotes a polynomial in $\nu$ of order $N$, with coefficients depending on $\gamma$. [The notation $[ \cdots]$ indicates
the integer part.]  The mass-polynomiality  of the radiation-reacted
impulse implies analog  restrictions on the  $\nu$-dependence of the other expansion coefficients 
$J_n$ and $P_n$ entering the radiation losses \eq{various_expansions}. This is seen by
using our previous results to write down explicit expressions for the PM-expansion
contributions to the radiation-reacted impulse coefficients  $c_b^{1\rm rr\, tot}$, 
 $c_{u_1}^{1\rm rr\, tot}$,  $c_{u_2}^{1\rm rr\, tot}$. 

Using Eqs. \eq{deltarrchirel}, \eq{chiexpj}, and \eq{various_expansions}, the PM-expansion coefficients of the radiation-reaction
contribution to the relative scattering angle, say
\beq
\delta^{\rm rr} \chi = \sum_{n=3}^\infty  \frac{ 2\chi^{\rm rr}_n}{j^n}\,,
\eeq
are found to read
\bea 
2\chi^{\rm rr}_3 &=& \nu \chi^{\rm cons}_1 J_2 \,, \nonumber\\
2\chi^{\rm rr}_4 &=& \nu \left(2 \chi^{\rm cons}_2 J_2+ \chi^{\rm cons}_1J_3  - h E_3  \frac{d\chi^{\rm cons}_1}{d\gamma} \right) \,,\nonumber\\
2\chi^{\rm rr}_5 &=&  \nu\left[ \left(\chi^{\rm cons}_1 J_4+2\chi^{\rm cons}_2 J_3+3\chi^{\rm cons}_3J_2\right) \right.\nonumber\\
&&\left. -h  \left( E_4 \frac{d\chi^{\rm cons}_1}{d\gamma}  
+ E_3  \frac{d\chi^{\rm cons}_2}{d\gamma} \right) \right] \,,
\eea
etc. 
To make these PM results explicit we need to insert the explicit values of the various
PM-expansion coefficients $ \chi^{\rm cons}_n$, $J_n$, and $E_n$. Among these PM
expansion coefficients the current knowledge concerns: (i) the values of the $ \chi^{\rm cons}_n$
up to $n=3$ included, e.g., the first two read,
\bea
\label{chi_1_2_cons}
\chi^{\rm cons}_1=\frac{2\gamma^2-1}{\sqrt{\gamma^2-1}}\,, \qquad
\chi^{\rm cons}_2=\frac{3}{8}\pi \frac{5\gamma^2-1}{h}\,;
\eea
and the explicit values of $J_2$ \cite{Damour:2020tta} (see Eqs. \eqref{various_expansions} and \eqref{JradbyJ} above), and of
$E_3$ \cite{Herrmann:2021tct} (see Eqs. 
\eqref{E3_and_P3} and \eqref{hat_E_def} above). Note, however, that the 3PM angular momentum loss
$J_3$ has not yet been exactly computed.  We, however,  derived its 4.5PN-accurate value.
See below and Appendix \ref{PNAppendix}.

The value of the radiation-reacted relative scattering angle $\chi^{\rm rr}$ 
yields only a partial contribution to the radiation-reacted impulse. 
One needs to add the radiative-recoil contributions
to get the total radiation-reacted impulse coefficients $c_b^{1\rm rr\, tot}$, 
 $c_{u_1}^{1\rm rr\, tot}$,  $c_{u_2}^{1\rm rr\, tot}$. Moreover, one needs to express their PM expansion
 in terms of the impact parameter $b$ in order to exhibit their polynomiality in the masses. 
 Starting from the expansion of the impulse coefficients in powers of $G/b$, say
 \bea
c_b^{1,\rm rr\, tot} &=&  \sum_{n=3}^\infty c_{b,G^n}^{\rm rr} \left(\frac{G}{b}\right)^n\,, \nonumber\\
c_{u_1}^{1,\rm rr\, tot} &=&  \sum_{n=3}^\infty c_{u_1,G^n}^{\rm rr} \left(\frac{G}{b}\right)^n\,, \nonumber\\
c_{u_2}^{1,\rm rr\, tot} &=&  \sum_{n=3}^\infty c_{u_2,G^n}^{\rm rr} \left(\frac{G}{b}\right)^n\,, 
\eea
 we can define dimensionless versions, $ {\hat c}_{b,G^n}^{\rm rr \, tot}, \cdots$, of the expansion 
 coefficients $ c_{b,G^n}^{\rm rr \, tot}, \cdots$,  by writing
\bea
c_{b,G^n}^{\rm rr \, tot} &=& m_1 m_2 M^{n -1} {\hat c}_{b,G^n}^{\rm rr \, tot} \,, \nonumber\\
c_{u_1,G^n}^{\rm rr \, tot} &=& m_1 m_2 M^{n -1} {\hat c}_{u_1,G^n}^{\rm rr \, tot} \,, \nonumber\\
c_{u_2,G^n}^{\rm rr \, tot} &=& m_1 m_2 M^{n -1} {\hat c}_{u_2,G^n}^{\rm rr \, tot} \,.
\eea
Let us start by discussing the expression of the 4PM impulse coefficient ${\hat c}_{b,G^4}^{\rm rr \, tot} $, which reads
\bea
\hat c_{b,G^4}^{\rm rr\, tot}&=& \frac{\nu m_1(m_1-m_2)(m_2\gamma+m_1)}{M^3 (\gamma^2-1)^2} \chi^{\rm cons}_1  h^2 P_3  \nonumber\\
&+&
\frac{\nu h^2}{(\gamma^2-1)^{3/2}}\left[h^2 \frac{d\chi^{\rm cons}_1}{d\gamma} \right.\nonumber\\
&+& \left.
\frac{2 (m_2\gamma+m_1)(m_1\gamma +m_2)}{M^2 (\gamma^2-1)}\chi^{\rm cons}_1\right] E_3 \nonumber\\
&-&
\frac{2\nu h^3 }{(\gamma^2-1)^{3/2}}\chi^{\rm cons}_2  J_2\nonumber\\
&-& \frac{\nu h^3 }{(\gamma^2-1)^{3/2}}\chi^{\rm cons}_1  J_3\,.
\eea
The structure of this quantity can be simplified by using the explicit expressions of $\chi^{\rm cons}_1$ 
and $\chi^{\rm cons}_2$ recalled above, together with the $\nu$-dependences of $J_2$ and 
$E_3$, as well as the relation between $E_3$ and $P_3$. Defining
\be
{\widehat J}_2(\gamma) \equiv 2(2\gamma^2-1)\sqrt{\gamma^2-1}{\mathcal I}(v) \,,
\ee
with  ${\mathcal I}(v)$ defined in Eq. \eqref{cal_I_di_v_def}, and 
using   Eq. \eqref{hat_E_def} for the definition of $\widehat {\mathcal E}$ and Eqs. \eqref{E3_and_P3} above, we get
\bea \label{J2E3P3}
h^2 J_2 &=&{\widehat J}_2(\gamma)  \,,\nonumber \\
h^4 E_3&=&\pi p_\infty^3 \widehat {\mathcal E}(\g)\,,\nonumber \\
\frac{P_3}{E_3}&=&\sqrt{\frac{\gamma-1}{\gamma+1}}\,.
\eea
Note that in these relations, the right-hand sides depend only on $\g$ (and not on $\nu$).

This leads to
\bea \label{hcb41}
\frac{{\hat c}_{b,G^4}^{\rm rr\, tot}}{\nu}
&=& \frac{\pi}{p_\infty^3}\left[
(6\gamma^2-5)\gamma\widehat {\mathcal E} 
-   \frac{3 (5\gamma^2-1) }{4} {\widehat J}_2 \right]
\nonumber\\
&&
-  \frac{m_1}{M}  \pi \frac{(2\gamma^2-1)}{p_\infty (\gamma+1) } \widehat {\mathcal E} 
\nonumber\\
&& -\frac{(2\gamma^2-1)}{p_\infty^4}\left(
\frac{\nu \pi p_\infty^3\widehat {\mathcal E}}{h^2}
   +  h^3 J_3    \right)\,.
\eea
The corresponding value of the dimensionfull impulse coefficient reads
\be
{ c}_{b,G^4}^{\rm rr\, tot}= m_1 m_2 M^3 {\hat c}_{b,G^4}^{\rm rr\, tot}=( m_1 m_2)^2  M \frac{{\hat c}_{b,G^4}^{\rm rr\, tot}}{\nu} \,.
\ee
As one sees on the last member of these equations, the fact that ${ c}_{b,G^4}^{\rm rr\, tot}$ should be 
polynomial (and actually quintic) in the masses implies that the product 
$M \frac{{\hat c}_{b,G^4}^{\rm rr\, tot}}{\nu}$ should be linear in $m_1$ and $m_2$ (with $\g$-dependent
coefficients). This property is clearly satisfied by the first three lines on the right-hand side of Eq. \eq{hcb41}.
By contrast, the last line  on the right-hand side of Eq. \eq{hcb41} will violate this property  because of its
dependence on $\nu = \frac{m_1 m_2 }{(m_1+m_2)^2}$, unless the function $ h^3 J_3(\g,\nu)$
has a special $\nu$-dependence which completely cancels the various $\nu$-dependences contained
in $\frac{\nu \pi p_\infty^3\widehat {\mathcal E}}{h^2}$. This proves that the quantity
\be
\widehat J_3 \equiv  h^3 J_3 + \frac{\nu \pi p_\infty^3\widehat {\mathcal E}}{h^2},
\ee
must be {\it independent of $\nu$}, and be only a function of $\g$. In other words, by considering the
limit $\nu \to 0$, we derive the remarkable identity
\be \label{J3identity}
 h(\g,\nu)^3 J_3(\g,\nu) + \frac{\nu \pi p_\infty^3\widehat {\mathcal E}(\g)}{h^2(\g,\nu)}=
 \widehat J_3(\g)= \lim_{\nu \to 0} J_3(\g,\nu).
\ee
Equivalently, one should have
\be
\frac{h^3 J_3-J_3|_{\nu=0}}{\nu}+\pi \frac{p_\infty^3}{h^2} {\widehat {\mathcal E}}\equiv 0\,.
\ee
In other words, the $\nu$-dependent terms in the function $h^3 J_3$ must be
entirely determined by the function $-\pi \frac{p_\infty^3}{h^2} {\widehat {\mathcal E}}$.

We have computed $J_3(\g,\nu)$ with  2PN fractional accuracy (see Table \ref{tab:table_EnJnPn}).
Our final result reads
\bea
\pi^{-1}J_3&=& \frac{28}{5} p_\infty^2+\left(\frac{739}{84} -\frac{163}{15}\nu\right)p_\infty^4\nonumber\\
&+&
\left(-\frac{5777}{2520} -\frac{5339}{420}\nu+\frac{50}{3}\nu^2\right)p_\infty^6\nonumber\\
&+&O(p_\infty^8)
\,.
\eea
We checked that it satisfies the rule \eq{J3identity}.

Let us now consider  the two other 4PM impulse coefficients ${\hat c}_{u_1,G^4}^{\rm rr\, tot}$, 
${\hat c}_{u_2,G^4}^{\rm rr\, tot}$. The initial expression for the first one reads
\bea
\frac{{\hat c}_{u_1,G^4}^{\rm rr\, tot}}{\nu}&=&-\frac{ \gamma(m_1\gamma +m_2) }{(\gamma^2-1)^3 M } h^3 E_4 \nonumber\\
&+&
\frac{m_1  (m_1-m_2) \gamma }{(\gamma^2-1)^{5/2} M^2}  h^3P_4\nonumber\\
&-&
\frac{2 (m_1\gamma +m_2)(\chi^{\rm cons}_1)^2 }{(\gamma^2-1)^2  M } h^2J_2\,.
\eea
As in the case of ${\hat c}_{b,G^4}^{\rm rr\, tot}$, the $\nu$-dependence of the various building blocks,
$E_4, P_4, J_2$, entering the latter expression is crucial to allow ${c}_{u_1,G^4}^{\rm rr\, tot}$ to be a
polynomial in the masses. We already know that the $\nu$-dependence of $J_2$ is determined
by the fact that 
\be
h^2J_2={\widehat J}_2(\gamma)\,,
\ee
where  ${\widehat J}_2(\gamma)$ is defined in the first line of Eq. \eqref{J2E3P3},
while the $\nu$-dependence of $E_4$ is restricted by Eq. \eq{Ennu}. In the
present case $n=4$, the latter restriction says that $h^5 E_4$ must be linear in $\nu$:
\be
h^5 E_4(\g,\nu)= \widetilde E_4^0(\g) + \nu \widetilde E_4^1(\g)\,.
\ee
Given these $\nu$-dependences of $J_2$ and $E_4$, the $\nu$-dependence of the remaining building block
$P_4$ must be such that it allows the corresponding dimensionfull impulse coefficient,
\be
{ c}_{u_1,G^4}^{1\,\rm rr\, tot}= m_1 m_2 M^3 {\hat c}_{u_1,G^4}^{\rm rr\, tot}=( m_1 m_2)^2  M \frac{{\hat c}_{u_1,G^4}^{\rm rr\, tot}}{\nu} \,,
\ee
to be a polynomial in the masses. In other words, $M \frac{{\hat c}_{u_1,G^4}^{\rm rr\, tot}}{\nu}$
must be linear in $m_1$ and $m_2$. 

Combining the simple $\nu$ dependences of $h^2 J_2$ and $h^5 E_4$
with the crucial identity \eq{identity}, one finds that  the mass-polynomiality of ${ c}_{u_1,G^4}^{1\, \rm rr\, tot}$
implies that the $\nu$-dependence of $P_4$ must be determined by the condition
\be \label{P4nu}
h^5 P_4  
=\frac{1}{\pinf} \left[(\g-1)\widetilde E_4^0(\g) -\frac12 \widetilde E_4^1(\g)  \right]\,.
\ee
This  identity can be conveniently rewritten by replacing 
$\tilde E_4^0=h^5E_4-\nu \tilde E_4^1$, 
and using $p_\infty = \sqrt{\g^2-1}$. This yields the following (double) identity
\bea
h^3 \left[P_4 -\sqrt{\frac{\gamma-1}{\gamma+1}}E_4 \right]&=&\left[P_4 -\sqrt{\frac{\gamma-1}{\gamma+1}}E_4 \right]_{\nu=0}\nonumber\\
&=&-\frac{\tilde E_4^1(\g)}{2p_\infty}
\,.
\eea
Let us mention in passing that the identities, between $E_n(\g,\nu)$ and $P_n(\g,\nu)$ (as well as the
$\nu$-structure of  $E_n$),
 derived here (in a somewhat indirect manner)
by using the mass-polynomiality of $\Delta p_{a\mu}^{\rm rr, tot}$ can also
be directly derived from the mass-polynomiality structure of the large-$b$ expansion
of the radiated 4-momentum $P_\mu^{\rm rad}$. This is done by combining:  (i) the $b^\mu$, $u_1^\mu$, $u_2^\mu$
decomposition of  $P_\mu^{\rm rad}$; (ii)  the property mentioned above that, in the c.m. frame,
${\bf P}^{\rm rad}$ is directed along the ${\bf e}_y$ axis; and (iii) the  $\nu$-structure of
the conservative scattering angle. We leave this derivation (and its extension
to higher PM orders) as an exercize to the reader. An extension of such a direct PM derivation to
the case of $J_n(\g,\nu)$ is more challenging in view of the delicate issues linked to defining the
radiation of relativistic angular momentum $J_{\mu \nu}^{\rm rad}$.

We checked that the 4.5PN-accurate PN expansions of $P_4$ and $E_4$ do
indeed satisfy the relation \eq{P4nu} to the corresponding PN accuracy, while the PN expressions for $\widetilde E_4^0$,  $\widetilde E_4^1$, and  $\widehat J_3$ are given below in Eq. \eq{tildeE401hatJ3}.

Concerning the third 4PM impulse coefficient, namely ${\hat c}_{u_2,G^4}^{\rm rr\, tot}$, one finds
that (as was the case at the 3PM level) it is simply connected to  ${\hat c}_{u_1,G^4}^{\rm rr\, tot}$
via
\bea
\nu^{-1}\left({\hat c}_{u_2,G^4}^{\rm rr\, tot}+\frac{1}{\gamma}{\hat c}_{u_1,G^4}^{\rm rr\, tot}\right)=\frac{2 (2\gamma^2-1)^2}{\gamma(\gamma^2-1)^2} \frac{m_2}{M}{\widehat J}_2\,.
\eea
The latter relation implies the needed mass polynomiality (namely the linearity in the masses
of the product $\frac{M}{\nu} {\hat c}_{u_2,G^4}^{\rm rr\, tot}$) without further restriction
on the building blocks $E_4$, $P_4$, $J_2$.

Gathering our results, we conclude that the radiation-reaction contribution to the 4PM
impulse can be written as (see Eq. \eqref{Delta_p1_con_u2perp1})
\be \label{4PMrrimpulse}
\Delta p_{1}^{\rm rr, 4PM}= c_b^{\rm rr , 4PM} \hat b 
+ {\widehat c}_{u_1}^{\rm rr\, 4PM} u_1+ { c}_{u_2}^{\rm rr\, 4PM}u_{2\perp 1}\,,
\ee
where the coefficients have the following simple dependence on the masses
\bea
{ c}_{b}^{\rm rr\, 4PM} &=& \frac{G^4 m_1^2 m_2^2}{b^4} \left[ C_{b M}^{\rm 4PM}(\g) M + C_{b m_1}^{\rm 4PM}(\g)m_1 \right] 
\,,\nonumber \\
{ \widehat c}_{u_1}^{\rm rr\, 4PM} &=& \frac{G^4 m_1^2 m_2^2}{b^4} \,{\widehat C}_{u_1, m_2}^{\rm 4PM}(\g)m_2   
\,,\nonumber \\
{ c}_{u_2}^{\rm rr\, 4PM} &=& \frac{G^4 m_1^2 m_2^2}{b^4} \left[ C_{u_2 M}^{\rm 4PM}(\g) M + C_{u_2 m_1}^{\rm 4PM}(\g)m_1 \right] 
\,.\nonumber\\
\eea
The $\g$-dependent coefficients entering the various impulse contributions read
\bea
 C_{b M}^{\rm 4PM}(\g) &=&   \pi \widehat{\mathcal E} \frac{\gamma(6\gamma^2-5)}{(\gamma^2-1)^{3/2}} 
-\pi \frac34 \widehat J_2  \frac{(5\gamma^2-1)}{(\gamma^2-1)^{3/2}} \nonumber\\
&&
-\widehat J_3 \frac{(2\gamma^2-1)}{(\gamma^2-1)^2}\,, \nonumber\\
C_{b m_1}^{\rm 4PM}(\g)&=& -\pi \frac{(2\gamma^2-1)\widehat{\mathcal E}}{(\gamma+1)\sqrt{\gamma^2-1}}   \,,
\eea
\bea
{\widehat C}_{u_1, m_2}^{\rm 4PM}(\g)&=& 2 \frac{(2\gamma^2-1)^2}{(\gamma^2-1)^2}\widehat J_2\,,
\eea
and 
\bea
C_{u_2 M}^{\rm 4PM}(\g)&=& \frac12  \frac{ 4\gamma (2\gamma^2-1)^2\widehat J_2+2\tilde E_{4}^0 }{(\gamma^2-1)^3} 
\,,\nonumber\\
C_{u_2 m_1}^{\rm 4PM}(\g)&=& \frac12 \frac{ -4 (\gamma-1) (2\gamma^2-1)^2\widehat J_2+\tilde E_{4}^1 }{(\gamma^2-1)^3}\,.\qquad 
\eea
Note that, in terms of our general notation \eq{various_expansions}, $\widehat{\mathcal E}$ could
be replaced by
\be
\widehat E_3(\g) \equiv h^4 E_3 = \pi \pinf^3 \widehat{\mathcal E}\,.
\ee 
Our general PM-expanded expressions
for the impulse $\Delta p_1$ are given, up to order $G^5$, in Table \ref{tab:impulse_coeffs3}.


\begin{table*}  
\caption{\label{tab:impulse_coeffs3} PM-expanded expression for $\Delta p_1$ up to 5PM. We use the
notation  $\check u_1=\frac{u_1^- - \g \, u_2^-}{\g^2-1}$, $\check u_2=\frac{u_2^- - \g \, u_1^-}{\g^2-1}$,  $\widehat J_4=h^4J_4+h^3\nu E_4$, $\tilde P_5 = \sqrt{\frac{\gamma-1}{\gamma+1}} \tilde E_5+\frac{C_5h^2}{(\gamma^2-1)^{1/2}}$, where $\tilde P_5\equiv h^6 P_5$, $\tilde E_5\equiv h^6 E_5$ and $C_5 = p_\infty\left(P_5-\sqrt{\frac{\gamma-1}{\gamma+1}}E_5\right)_{\nu=0}
=\pi\left[\frac{73}{12}p_\infty^4-\frac{69707}{10080}p_\infty^6+\frac{939}{280}\pi^2p_\infty^7-\frac{97951}{5760}p_\infty^8+O(p_\infty^9)\right]$.}
\begin{ruledtabular}
\begin{tabular}{lll}
 1PM & $ -\frac{G m_1 m_2}{b} \frac{2(2\gamma^2-1)}{\sqrt{\gamma^2-1}} \hat b $\\
\hline
 2PM  
& $ \frac{Gm_1m_2}{b} \left[
-\frac{GM}{b} \frac{3\pi (5\gamma^2-1)}{4\sqrt{\gamma^2-1}}
\hat b 
+\frac{G}{b}\frac{2(2\gamma^2-1)^2}{\gamma^2-1} (m_1\check u_2-m_2\check u_1)
\right]$\\
\hline
3PM \, cons & $\frac{Gm_1m_2}{b}\left[-\frac{G^2m_1 m_2 }{b^2}\frac{2(2\gamma^2-1)^2}{\gamma^2-1}{\mathcal I}(v)\hat b + (m_1+m_2)  \frac{G^2}{b^2}\frac{3\pi (5\gamma^2-1)(2\gamma^2-1)}{(\gamma^2-1)}(m_1\check u_2-m_2\check u_1)\right]$\\
3PM \, rr, tot & $\frac{G^3 m_1^2 m_2^2}{b^3} \left[ -  \frac{2 (2 \g^2-1)^2}{(\g^2-1)}\,    {\cI}(v)\hat b  + \pi (\gamma-1) \widehat {\mathcal E}(\gamma) \check u_2 \right]$\\
\hline
4PM\, cons &$\frac{Gm_1m_2}{b}\left[   \frac{G^3M^3}{b^3} \frac{2h^3\left(2\chi_2^{\rm cons}(\chi_1^{\rm cons}) ^2-\chi_4^{\rm cons}\right) }{(\gamma^2-1)^{3/2}}\hat b +  \frac{G^3M^2}{b^3} \frac{2h^2\left(-(\chi_1^{\rm cons})^4+3(\chi_2^{\rm cons})^2+6\chi_1^{\rm cons}\chi_3^{\rm cons}\right)}{3 (\gamma^2-1) }(m_1\check u_2-m_2\check u_1)\right]$\\ 
4PM \, rr, tot & $\frac{G^4m_1^2m_2^2}{b^4}\left\{\left[\frac{\pi}{(\gamma^2-1)^{3/2}}M\left[
(6\gamma^2-5)\gamma\widehat {\mathcal E} 
-   \frac{3 (5\gamma^2-1) }{4} {\widehat J}_2 \right]
-   m_1  \pi \frac{(2\gamma^2-1)}{\sqrt{\gamma^2-1} (\gamma+1) } \widehat {\mathcal E} 
  -\frac{(2\gamma^2-1)}{(\gamma^2-1)^2}M\widehat J_3 \right]\hat b\right.$\\
&$\left.+\frac{1}{(\gamma^2-1)^2}\left[ \frac12 (2M \tilde E_4^0 +m_1\tilde E_4^1)\check u_2 +2(2\gamma^2-1)\widehat J_2  (m_1\check u_2-m_2\check u_1) \right]\right\}$\\
\hline
5PM\, cons &$\frac{Gm_1m_2}{b}\left\{-\frac{2}{15}\frac{G^4M^4}{b^4(\gamma^2-1)^2}h^4(-30\chi_3^{\rm  cons}(\chi_1^{\rm  cons})^2+15\chi_5^{\rm  cons}+2(\chi_1^{\rm  cons})^5-30(\chi_2^{\rm  cons})^2\chi_1^{\rm  cons} ) \hat b\right.$\\
&$\left. 
-\frac43 \frac{G^4M^3 h^3( 2\chi_2^{\rm  cons} (\chi_1^{\rm  cons})^3-3\chi_3^{\rm  cons} \chi_2^{\rm  cons} -3\chi_4^{\rm  cons} \chi_1^{\rm  cons} )}{b^4(\gamma^2-1)^{3/2}}
(m_1\check u_2-m_2\check u_1)
\right\}
$\\
5PM \, rr, tot & $\frac{G^5 m_1^2m_2^2}{b^5}\left\{
M^2\left[
\frac{3\pi^2\gamma(5\gamma^2-3)}{2(\gamma^2-1)^{3/2}}\widehat {\mathcal E}
-\frac{3\tilde \chi_3^{\rm  cons}}{(\gamma^2-1)^2}{\widehat J}_2
-\frac{3\pi (5\gamma^2-1)}{4(\gamma^2-1)^2}{\widehat J}_3
-\frac{(2\gamma^2-1)}{(\gamma^2-1)^{5/2}}{\widehat J}_4 
+\frac{\gamma(6\gamma^2-5)}{(\gamma^2-1)^{7/2}}\tilde E_4^0
\right]\right. 
$\\
& $+\frac{2(2\gamma^2-1)^3}{(\gamma^2-1)^{7/2}}(m_1^2+m_2^2+2m_1m_2\gamma){\widehat J}_2$\\
& $\left.
+Mm_1\left[
-\frac{3\pi^2(5\gamma^2-1)}{8(\gamma^2-1)^{1/2}(\gamma+1)}\widehat {\mathcal E}
-\frac{2\gamma^2-1}{(\gamma^2-1)^{5/2}(\gamma+1)}\tilde E_4^0
+\frac{\gamma(10\gamma^2-9)}{2(\gamma^2-1)^{7/2}}\tilde E_4^1
\right]
-m_1^2\frac{10\gamma^2+8\gamma-1}{2(\gamma^2-1)^{5/2}(\gamma+1)}\tilde E_4^1
\right\}\hat b$\\
& $+\frac{G^5 m_1^2m_2^2}{b^5}\left\{ \frac{M(2\gamma^2-1)}{(\gamma^2-1)^2}\left[\frac94 \pi (5\gamma^2-1)\widehat J_2 +\frac{2(2\gamma^2-1)}{\sqrt{\gamma^2-1}}\widehat J_3-8\pi \gamma (\gamma^2-1)\widehat {\mathcal E}\right](m_1\check u_2-m_2\check u_1) \right.$\\
& $\left.+\left[\frac{m_1(m_1-m_2)}{\gamma^2-1}\left(\pi\frac{(2\gamma^2-1)^2}{2(\gamma+1)}\widehat {\mathcal E}-\frac{C_5 }{(\gamma^2-1)^{3/2}} \right) +\tilde E_5 \frac{M^2}{(\gamma^2-1)^{5/2}}\right]\check u_2   \right\}$\\
 \end{tabular}
\end{ruledtabular}
\end{table*}

\section{PN-expanded impulse coefficients at $O(G^4)$ and $O(G^5)$}
\label{PN_impulse_coeffs}

Our general results, displayed in Table \ref{tab:impulse_coeffs3}, for the radiation-reacted impulses
 involve the radiative losses (in the c.m. frame) of energy, angular momentum
and linear momentum. These losses admit a double PM and PN expansion,
which can be expressed as (with $j = J/(G m_1 m_2)$, $\pinf = \sqrt{\g^2-1}$
and ${\bf P}^{\rm rad}= {P}^{\rm rad}_y {\mathbf e}_y$)
\bea
\frac{E^{\rm rad}}{M} &=&  + \nu^2  \left[ \frac{{ E}_{3}(\pinf)}{j^3}+ \frac{{ E}_{4}(\pinf)}{j^4}+  \cdots\right]  \,, \nonumber\\
\frac{ J^{\rm rad}}{J_{\rm c.m.}} &=& + \nu^1 \left[ \frac{{ J}_{2}(\pinf)}{j^2}+ \frac{{ J}_{3}(\pinf)}{j^3} + \cdots\right]\,, \nonumber\\
\frac{P_y^{\rm rad}}{M}&=& + \frac{m_2-m_1}{M} \nu^2 \left[ \frac{{ P}_{3}(\pinf)}{j^3}+ \frac{{ P}_{4}(\pinf)}{j^4}+  \cdots\right]\,.\nonumber\\
\eea
Here the subscripts $n$ (e.g., in $E_n$) label  the $n$PM order, i.e. $O(G^n)$.
The subsequent expansion of the various PM coefficients, $E_n(\pinf)$, $J_n(\pinf)$, $P_n(\pinf)$
in powers of $\pinf$ then corresponds to the usual PN expansion.

The only radiative losses that are known in a PM-exact way are $J_2$ \cite{Damour:2020tta}
and, $E_3$ and $P_3$ \cite{Herrmann:2021tct}. In order to compute the radiation-reacted
 impulses at PM orders $G^4$ and $G^5$, we also need to know the values of $J_3$, $E_4$ and $P_4$.
 Existing results in the PN literature (notably \cite{Blanchet:1993ec,Junker:1992kle})
provide the values of the needed radiative losses only at the NLO PN accuracy, i.e., with 1PN fractional
accuracy beyond the LO PN result. 
Using techniques we already used in our previous works,
we have computed the various needed radiative losses along hyperbolic motions at the NNLO 
accuracy, i.e. with 2PN fractional accuracy. In addition, we also evaluated the fractional 1.5PN
correction coming from the {\it time-antisymmetric tail} contribution to radiation reaction.
Details on our computations are given in Appendices \ref{App_tail_E}, \ref{App_tail_J}, and \ref{App_tail_P}.
We computed the complete, 2PN expressions for the radiative losses of energy, angular momentum and linear momentum in terms of two independent orbital parameters (namely, $\bar a_r$ and $e_r$).
These results can then be easily expanded (to any needed order) in powers
of $\frac1j$, i.e. in powers of $G$. The explicit 2PN-accurate results for $E_n(\pinf)$, $J_n(\pinf)$ and $P_n(\pinf)$ up to $n=7$ will be found in Appendix \ref{PNAppendix}, Table \ref{tab:table_EnJnPn}.  Let us only display here our final results
for the PN expansion of the three functions of $\g$: $\widetilde E_4^0(\g)$, $\widetilde E_4^1(\g)$
and $ \widehat J_3(\g)$
\bea
\label{tildeE401hatJ3}
\widetilde E_4^0(\g) &=& \frac{1568}{45}p_\infty^3+\frac{18608}{525}p_\infty^5+\frac{3136}{45}p_\infty^6+\frac{220348}{11025}p_\infty^7\,,\nonumber\\
\widetilde E_4^1(\g) &=& -\frac{352}{45}p_\infty^5+\frac{1736}{225}p_\infty^7 -\frac{96}{5}p_\infty^8 \,,\nonumber\\
\widehat J_3(\g) &=& \pi\left(\frac{28}{5}\pinf^2+\frac{739}{84} \pinf^4-\frac{5777}{2520} \pinf^6 \right)\,.  
\eea
Here, we have used  the PM-exact relation  \eqref{P4nu}, and our computation of the
tail contribution to $P_4$ (see Appendix \ref{App_tail_P}) to add the 2.5PN tail contribution 
$-\frac{96}{5}p_\infty^8\,,$ to $\widetilde E_4^1$.

Let us explicitly display here our fractionally 2PN-accurate $O(G^4)$ and $O(G^5)$ results for the various impulse coefficients. [The corresponding  results at orders $O(G^{\le 3})$ will be found in Appendix \ref{PNAppendix}.] Let us emphasize  that, as radiation-reaction effects start at the 2.5PN
level ($\eta^5= \frac1{c^5}$), our radiation-reaction contributions to the impulses reach the
absolute 4.5PN accuracy ($\eta^9= \frac1{c^9}$), say
\be \label{PNdprrG4}
\Delta p_{a \mu}^{\rm rr, tot 4PM} \sim \frac1{c^5} + \frac1{c^7} +\frac1{c^8} +\frac1{c^9}.
\ee
Let us decompose the generic impulse coefficient $c_X$, $X=b, u_1, u_2$ (as well as $\widehat c_{u_1}$)
 in the form 
\bea
c_X=\mu \sum_{n=1}^\infty \left( \frac{GM}{b}\right)^n c_{X,G^n}\,,
\eea
where $X=b, u_1, u_2$, and
\bea
c_{X,G^n}&=&c_{X,G^n}^{\rm cons}+c_{X,G^n}^{\rm rr, rel}+c_{X,G^n}^{\rm rr, rec}\nonumber\\
&\equiv& c_{X,G^n}^{\rm cons}+c_{X,G^n}^{\rm rr, tot}\,,
\eea
where $c_{X,G^n}^{\rm rr, tot}\equiv c_{X,G^n}^{\rm rr, rel}+c_{X,G^n}^{\rm rr, rec}$.
\begin{widetext}
For $X=b$ at $O(G^4)$ we find (recalling the notation $\Delta \equiv \sqrt{1-4\nu}$)
\bea
c_{b,G^4}^{\rm cons}&=& \pi\left[\frac{6}{p_\infty^5}+\frac{\left(\frac{21}{4}+\frac{27}{2}\nu\right)}{p_\infty^3}
+\left(-\frac{99}{4}+\left(\frac{797}{8}-\frac{123}{128}\pi^2\right)\nu  \right)\frac{1}{p_\infty}
+O(p_\infty)\right]
\,,\nonumber\\
c_{b,G^4}^{\rm rr, rel}&=& \pi \nu \left[-\frac{47}{15p_\infty^2}-\frac{37}{15}\nu -\frac{10037}{840} +\left(-\frac{5501}{840}\nu +\frac{37}{15} \nu^2-\frac{221993}{10080}\right)p_\infty^2+O(p_\infty^4)\right]
\,,\nonumber\\
c_{b,G^4}^{\rm rr, rec}&=& -\pi\nu\left[-\frac{37}{60} \Delta -\frac{37}{15}\nu +\frac{37}{60} 
+\left(-\frac{5501}{840}\nu +\frac{1661}{1120} +\frac{37}{15}\nu^2-\frac{1661}{1120} \Delta\right) p_\infty^2+O(p_\infty^4)\right]
\,,\nonumber\\
c_{b,G^4}^{\rm rr, tot}&=&\pi\nu\left[\frac{37}{60}\Delta -\frac{2111}{168}-\frac{47}{15 p_\infty^2}+\left(-\frac{118471}{5040}+\frac{1661}{1120}\Delta \right) p_\infty^2
+O(p_\infty^4)\right]\,,
\eea
while at $O(G^5)$
\bea
c_{b,G^5}^{\rm cons}&=&  -\frac{2}{p_\infty^9}+\frac{(12-4\nu)}{p_\infty^7}+\frac{(9\pi^2+80+9\nu-2\nu^2)}{p_\infty^5}+O\left(\frac1{p_\infty^3}\right) 
\,,\nonumber\\
c_{b,G^5}^{\rm rr, rel}&=&\nu\left[ \frac{416}{45p_\infty^4} +\frac{(\frac{203264}{1575} -\frac{47}{5}\pi^2-\frac{96}{5} \nu)}{p_\infty^2}
-\frac{896}{45p_\infty}
+\frac{128}{3} \nu^2-\frac{37}{10}\nu\pi^2-\frac{5697}{280}\pi^2-\frac{73592}{1575} \nu +\frac{116992}{1225} 
+O(p_\infty) \right]  
\,,\nonumber\\
c_{b,G^5}^{\rm rr, rec}&=& -\nu \left[\left(\frac{32}{3} -\frac{32}{3}\Delta -\frac{128}{3}\nu \right)\frac{1}{p_\infty^2}
-\frac{13696}{525}\Delta +\frac{128}{3}\nu^2+\frac{13696}{525} -\frac{37}{10}\nu \pi^2+\frac{37}{40} \pi^2-\frac{20128}{175}\nu -\frac{37}{40} \pi^2\Delta\right.\nonumber\\
&&\left.
+O(p_\infty)\right] 
\,,\nonumber\\
c_{b,G^5}^{\rm rr, tot}&=&\nu \left[\frac{416}{45 p_\infty^4}+\left(-\frac{47}{5}\pi^2+\frac{186464}{1575}+\frac{352}{15}\nu+\frac{32}{3}\Delta\right)\frac{1}{p_\infty^2}
-\frac{896}{45 p_\infty}+\left(\frac{37}{40}\pi^2+\frac{13696}{525}\right)\Delta-\frac{1489}{70}\pi^2+\frac{255104}{3675}\right.\nonumber\\
&&\left.
+\frac{21512}{315}\nu+O(p_\infty)\right]
\,.
\eea

For $X=u_1$ at $O(G^4)$ we find
\bea
c_{u_1,G^4}^{\rm cons}&=& \frac{2}{p_\infty^8}+\frac{\left(2\nu-\frac12 \Delta-\frac{15}{2}\right)}{p_\infty^6}+\frac{(24\nu-\frac{913}{8}-\frac{9}{2}\pi^2+\frac{17}{8}\Delta-\frac12 \Delta\nu)}{p_\infty^4}+O\left(\frac1{p_\infty^2}\right)  
\,,\nonumber\\
c_{u_1,G^4}^{\rm rr, rel}&=& \nu\left[-\frac{1856}{ 45 p_\infty^3}
+\frac{\left(\frac{464}{45} \Delta+\frac{128}{3} \nu -\frac{17296}{175} \right)}{p_\infty}
-\frac{3136}{45} 
+\left(-\frac{32}{3} \Delta \nu -\frac{1266484}{11025} -\frac{128}{3}\nu^2+\frac{30796}{1575} \Delta+\frac{8928}{175}\nu \right) p_\infty\right.\nonumber\\
&&\left.
+O(p_\infty^2)\right]  
\,,\nonumber\\
c_{u_1,G^4}^{\rm rr, rec}&=& -\nu \left[ \frac{(-\frac{32}{3} +\frac{128}{3} \nu +\frac{32}{3}\Delta )}{ p_\infty}
+ \left(\frac{8928}{175}\nu -\frac{128}{3}\nu^2-\frac{5296}{525} +\frac{5296}{525}\Delta -\frac{32}{3}\Delta\nu \right) p_\infty 
+O(p_\infty^2)\right] 
\,,\nonumber\\
c_{u_1,G^4}^{\rm rr, tot}&=&\nu\left[-\frac{3136}{45}-\frac{1856}{45 p_\infty^3}+\left(-\frac{46288}{525}-\frac{16}{45}\Delta\right)\frac{1}{p_\infty}+\left(-\frac{1155268}{11025}+\frac{14908}{1575}\Delta \right) p_\infty+O(p_\infty^2)\right] 
\,,
\eea
while at $O(G^5)$
\bea
c_{u_1,G^5}^{\rm cons}&=& \pi \left[\frac{6}{p_\infty^8}+\frac{(-\frac{87}{2} +21\nu-\frac{3}{2} \Delta)}{p_\infty^6} 
+\left(-\frac{2961}{8}+\frac{93}{8}\Delta+\frac{457}{2}\nu-\frac{21}{4}\Delta \nu-\frac{123}{64}\nu\pi^2 \right)\frac{1}{p_\infty^4}
+O\left(\frac1{p_\infty^2}\right)\right]  
\,,\nonumber\\
c_{u_1,G^5}^{\rm rr, rel}&=& \pi \nu \left[-\frac{178}{5p_\infty^5}+\frac{\left(\frac{89}{2}\nu +\frac{89}{10}\Delta-\frac{3407}{24}\right)}{p_\infty^3}
-\frac{297\pi^2}{20p_\infty^2}   
+\frac{\left(-\frac{89}{8}\Delta\nu +\frac{14899}{480}\Delta-\frac{89}{2}\nu^2+\frac{98153}{1680} \nu -\frac{254319}{1120}\right)}{p_\infty} 
+O(p_\infty^0)\right] 
\,,\nonumber\\
c_{u_1,G^5}^{\rm rr, rec}&=&  -\pi\nu \left[\frac{\left(\frac{106}{3}\nu -\frac{53}{6} +\frac{53}{6}\Delta\right)}{p_\infty^3}+\frac{\left(\frac{105083}{10080}\Delta-\frac{89}{8}\Delta\nu 
+\frac{66559}{1260}\nu -\frac{89}{2}\nu^2-\frac{105083}{10080} \right)}{p_\infty}
+O(p_\infty^0)\right] 
\,,\nonumber\\
c_{u_1,G^5}^{\rm rr, tot}&=& \pi \nu \left[-\frac{178}{5 p_\infty^5}+\left(-\frac{1065}{8} +\frac{1}{15} \Delta +\frac{55}{6} \nu\right)\frac{1}{p_\infty^3}
-\frac{297}{20 p_\infty^2}\pi^2+\left(\frac{51949}{2520} \Delta+\frac{28223}{5040} \nu-\frac{545947}{2520} \right)\frac{1}{p_\infty} +O(p_\infty^0)\right]
\,.\nonumber\\
\eea

For $X=u_2$ at $O(G^4)$ we find
\bea
c_{u_2,G^4}^{\rm cons}&=& -\frac{2}{p_\infty^8}+\frac{\left(\frac{15}{2}-\frac12 \Delta-2\nu\right)}{p_\infty^6}
+\frac{\left(\frac{17}{8}\Delta+\frac{9}{2}\pi^2-\frac12 \Delta\nu-24\nu+\frac{913}{8}\right)}{p_\infty^4}
+O\left(\frac1{p_\infty^2}\right) 
\,,\nonumber\\
c_{u_2,G^4}^{\rm rr, rel}&=&\nu \left[\frac{1856}{45p_\infty^3}+\frac{\left(-\frac{128}{3}\nu +\frac{128224}{1575} -\frac{64}{9}\Delta \right)}{p_\infty} 
+\frac{3136}{45} 
+\left(\frac{128}{3}\nu^2+\frac{42844}{441} -\frac{15584}{525}\nu +\frac{32}{3}\Delta\nu +\frac{412}{225}\Delta \right) p_\infty\right.\nonumber\\
&&\left. 
+O(p_\infty^2)\right]  
\,,\nonumber\\
c_{u_2,G^4}^{\rm rr, rec}&=& -\nu \left[\frac{\left(-\frac{32}{3}\Delta -\frac{128}{3}\nu +\frac{32}{3} \right)}{p_\infty}+\left(\frac{832}{175} +\frac{32}{3}\Delta\nu +\frac{128}{3}\nu^2-\frac{832}{175}\Delta -\frac{15584}{525}\nu \right) p_\infty
+O(p_\infty^2)\right] 
\,,\nonumber\\
c_{u_2,G^4}^{\rm rr, tot}&=&\nu \left[\frac{1856}{45p_\infty^3}+\left(\frac{32}{9}\Delta +\frac{111424}{1575}\right)\frac{1}{p_\infty}+\frac{3136}{45}+\left(\frac{1018684}{11025}+\frac{10372}{1575}\Delta \right) p_\infty+O(p_\infty^2)\right]
\,,
\eea
while at $O(G^5)$
\bea
c_{u_2,G^5}^{\rm cons}&=&  \pi\left[-\frac{6}{p_\infty^8}-\frac32 \frac{(-29+14\nu+\Delta)}{p_\infty^6}+ \frac{\left(\frac{2961}{8}+\frac{93}{8}\Delta-\frac{457}{2}\nu-\frac{21}{4}\Delta \nu +\frac{123}{64}\pi^2\nu \right)}{p_\infty^4}
+O\left(\frac1{p_\infty^2}\right)\right] 
\,,\nonumber\\
c_{u_2,G^5}^{\rm rr, rel}&=& \pi \nu \left[ \frac{178}{ 5 p_\infty^5}
+\frac{\left(-\frac{89}{2} \nu -\frac{33}{10}\Delta+\frac{15571}{120} \right)}{p_\infty^3}
+\frac{297}{20p_\infty^2} \pi^2  
+\frac{(-\frac{60773}{1680}\nu +\frac{89}{8} \Delta\nu +\frac{21307}{3360} \Delta+\frac{226657}{1120} +\frac{89}{2}\nu^2)}{p_\infty}
+O(p_\infty^0)\right] 
\,,\nonumber\\
c_{u_2,G^5}^{\rm rr, rec}&=& -\pi \nu \left[\frac{\left(\frac{53}{6} -\frac{106}{3}\nu -\frac{53}{6}\Delta\right)}{p_\infty^3}
+\frac{\left(\frac{89}{2}\nu^2-\frac{44299}{1260}\nu +\frac{89}{8}\Delta\nu -\frac{60563}{10080}\Delta+\frac{60563}{10080}\right)}{p_\infty}
+O(p_\infty^0)\right] 
\,,\nonumber\\
c_{u_2,G^5}^{\rm rr, tot}&=& \pi \nu\left[\frac{178}{5p_\infty^5}+\left(\frac{4837}{40}+\frac{83}{15}\Delta -\frac{55}{6} \nu\right)\frac{1}{p_\infty^3}
+\frac{297}{20 p_\infty^2}\pi^2+\left(\frac{31121}{2520}\Delta -\frac{5123}{5040}\nu+\frac{197935}{1008}\right)\frac{1}{p_\infty}+O(p_\infty^0)\right]
\,.
\eea

 Finally, for $\widehat c_{u_1}= c_{u_1} + \g c_{u_2}$ at $O(G^4)$ we find
\bea
\widehat c_{u_1,G^4}^{\rm cons}&=&  (1+\Delta)\left[-\frac{1}{p_\infty^6}+\frac{(4-\nu)}{p_\infty^4}+\frac{\left(\frac{9}{4}\pi^2+56-\frac{47}{4}\nu  \right)}{p_\infty^2}
	+O(p_\infty^0)\right]
	\,,\nonumber\\
\widehat c_{u_1,G^4}^{\rm rr, rel}&=& (1+\Delta)\nu \left[\frac{16}{5 p_\infty}+\frac{624}{35} p_\infty+\frac{1952}{63}  p_\infty^3
+O(p_\infty^5)\right] 
\,,\nonumber\\
\widehat c_{u_1,G^4}^{\rm rr, rec}&=& 0  
\,,\nonumber\\
\widehat c_{u_1,G^4}^{\rm rr, tot}&=& \widehat c_{u_1,G^4}^{\rm rr, rel}
\,,
\eea
while at $O(G^5)$
\bea
\widehat c_{u_1,G^5}^{\rm cons}&=& \pi (1 +\Delta)\left[ -\frac{3}{p_\infty^6}+\frac{\left(-\frac{21}{2}\nu+\frac{45}{2}\right)}{p_\infty^4}
+\frac{\frac{717}{4}-\frac{893}{8}\nu +\frac{123}{128}\nu\pi^2}{p_\infty^2}
+O(p_\infty^0)\right]  
\,,\nonumber\\
\widehat c_{u_1,G^5}^{\rm rr, rel}&=& \pi (1+\Delta)\nu \left( \frac{28}{5 p_\infty^3}+\frac{15007}{420 p_\infty}+\frac{235867}{2520} p_\infty
+O(p_\infty^3)\right) 
\,,\nonumber\\
\widehat c_{u_1,G^5}^{\rm rr, rec}&=& 0 
\,,\nonumber\\
\widehat c_{u_1,G^5}^{\rm rr, tot}&=& \widehat c_{u_1,G^5}^{\rm rr, rel} 
\,.
\eea

\end{widetext}
It is easily checked that all the above coefficients are (when multiplied by $( G M)^n$) polynomials in the masses.

\section{Conservative scattering beyond $O(G^4)$ at 5PN accuracy; comparison with EFT results} 

To complete the results given in previous sections, we shall now discuss the (conservative) radiation-graviton contribution
to scattering at PM orders $G^4$, $G^5$ and $G^6$, when working at the 5PN accuracy, ie. the next-to-leading-order (NLO)
in the tail-related, radiation-graviton contribution. Indeed, on the one hand, a recent EFT study of Bl\"umlein et al. \cite{Blumlein:2020pyo}
has computed the potential-graviton contribution to the 5PN two-body Hamiltonian, while, on the other hand,
 a pioneering work of  Foffa and Sturani \cite{Foffa:2019eeb,Foffa:2021pkg}
has derived,  in the EFT approach, semi-explicit\footnote{The qualification ``semi-explicit" here refers to  the fact that
the results of Refs. \cite{Foffa:2019eeb,Foffa:2021pkg} are expressed in terms of PN-corrected, dimensionally-continued multipole moments,
which need to be separately computed. See below.} expressions for the nonlocal-in-time NLO (5PN-level) conservative contributions 
coming from the exchange of radiation gravitons.  The PN-expanded, EFT computation of potential gravitons to 
the conservative dynamics was found to agree, both at 5PN \cite{Blumlein:2020pyo} and at 6PN \cite{Blumlein:2021txj} with the
resummed, $O(G^4)$ potential-graviton dynamics of Ref. \cite{Bern:2021dqo}.

On the other hand, the TF approach has independently computed the full, local-plus-nonlocal, conservative dynamics
at the 6PN accuracy, in terms of a small number of undetermined coefficients that enter at order $G^5$, $G^6$ and $G^7$.
In particular, if we focus on the 5PN accuracy, the TF approach involves only two undetermined coefficients:
one of them, denoted ${\bar d}_5^{\nu^2}$, enters at the $G^5$ level, while the other one, ${a}_6^{\nu^2}$,
enters at the $G^6$ level. 
In addition, the TF approach has  allowed one to compute the (full, local-plus-nonlocal)
conservative scattering angle at the 5PN accuracy, and up to the sixth PM order, i.e, the 5PN-accurate values of
the coefficients $ \chi_4^{\rm cons, tot}$, $ \chi_5^{\rm cons, tot}$, $ \chi_6^{\rm cons, tot}$ in
 \be
 \frac12 \chi^{\rm cons, tot}(j, \g, \nu)= \sum_{n \geq1} \frac{\chi_n^{\rm cons, tot}(\g, \nu)}{j^n}\,.
 \ee 
We can then compare the TF-computed values of $ \chi_4^{\rm cons, tot}$, $ \chi_5^{\rm  cons, tot}$, $ \chi_6^{\rm  cons, tot}$ to
the corresponding results obtained by adding to the corresponding (local) potential-graviton contributions $ \chi_n^{\rm cons, pot}$,
computable from Ref. \cite{Blumlein:2020pyo}, the additional, radiation-graviton contributions $ \chi_n^{\rm cons, rad}$, computable
from the results of Ref. \cite{Foffa:2019eeb}. Let us briefly sketch how we performed these various computations,
and what results we got.

Using results already derived in our previous TF  work \cite{Bini:2019nra,Bini:2020wpo,Bini:2020nsb,Bini:2020hmy,Bini:2020rzn}, we find that
the PN expansions of the energy-rescaled scattering coefficients $\tilde \chi_n^{\rm cons} = h^{n-1} \chi_n^{\rm cons}$ read
\begin{widetext}
\bea
\label{tildechi4consminschwTF}
\tilde \chi_4^{\rm cons, TF}- \chi_4^{\rm Schw}&=&
\pi\nu\left[-\frac{15}{4}+\left(\frac{123}{256}\pi^2-\frac{557}{16}\right)p_\infty^2
+\left(-\frac{6113}{96}+\frac{33601}{16384}\pi^2-\frac{37}{5}\ln\left(\frac{p_\infty}{2}\right)\right)p_\infty^4\right.\nonumber\\
&&\left.
+\left(-\frac{615581}{19200}+\frac{93031}{32768}\pi^2-\frac{1357}{280}\ln\left(\frac{p_\infty}{2}\right)\right)p_\infty^6
+O(p_\infty^8)
\right]
\,,
\eea

\bea
\label{tildechi5consminschwTF}
\tilde \chi_5^{\rm cons, TF}- \chi_5^{\rm Schw}&=&\nu\left[
\frac{2}{ 5 p_\infty^3}
+\left(-\frac{121}{10}+\frac{1}{5}\nu\right)\frac{1}{p_\infty}
+\left(\frac{41}{8}\pi^2-\frac{19457}{60}+\frac{59}{10}\nu\right)p_\infty\right.\nonumber\\
&&
+\left(-\frac{41}{24}\nu\pi^2+\frac{5069}{144}\pi^2-\frac{6272}{45}\ln\left(\frac{p_\infty}{2}\right)-\frac{12544}{45}\ln(2)+\frac{10681}{144}\nu-\frac{5211479}{4320}\right)p_\infty^3\nonumber\\
&&
+\left(-\frac{782142451}{504000}-\frac{365555}{6048}\nu+\frac{2816}{45}\nu\ln(2)+\frac{111049}{960}\pi^2+\frac{23407}{5760}\nu\pi^2
+\frac{1408}{45}\nu\ln\left(\frac{p_\infty}{2}\right)\right.\nonumber\\
&&\left.\left.-\frac{148864}{525}\ln(2)-\frac{4}{15}\nu\bar d_5^{\nu^2}-\frac{74432}{525}\ln\left(\frac{p_\infty}{2}\right)\right)p_\infty^5
+O(p_\infty^7)
\right]\,,
\eea
and
\bea
\label{tildechi6consminschwTF}
\tilde \chi_6^{\rm cons, TF}- \chi_6^{\rm Schw}&=&
\pi\nu\left[
\frac{105}{16}\nu+\frac{615}{256}\pi^2-\frac{625}{4}\right.\nonumber\\
&&+\left(-\frac{228865}{192}-\frac{1845}{512}\nu\pi^2+\frac{257195}{8192}\pi^2
-\frac{2079}{8}\zeta(3)+\frac{10065}{64}\nu-122\ln\left(\frac{p_\infty}{2}\right)\right)p_\infty^2\nonumber\\
&&
+\left(\frac{2817}{16}\nu\zeta(3)+\frac{2321185}{16384}\pi^2-\frac{15}{32}\nu \bar d_5^{\nu^2} -\frac{13831}{56}\ln\left(\frac{p_\infty}{2}\right)
+\frac{30161}{192}\nu+\frac{201}{2}\nu \ln\left(\frac{p_\infty}{2}\right)\right.\nonumber\\
&&\left.\left.
-\frac{61855}{32768}\nu\pi^2+\frac{49941}{64}\zeta(3)-\frac{15}{32}\nu a_6^{\nu^2} -\frac{9216}{7}\ln(2)-\frac{22898669}{6720}\right)p_\infty^4
+O(p_\infty^6)
\right]\,,
\eea
at the 5PN accuracy.

\end{widetext}
The TF result for $\tilde \chi_4^{\rm cons, TF}$ is fully
explicit and does not involve any undetermined coefficient. By contrast, the TF result for $\tilde \chi_5^{\rm cons, TF}$
involve the yet-undetermined $G^5$-level theory coefficient ${\bar d}_5^{\nu^2}$.
Similarly $\tilde \chi_6^{\rm cons, TF}$ involves both ${\bar d}_5^{\nu^2}$
and the yet-undetermined $G^6$-level theory coefficient ${a}_6^{\nu^2}$. 

Concerning the EFT side of the calculation, we proceeded as follows. The additional term in the conservative
effective action due to radiation-graviton exchange
was derived by Foffa and Sturani \cite{Foffa:2019eeb,Foffa:2021pkg}   in the form (using the notation $D = d+1=4+ \varepsilon$)\footnote{We use here the script  $\varepsilon$ for $D=4+\varepsilon$, and reserve $\epsilon$ for  $D= 4- 2 \epsilon$.}
\bea \label{Srad}
S_{\rm rad} &=& \eta^8 S_{I_2} +  \eta^{10} S_{I_3}+  \eta^{10} S_{J_2}\nonumber\\
&& + \eta^{10}[S_{_{QQL}}+ S_{_{QQQ_1}} + S_{_{QQQ_2}}]\,,
\eea
where (with $\eta\equiv\frac1c$)  $S_{I_2}$ is the contribution of the (1PN-accurate, $d$-dimensional) mass-type (or electric-type) quadrupole moment $I_{ij}^{(d)}$,
$S_{I_3}$ is the contribution of the (0PN-accurate, $d$-dimensional) mass-type (electric-type) octupole moment $I_{ijk}^{(d)}$, and $S_{J_2}$ is the contribution of the (0PN-accurate, $d$-dimensional) spin-type 
(or magnetic-type) quadrupolar moment $J_{ij}^{(d)}$. When working in the frequency domain, the various (nonlocal) action contributions
read (with $E\equiv E_{\rm c.m.}$)
\bea
\label{S_Is}
S_{I_2}&=&-\frac{G E}{5} \int \frac{d\omega}{2\pi} \left(\frac{1}{\varepsilon}-R_{\rm quad, e}+{\mathcal L}\right) \left(I_{ij}^{d,(3)}\right)^2\,, \nonumber\\
S_{I_3}&=& -\frac{G E}{189} \int \frac{d\omega}{2\pi} \left(\frac{1}{\varepsilon}-R_{\rm oct, e}+{\mathcal L}\right) \left(I_{ijk}^{d,(4)}\right)^2\,, \nonumber\\  
S_{J_2}&=&-\frac{16G E}{45} \int \frac{d\omega}{2\pi} 
\left(\frac{1}{\varepsilon}-R_{\rm quad, m}+{\mathcal L}\right) \left(J_{ij}^{d,(3)}\right)^2
 \,,\nonumber\\
\eea
where $\left(I_{L}^{d,(3)}\right)^2 \equiv I_{L}^{d,(3)}(\omega)I_{L}^{d,(3)}(-\omega)$ and
\bea
{\mathcal L}&\equiv&\ln\left(\frac{\omega^2 e^{\gamma_E}}{\pi \mu_0^2} \right)\,.
\eea
Here, $\gamma_E$ denotes Euler's constant, $\mu_0=\frac{1}{\ell_0}$ is the mass scale entering dimensional regularization, i.e. $G_{[d]}=G \ell_0^{\varepsilon}$, where $G$ denotes the 4-dimensional gravitational constant. The terms $R_{\rm quad, e}$, $R_{\rm oct, e}$, $R_{\rm quad, m}$ denote
some {\it rational} numbers. These numbers have been evaluated by Foffa and Sturani to be
\beq 
R_{\rm quad, e}=\frac{41}{30}\,,\qquad
R_{\rm oct, e}=\frac{82}{35}\,,\qquad
R_{\rm quad, m}=\frac{127}{60}\,. 
\eeq
The first, electric-quadupole,  term $ R_{\rm quad, e}$ was first derived in Ref. \cite{Foffa:2011np},
and has been verified via the recent re-derivations of the 4PN dynamics 
\cite{Marchand:2017pir,Foffa:2019yfl}.
 By contrast, there are no published, detailed rederivations of the values
 of the (5PN-level) electric-octupole and magnetic-quadrupole rational terms 
$R_{\rm oct, e}$ and  $R_{\rm quad, m}$.  
In our computations, we shall therefore use the numerical value $R_{\rm quad, e}=\frac{41}{30}$
of $ R_{\rm quad, e}$, but leave the two other rational terms  as unspecified parameters.

The additional (local-in-time) contributions in the second line of the effective action Eq. \eq{Srad}, 
derived in Refs. \cite{Foffa:2019eeb,Foffa:2021pkg}, are defined  in dimension $d = 3$, and read
(with $\varepsilon_{ijk}= \varepsilon_{[ijk]}$, $\varepsilon_{123}=+1$,  and $L_i$ denoting the total
Newtonian angular momentum of the system)
\bea \label{SQQ}
S_{_{QQL}}&=& -C_{_{QQL}} G^2 \int dt I_{i s}^{(4)}  I_{js}^{(3)} \varepsilon_{ijk} L_k\,,\\
S_{_{QQQ_1}}&=& -C_{_{QQQ_1}} G^2 \int dt I_{i s}^{(4)}  I_{js}^{(4)} I_{ij}\,,\\
S_{_{QQQ_2}}&=& -C_{_{QQQ_2}} G^2 \int dt I_{i s}^{(3)}  I_{js}^{(3)}  I_{ij}^{(2)}\,.
\eea
Here, similarly to what we did for the rational coefficients
$R_{\rm quad, e}$, $R_{\rm oct, e}$ and $
R_{\rm quad, m}$, we have not specified the values of the  {\it rational} numerical coefficients $C_{_{QQL}},  C_{_{QQQ_1}},  C_{_{QQQ_2}}$. 
The estimated values of $C_{_{QQQ_1}}$,  $C_{_{QQQ_2}}$
 have varied between the published and  erratum versions of Ref. \cite{Foffa:2019eeb}.
The final estimates  given by Foffa and Sturani for these coefficients are
\beq \label{CFS}
C_{_{QQL}}=\frac{8}{15}\,,\qquad
C_{_{QQQ_1}}= \frac{1}{15}\,,\qquad
C_{_{QQQ_2}}= \frac{4}{105}\,.
\eeq
[Here we took into account the unconventional definition
$L_i^{\mbox{ \scriptsize\rm Ref. \cite{Foffa:2019eeb}}}= - L_i^{\rm standard}$,
and the sign conventions of Eqs. (9.9), (9.10), (9.11).] 
We shall later compare these values to the results derived from the TF approach.

The presence of  $\frac1{\varepsilon}$ (UV) poles in the first three contributions
 $S_{I_2}$, $  S_{I_3}$, $  S_{J_2}$ require that the multipole moments $I_{ij}^{(d)}$, $I_{ijk}^{(d)}$, $J_{ij}^{(d)}$
 be computed in $d$ dimensions. General $d$-dimensional expressions for the mass-type mutipole moments 
 $I_{L}^{(d)}$ have been derived some time ago in Ref. \cite{Blanchet:2005tk}, see Eq. (3.50) there.
We have used the latter expressions (together with the $d$-dimensional gravitational field expressions
derived in Refs. \cite{Blanchet:2005tk,Blanchet:2003gy}) to derive the 1PN-accurate values of the
$d$-dimensional mass quadrupole and mass octupole of a binary system. For instance, we find,
in the ($d$-dimensional) c.m. frame and at the 1PN level of accuracy (denoting $x^i=x_1^i-x_2^i$)
\beq
\mu^{-1}I_{ij}^{[d]}=  (
1+\eta^2 C_{1} ) x_{\langle ij\rangle }+\eta^2 C_2 v_{\langle ij\rangle}+\eta^2 C_3 x_{\langle i} v_{j\rangle}
\,,
\eeq
where  $C_i=C_i(r,v^2,({\mathbf v}\cdot {\mathbf n}))$ [using the standard  notation $r^2=x^2$, $({\mathbf v}\cdot {\mathbf n})=\dot r$, etc.] are given by
\bea
C_{1}&=&  (1-3\nu)\frac{ d^2(d+2)-8 (d-1) }{2 d (d+4) (d-2)} v^2\nonumber\\
&&+\frac{-(d+2)+2( d+1)\nu }{(d+4)}{\mathcal U}\,, \nonumber\\
C_2&=& (1-3\nu)r^2 \frac{(2 d+d^2-4) }{ d (d+4) (d-2)}\,, \nonumber\\  
C_3&=&  -4 (1-3\nu) r\frac{({\mathbf v}\cdot {\mathbf n}) }{ (d+4) (d-2)}\,,
\eea
where 
\beq
{\mathcal U}=f G_{[d]}M \tilde k |{\mathbf y}_1-{\mathbf y}_2|^{2-d}= G_{[d]}f\tilde k M   r^{2-d}\,,
\eeq
with
\beq
G_{[d]}=G \, \ell_0^{d-3}=G (1+\epsilon \ln (\ell_0))+O(\epsilon^2)\,,
\eeq
[The relations among the various scale factors is given in  footnote \ref{foot2}.]
and
\beq
\tilde k=\frac{\Gamma \left( \frac{d-2}{2}\right)}{\pi \left( \frac{d-2}{2}\right)}\,,\qquad
f=\frac{2 (d-2)}{(d-1)}\,.
\eeq
When $d=3$, ${\mathcal U}=GM/r$ and the coefficients $C_i$ reduce to
\bea
C_{1}&=&  -\frac{29}{42} (-1+3\nu) v^2+\frac17  (-5+8\nu) {\mathcal U}\,,\nonumber\\  
C_2&=& -\frac{11}{21} r^2 (-1+3\nu)\,,  \nonumber\\  
C_3&=&  \frac{4}{7} (-1+3\nu) r ({\mathbf v}\cdot {\mathbf n})\,.
\eea
Note that the needed third time derivative of $I_{ij}^{(d)}$ must also be computed by using the 1PN-accurate,
$d$-dimensional equations of motion (see Refs. \cite{Blanchet:2005tk,Blanchet:2003gy}).

The Newtonian-level accurate expression for $I_{ijk}^{[d]}$ is simply (in the c.m. frame)
\beq
I_{ijk}^{[d]}(t) =\nu (m_2-m_1) x_{\langle ijk\rangle}\,.
\eeq
It was shown in Ref. \cite{Blanchet:2005tk} that spin-type multipoles in $d$ dimensions
must be described by non-symmetric, mixed Young tableaux. In particular, the $d$-dimensional avatar 
of the spin-type quadrupole must be described by a rank-three tensor $J_{i|ab}$ that
is antisymmetric with respect to (say) $i$ and $b$, and that satisfies the cyclic identity
$J_{i|ab}+ J_{a|bi} +J_{b|ia}=0$. The explicit $d$-dimensional value of 
the spin-type quadrupole $J_{ij}$ has only been computed  very recently \cite{Henry:2021cek}.
Here, we only need its value at the Newtonian level. It reads 
\bea
\label{finale}
J_{i|ab}&=&\nu (m_2-m_1)\left[\left(x^{ia}-\frac{x^2}{d-1}\delta^{ia}\right)v_b\right.\nonumber\\
&&-\left(x^{ab}-\frac{x^2}{d-1}\delta^{ab}\right)v_i \nonumber\\
&&\left. -\frac{x\cdot v}{d-1}(x^i \delta^{ab}-x^b\delta^{ia}) \right]\,.
\eea
Though Ref. \cite{Foffa:2019eeb} did not use the needed $d$-dimensional magnetic quadrupole 
$J_{i|ab}$, their derivation rests on the coupling $\frac12 R_{0iab} \epsilon_{abj} J_{ij}$ which
could have been expressed (in $d$ dimensions) in terms of the coupling of $R_{0iab}$
to the relevant spin quadrupole 
$\epsilon_{abj} J_{ij} \equiv J_{b|ia}$. 
This implies that one should simply
re-interpret their action contribution $J_{ij}^{(3)} J_{ij}^{(3)}$ by the  replacement
\be
J_{ij}^{(3)} J_{ij}^{(3)} \to \frac12 J_{i|ab}^{(3)} J_{i|ab}^{(3)}\,,
\ee
where the right-hand side is to be evaluated (including the time derivatives) in $d$ dimensions.

A simple way to by-pass the subtle issues linked to the spin quadrupole
would be  to focus on the equal-mass case, $m_1=m_2$, or $\nu=\frac14$. For symmetry reasons,
the spin-quadrupole (as well as the mass-octupole) would then not contribute, in any dimension,
to the problematic nonlocal terms entering $S_{J_2}$ (and $S_{I_3}$).
Both terms are proportional to $\nu^2 (1-4 \nu)$.
In focussing on  the effective action for the specific value $\nu=\frac14$.
we would, however, reduce the number of constraints obtained below by comparing TF and EFT. 

The multipolar terms entering Eq. \eqref{S_Is} should be time-differentiated an appropriate number of times.
This is done by using the 1PN accurate $d$-dimensional Hamiltonian
\bea
\hat H^{[d]}_{\rm 1PN, h} &=& \frac12 p^2-{\cal U}-\eta^2\left\{
\frac18 (1-3\nu)p^4 \right.\nonumber\\
&+& \frac12 {\cal U} \left[
-{\cal U}+\left(\frac{d}{(d-2)}+\nu  \right) p^2 \right. \nonumber\\
&+&\left.\left.
\nu(d-2) ({\mathbf p}\cdot {\mathbf n})^2
\right]\right\}\,.
\eea
Introducing the notation
\bea
\label{calFdefs}
{\mathcal F}_{I_2}&=&\left(I_{ij}^{d,(3)}\right)^2
\,,\nonumber\\
{\mathcal F}_{I_3}&=&\left(I_{ijk}^{d,(3)}\right)^2
\,,\nonumber\\
{\mathcal F}_{J_2}&=&\frac12 \left(J_{i|ab}^{d,(3)}\right)^2\,,
\eea
the structure of these terms, after  replacing  $d=3+\varepsilon$ and expanding in $\varepsilon$ is such that
\bea
\label{calFexp}
{\mathcal F}_{I_2}&=&\nu^2\frac{G^2 M^4}{r^4}\left[{\mathcal F}_{I_2}^{00}+{\mathcal F}_{I_2}^{01}\varepsilon
+\eta^2 \left({\mathcal F}_{I_2}^{20} +{\mathcal F}_{I_2}^{21}\varepsilon\right) \right]
\,,\nonumber\\
{\mathcal F}_{I_3}&=&\nu^2\frac{G^2 M^4}{r^4}(1-4\nu)\left({\mathcal F}_{I_3}^{00}+{\mathcal F}_{I_3}^{01}\varepsilon\right)
\,,\nonumber\\
{\mathcal F}_{J_2}&=&\nu^2\frac{G^2 M^4}{r^4}(1-4\nu)\left({\mathcal F}_{J_2}^{00}+{\mathcal F}_{J_2}^{01}\varepsilon\right)\,,
\eea
with each $(\varepsilon)$ term containing a no-log part and a $\ln (r/r_0)$ part, e.g., 
\beq
{\mathcal F}_{I_L}^{i1}={\mathcal F}_{I_L}^{\rm (no-log)}{}^{i1}+{\mathcal F}_{I_L}^{(\ln )}{}^{i1}\ln \left(\frac{r}{r_0}\right)\,,
\eeq
where we recall the notation $ r_0 \equiv \frac{e^{-\gamma/2}}{2\mu_0 \sqrt{\pi}}$.
The complete results are listed in Table  \ref{FX_eta_eps} below.


\begin{table*}[ht]
\caption{ \label{FX_eta_eps}  
Coefficients of the first order $\varepsilon$-expansion \eq{calFexp} of the square of the $d$-dimensional multipoles \eq{calFdefs} with  $d=3+\varepsilon$. 
}
\begin{ruledtabular}
\begin{tabular}{l|l}
${\mathcal F}_{I_2}^{00}$&$32 p^2-\frac{88}{3} p_r^2$\\
${\mathcal F}_{I_2}^{01}$&$\left(-64 p^2+\frac{176}{3}p_r^2\right)\ln \left(\frac{r}{r_0}\right)
+ 96 p^2-\frac{836}{9} p_r^2 $\\ 
${\mathcal F}_{I_2}^{20}$&$
\left(\frac{9808}{21}-\frac{80}{21}\nu\right)\frac{G M}{r} p_r^2
+(168-64\nu)p_r^4
+\left(-\frac{9680}{21} -\frac{64}{7}\nu\right)\frac{G M}{r} p^2
+\left(\frac{208}{21}+\frac{1024}{7}\nu \right)p^4
+\left(-\frac{1352}{7} -\frac{1384}{21}\nu  \right)p^2p_r^2
$\\
${\mathcal F}_{I_2}^{21}$& $\left[
\left(\frac{9680}{7}+\frac{192}{7}\nu  \right) \frac{G M}{r} p^2
+\left(-\frac{416}{21}-\frac{2048}{7}\nu \right)p^4
+\left(\frac{2704}{7}+\frac{2768}{21}\nu \right)p^2p_r^2
+\left(-\frac{9808}{7}+\frac{80}{7}\nu\right)\frac{G M}{r} p_r^2
\right.$\\
&$\left.
+(-336+128\nu)p_r^4 \right]
\ln \left(\frac{r}{r_0}\right)$\\
& $+\left(\frac{3104}{7}-\frac{1752}{7}\nu\right)p_r^4
+\left(-\frac{595528}{441}-\frac{704}{147}\nu\right)\frac{G M}{r} p^2
+\left(-\frac{22224}{49} -\frac{74780}{441}\nu \right) p^2 p_r^2
+\left(\frac{644648}{441}-\frac{19024}{441}\nu\right)\frac{G M}{r} p_r^2
$\\
&$ 
+\left(-\frac{1744}{441}+ \frac{65248}{147}\nu \right)p^4$ \\
\hline
${\mathcal F}_{I_3}^{00}$&$1170 p_r^4+\left(-432\frac{G M}{r} -2364 p^2\right)p_r^2+\frac{288}{5}\frac{G^2 M^2}{r^2} +\frac{6042}{5} p^4+\frac{1872}{ 5 }\frac{G M}{r} p^2$\\
${\mathcal F}_{I_3}^{01}$&$\left( 1296 \frac{G M}{r} p_r^2-\frac{1152}{5} \frac{G^2 M^2}{r^2} -\frac{12084}{5}p^4-\frac{5616}{5} \frac{G M}{r} p^2
+4728 p^2p_r^2-2340p_r^4 \right)\ln \left(\frac{r}{r_0}\right)$\\
&$+
\frac{93333}{25} p^4-1764 \frac{G M}{r} p_r^2+\frac{8352}{25} \frac{G^2 M^2}{r^2} -7494 p^2p_r^2+3765p_r^4+\frac{39708}{25} \frac{G M}{r} p^2$\\ 
\hline
${\mathcal F}_{J_2}^{00}$&$\frac12 p^4+p^2 p_r^2-\frac{3}{2}p_r^4$\\
${\mathcal F}_{J_2}^{01}$&$\left( -p^4+3p_r^4-2 p^2p_r^2 \right)\ln \left(\frac{r}{r_0}\right)-\frac{29}{4}p_r^4+\frac{11}{2}p^2p_r^2+\frac{7}{4} p^4 $\\ 
\end{tabular}
\end{ruledtabular}
\end{table*}

It is easily seen that the (frequency-domain) terms involving the integral (in $d=3$) of 
${\mathcal L}=\ln\left(\frac{\omega^2 e^\gamma}{\pi \mu_0^2} \right)$
exactly correspond to the part of the nonlocal action denoted 
$- W_1^{\rm nonloc}$ in Ref. \cite{Bini:2020hmy}, i.e. the time-domain contribution
\be
- W_1^{\rm nonloc}= +2GE \int dt\,  {\rm Pf}_{2 r_0}\, \int \frac{dt'}{|t-t'|}{\mathcal F}_{\rm split, GW}^{[d=3]}(t,t').
\ee

When adding the (UV-divergent) 5PN-accurate radiation-graviton EFT action \eq{Srad} 
\beq
S_{\rm rad} =-\int dt H^{\rm rad},
\eeq
to the  (IR-divergent) 5PN-accurate 
potential-graviton (Hamiltonian) action derived in Ref. \cite{Blumlein:2020pyo}, 
\beq
S_{\rm pot} =-\int dt H^{\rm pot}\,,
\eeq
the $\frac1{\varepsilon}$ poles cancel
and one can (using the techniques explained  in our previous works) compute the large-impact-parameter
(or large-$j$) expansion of the (conservative) scattering angle predicted by the EFT dynamics.
Namely, one uses the $d$-dimensional energy conservation
\beq
\bar E=\hat H^{[d]}_{\rm pot} + \hat H^{[d]}_{\rm rad} 
\eeq
to obtain the radial momentum $p_r=p_r(\bar E, j; r)$ (expanded to the appropriate PN order,
and also expanded to first order in $\varepsilon$). Here $H^{[d]}_{\rm pot}$ is the 
(harmonic gauge) potential-graviton Hamiltonian $H^{\rm pot}$ of Ref. \cite{Blumlein:2020pyo},
while $H^{[d]}_{\rm rad} $ is the additional radiation-graviton Hamiltonian, 
deduced from the nonlocal action \eq{Srad}. 
The scattering angle is then obtained as the sum
of the potential-graviton (half) scattering angle $\frac{\chi^{\rm pot}}{2}$ (derived from 
the local Hamiltonian $H^{[d]}_{\rm pot}$) and the correction
coming from the nonlocal term $ H^{[d]}_{\rm rad} $, obtained as
\bea
\frac{\chi^{\rm rad}}{2}&=&\frac{1}{2\nu}\partial_j \int dt \hat H^{[d=3+\varepsilon]}_{\rm rad}|_{p_r=p_r(\bar E,j;r)}.
\eea
This method gave us the following  results for the PM expansion of the 5PN-accurate total scattering angle,
\be
 \chi^{\rm cons, EFT}=\chi^{\rm pot}+\chi^{\rm rad}= \sum_n \frac{2 \, \tilde \chi_n^{\rm cons, EFT}}{h^{n-1} j^n}\,,
 \ee
 predicted by the EFT approach, in terms of the various coefficients entering \eq{Srad}
 (namely  $R_{\rm oct, e}$, $R_{\rm quad, m}$,  $C_{_{QQL}},  C_{_{QQQ_1}},  C_{_{QQQ_2}}$):
\begin{widetext}
\bea
\label{tildechi4consminschwEFT}
\tilde \chi_4^{\rm cons, EFT}- \chi_4^{\rm Schw}&=&
\pi\nu\left\{-\frac{15}{4}+\left(\frac{123}{256}\pi^2-\frac{557}{16}\right)p_\infty^2
+\left(-\frac{6113}{96}+\frac{33601}{16384}\pi^2-\frac{37}{5}\ln\left(\frac{p_\infty}{2}\right)\right)p_\infty^4\right.\nonumber\\
&+&
\left(\left(-\frac{1697}{140} R_{\rm oct, e}-\frac{85}{12}C_{_{QQQ_2}} -\frac{253}{24} C_{_{QQQ_1}}  -\frac{3}{5} R_{\rm quad, m}+\frac{207}{8}C_{_{QQL}}+\frac{230281}{9800}\right)\nu\right.\nonumber\\
&-&\left.\left.
\frac{1357}{280}\ln\left(\frac{p_\infty}{2}\right)-\frac{7437721}{188160}+\frac{1697}{560} R_{\rm oct, e}+\frac{3}{20} R_{\rm quad, m}+\frac{93031}{32768}\pi^2\right)p_\infty^6\right\}\,,
\eea
\bea
\label{tildechi5consminschwEFT}
\tilde \chi_5^{\rm cons, EFT}- \chi_5^{\rm Schw}&=&\nu\left[
\frac{2}{5p_\infty^3}+\left(-\frac{121}{10}+\frac{1}{5}\nu \right)\frac{1}{p_\infty}
+\left(\frac{41}{8}\pi^2-\frac{19457}{60}+\frac{59}{10}\nu \right)p_\infty\right.\nonumber\\
&+&\left(-\frac{41}{24}\nu\pi^2+\frac{5069}{144}\pi^2-\frac{6272}{45}\ln\left(\frac{p_\infty}{2}\right)-\frac{12544}{45}\ln(2)+\frac{10681}{144}\nu-\frac{5211479}{4320}\right)p_\infty^3\nonumber\\
&+&\left(\left(-\frac{327424}{945} R_{\rm oct, e}+\frac{1408}{45}\ln\left(\frac{p_\infty}{2}\right)
+\frac{2176}{3}C_{_{QQL}}-\frac{38272}{135} C_{_{QQQ_1}}-\frac{7168}{45}C_{_{QQQ_2}}-\frac{2048}{135} R_{\rm quad, m}\right.\right.\nonumber\\
&+&\left.
\frac{2816}{45}\ln(2)-\frac{224057}{1440}\pi^2+\frac{2423563619}{1058400}\right)\nu \nonumber\\
&-&\left.\left.\frac{148864}{525}\ln(2)+\frac{512}{135} R_{\rm quad, m}-\frac{6223745957}{3528000}+\frac{81856}{945} R_{\rm oct, e}-\frac{74432}{525}\ln\left(\frac{p_\infty}{2}\right)+\frac{111049}{960}\pi^2\right)p_\infty^5\right]\,,\nonumber\\
\eea
and
\bea
\label{tildechi6consminschwEFT}
\tilde \chi_6^{\rm cons, EFT}- \chi_6^{\rm Schw}&=&\pi\nu\left[
\frac{615}{256}\pi^2-\frac{625}{4}+\frac{105}{16}\nu\right.\nonumber\\
&+&
\left(-\frac{228865}{192}-\frac{1845}{512}\nu\pi^2+\frac{257195}{8192}\pi^2-\frac{2079}{8}\zeta(3)+\frac{10065}{64}\nu-122\ln\left(\frac{p_\infty}{2}\right)\right)p_\infty^2\nonumber\\
&+&
\left(\left(-\frac{55}{3} R_{\rm quad, m}+\frac{201}{2}\ln\left(\frac{p_\infty}{2}\right)-\frac{8645}{24} C_{_{QQQ_1}} +\frac{7575}{8} C_{_{QQL}}+\frac{2817}{16}\zeta(3)-\frac{38621}{84} R_{\rm oct, e}\right.\right.\nonumber\\
&-&\left.
\frac{10812865}{32768}\pi^2+\frac{14120063}{3136}-\frac{695}{4} C_{_{QQQ_2}} \right)\nu
\nonumber\\
&+&
\frac{38621}{336}R_{\rm oct, e}+\frac{2321185}{16384}\pi^2+\frac{55}{12} R_{\rm quad, m}-\frac{173486591}{47040}-\frac{13831}{56}\ln\left(\frac{p_\infty}{2}\right)\nonumber\\
&+&\left.\left. \frac{49941}{64}\zeta(3)-\frac{9216}{7}\ln(2)\right)p_\infty^4\right]\,.
\eea
\end{widetext}

We can now compare, term by term, these expressions (which are 5PN accurate and
respectively belong to the 4PM, 5PM and 6PM approximations)
with the corresponding ones derived from the TF approach, and displayed in Eqs. \eq{tildechi4consminschwTF}, \eq{tildechi5consminschwTF} and \eq{tildechi6consminschwTF} above.

Let us denote as $\Delta C_{\chi_n}(\nu^p \pinf^q)$ the coefficient of $\nu^p \pinf^q$
in the difference $\tilde \chi_n^{\rm cons, EFT}- \tilde \chi_n^{\rm cons, TF}$. 
The comparison, at the 5PN level, between
the TF and EFT scattering angles yields six equations, namely
\bea \label{TFvsEFT}
&&\Delta C_{\chi_4}(\nu^1 \pinf^6) =0\,,\qquad \Delta C_{\chi_4}(\nu^2 \pinf^6) =0\,, \nonumber\\
&&\Delta C_{\chi_5}(\nu^1 \pinf^5) =0\,,\qquad \Delta C_{\chi_5}(\nu^2 \pinf^5) =0\,, \nonumber\\
&&\Delta C_{\chi_6}(\nu^1 \pinf^4) =0\,,\qquad \Delta C_{\chi_6}(\nu^2 \pinf^4) =0\,. 
\eea
These six equations are linear in the seven variables $R_{\rm oct, e}$, $R_{\rm quad, m}$,
$C_{_{QQL}}$ $ C_{_{QQQ_1}}$ $ C_{_{QQQ_2}}$, $\bar d_5^{\nu^2}$ and $a_6^{\nu^2}$.
Their explicit form reads
\begin{widetext}
\bea
\Delta C_{\chi_4}(\nu^1 \pinf^6) &=&
\pi  \left(\frac{1697 }{560}R_{\rm oct, e}+\frac{3
   }{20}R_{\rm quad, m}-\frac{146357}{19600}\right)
\,,\nonumber\\
\Delta C_{\chi_4}(\nu^2 \pinf^6) &=&
\pi  \left(-\frac{253 }{24}C_{_{QQQ_1}}-\frac{85 }{12}C_{_{QQQ_2}}+\frac{207
   }{8}C_{_{QQL}}-\frac{1697 }{140}R_{\rm oct, e}-\frac{3
   }{5}R_{\rm quad, m}+\frac{230281}{9800}\right)
\,,\nonumber\\
\Delta C_{\chi_5}(\nu^1 \pinf^5) &=&
\frac{81856 }{945}R_{\rm oct, e}+\frac{512 }{135}R_{\rm quad, m}-\frac{467968}{2205}
\,,\nonumber\\
\Delta C_{\chi_5}(\nu^2 \pinf^5) &=&
\frac{4 }{15}\bar d_5^{\nu^2}-\frac{38272 }{135}C_{_{QQQ_1}}-\frac{7168
   }{45}C_{_{QQQ_2}}+\frac{2176 }{3}C_{_{QQL}}-\frac{327424
  }{945} R_{\rm oct, e}-\frac{2048 }{135}R_{\rm quad, m}\nonumber\\
&&-\frac{61309 \pi
   ^2}{384}+\frac{77735492}{33075}
\,,\nonumber\\
\Delta C_{\chi_6}(\nu^1 \pinf^4) &=&
\pi  \left(\frac{38621 }{336}R_{\rm oct, e}+\frac{55
   }{12}R_{\rm quad, m}-\frac{1099659}{3920}\right)
\,,\nonumber\\
\Delta C_{\chi_6}(\nu^2 \pinf^4) &=&
\pi  \left(\frac{15 }{32}a_6^{\nu^2}+\frac{15 }{32}\bar d_5^{\nu^2}-\frac{8645
   }{24}C_{_{QQQ_1}}-\frac{695 }{4}C_{_{QQQ_2}}+\frac{7575
  }{8} C_{_{QQL}}-\frac{38621 }{84}R_{\rm oct, e}-\frac{55
   }{3}R_{\rm quad, m}\right.\nonumber\\
&&\left.
-\frac{5375505 \pi ^2}{16384}+\frac{10220575}{2352}\right)
\,.
\eea
\end{widetext}
It is important to keep in mind the physical origin of these equations. The three equations involving
the first power of $\nu$ correspond to the first-order self-force computations that constitute
one of the important (and well tested) building blocks of the TF approach. The TF side
of these equations is therefore fully predicted. Their EFT side contain only
 $R_{\rm oct, e}$ and $R_{\rm quad, m}$. We therefore obtain three linear equations
 involving $R_{\rm oct, e}$ and $R_{\rm quad, m}$. One finds that there are only two
 independent equations among these three $O(\nu^1)$ equations. This allows us
 to uniquely determine the values of $R_{\rm oct, e}$ and $R_{\rm quad, m}$
 from the TF results. We find
 \be \label{oct}
\left[ R_{\rm oct, e} \right]^{\rm from \; TF} = \frac{82}{35}\,,
\ee
and
\beq \label{quad_new}
\left[R_{\rm quad, m} \right]^{\rm from \; TF}=\frac{49}{20}= \frac{147}{60}\,.
\eeq
The TF-deduced value of $ R_{\rm oct, e}$, Eq. \eq{oct}, satisfactorily coincides with the result 
of Foffa and Sturani \cite{Foffa:2019eeb}.
By contrast, the  TF-deduced value of the magnetic-quadrupole term, $ R_{\rm quad, m}$,
Eq. \eq{quad_new},  disagrees with  the one given in Ref. \cite{Foffa:2019eeb}, 
which is instead $\left[R_{\rm quad, m} \right]^{\mbox{\scriptsize\rm Ref. \cite{Foffa:2019eeb}}}=\frac{127}{60}$. The difference is
\beq \label{diffquad}
\left[R_{\rm quad, m} \right]^{\rm from \; TF}- \left[R_{\rm quad, m} \right]^{\mbox{\scriptsize\rm Ref. \cite{Foffa:2019eeb}}}=\frac{49}{20}-\frac{127}{60}= + \frac13\,.
\eeq

 Ref. \cite{Blumlein:2020pyo} similarly reported the need for correcting the spin-quadrupole contribution
$\left(J_{ij}^{d,(3)}\right)^2$ derived in Ref. \cite{Foffa:2019eeb} by a finite-renormalization factor
 $Z_J=1+ \frac16 \varepsilon$ (in our notation).  A direct translation of  this factor in terms of a change in
$R_{\rm quad, m}$, would correspond to the change
\bea \label{diffquadBlum}
\left[R_{\rm quad, m} \right]^{\mbox{\scriptsize\rm Ref.~\cite{Blumlein:2020pyo}}}- \left[R_{\rm quad, m} \right]^{\mbox{\scriptsize\rm Ref.~\cite{Foffa:2019eeb}}}&=&  - \frac16 \,,
\eea
which differs from  our TF-derived result, Eq. \eq{diffquad}.
This difference is expected to come from the fact that Ref. \cite{Blumlein:2020pyo}
used a different estimate for the $d$-dimensional value of  $\left(J_{ij}^{d,(3)}\right)^2$.
Their final result for the nonlocal contributions, $S_{I_3}$ and $S_{J_2}$ in Eq. \eqref{S_Is},  to the 5PN dynamics must, however, agree with our corresponding result, both results being, essentially, calibrated on the 5PN $O(e^2)$ self-force result first
derived in Ref. \cite{Bini:2015bfb}.

Let us now go beyond the terms linear in $\nu$ in the TF/EFT comparison. At the $\nu^2$ level
we get three new equations among Eqs. \eq{TFvsEFT}. Of particular importance among
these equations is the equation $ \Delta C_{\chi_4}(\nu^2 \pinf^6) =0$.

From the TF side, this equation contains no undetermined parameters because the TF approach
reached a complete determination of the 5PN (and 6PN) dynamics at the $G^4$ level.
On the other hand, from the EFT side this equation involves a linear
combination of $R_{\rm oct, e}$, $R_{\rm quad, m}$, $C_{_{QQL}}$, $ C_{_{QQQ_1}}$ and $ C_{_{QQQ_2}}$. After inserting the $O(\nu^1)$-based unique determinations of  $R_{\rm oct, e}$
and $R_{\rm quad, m}$, Eqs. \eq{oct}, \eq{quad_new}, one gets one linear  equation involving
$C_{_{QQL}}$, $ C_{_{QQQ_1}}$ and $ C_{_{QQQ_2}}$. Namely we get the
following constraint on  the coefficients $C_{_{QQL}}$, $ C_{_{QQQ_1}}$ and $ C_{_{QQQ_2}}$
of the 5PN EFT action Eq. \eq{SQQ}:
\beq \label{constr_eq}
0=\frac{2973}{350}-\frac{69}{2}C_{_{QQL}}+\frac{253}{18}C_{_{QQQ_1}}+\frac{85}{9}C_{_{QQQ_2}}\,.
\eeq
When inserting the  rational values of  $ C_{_{QQL}}, C_{_{QQQ_1}}, C_{_{QQQ_2}}$ derived in Ref. \cite{Foffa:2019eeb}, namely Eqs. \eq{CFS}, one finds
that the constraint \eq{constr_eq} is {\it not satisfied}. 
 [Nor is Eq. \eq{constr_eq}  satisfied
when using the  values of the published version of
Ref. \cite{Foffa:2019eeb}.] For instance, when using Eqs. \eq{CFS},
 the right-hand side of Eq. \eq{constr_eq} is equal to $-1937/225\approx -8.6$. [The corresponding value when using
 the published values  $ C_{_{QQL}}=\frac{8}{15}$,
 $ C_{_{QQQ_1}}= \frac{11}{14}, C_{_{QQQ_2}}=\frac15$ is equal to $19069/6300\approx 3.0$.]
We shall discuss below possible subtleties underlying this discrepancy.

Our TF/EFT comparison \eq{TFvsEFT} yields two more equations at the $\nu^2$ level,
namely $\Delta C_{\chi_5}(\nu^2 \pinf^5) =0$ and $\Delta C_{\chi_6}(\nu^2 \pinf^4) =0$.
These two equations now involve, besides the EFT coefficients 
$R_{\rm oct, e}$, $R_{\rm quad, m}$,
$C_{_{QQL}}$ $ C_{_{QQQ_1}}$ $ C_{_{QQQ_2}}$, 
the  two yet-undetermined 5PN-level TF parameters ${\bar d}_5^{\nu^2}$ and ${a}_6^{\nu^2}$.
[As indicated by their subscripts, the latter EOB parameters respectively belong to the $G^5$
and $G^6$ PM levels.] One can solve these two constraints in terms of 
${\bar d}_5^{\nu^2}$ and ${a}_6^{\nu^2}$. This yields the two equations
\bea \label{a6d5}
a_6^{\nu^2}&=&\frac{25911}{256}\pi^2+ R_{a_6}(C_{_{QQL}},C_{_{QQQ_1}},C_{_{QQQ_2}})\,, \nonumber\\
\bar d_5^{\nu^2}&=&\frac{306545}{512}\pi^2 + R_{\bar d_5}(C_{_{QQL}},C_{_{QQQ_1}},C_{_{QQQ_2}})\,,\nonumber\\
\eea
where $R_{a_6}$ and $R_{\bar d_5}$ denote the following inhomogeneous linear combinations of 
$C_{_{QQL}}$ $ C_{_{QQQ_1}}$ and $ C_{_{QQQ_2}}$:
\bea \label{R6R5}
R_{a_6}&=&   -\frac{654389}{525}+700 C_{_{QQL}}-\frac{884}{3} C_{_{QQQ_1}}\nonumber\\
&&-\frac{680}{3} C_{_{QQQ_2}}\,, \nonumber\\ 
R_{\bar d_5}&=&  -\frac{1773479}{315}-2720 C_{_{QQL}}+\frac{9568}{9} C_{_{QQQ_1}}\nonumber\\
&&+\frac{1792}{3} C_{_{QQQ_2}}\,.
\eea  
As the work of Ref. \cite{Foffa:2019eeb} establishes that
the coefficients $ C_{_{QQL}}, C_{_{QQQ_1}}, C_{_{QQQ_2}}$ must be rational, 
the results  Eqs. \eq{a6d5} uniquely determine the irrational
contributions to ${a}_6^{\nu^2}$ and ${\bar d}_5^{\nu^2}$, namely the $\pi^2$ terms
entering Eqs. \eq{a6d5}. These $\pi^2$ terms, here directly derived by comparing TF results 
to the combination of EFT results in Refs. \cite{Blumlein:2020pyo} and \cite{Foffa:2019eeb},
satisfactorily agree with the analog results first derived by Bl\"umlein et al. \cite{Blumlein:2020pyo}. 
[However, Ref. \cite{Blumlein:2020pyo} did not provide any results similar to our explicit
expressions \eq{R6R5} for the rational contributions $R_{a_6}$ and $R_{\bar d_5}$.]

As the currently computed values for the rational coefficients  
$ C_{_{QQL}}, C_{_{QQQ_1}}, C_{_{QQQ_2}}$ are not consistent with 
our constraint \eq{constr_eq}, we
looked for rational solutions of the constraint \eq{constr_eq} having smallish
denominators involving only small powers of 2, 3, 5 and 7 (as suggested by the denominators
appearing in Appendix B of \cite{Foffa:2019eeb}). The simplest\footnote{In the sense of having
small denominators.} such solutions we found, 
for possible triplets $ C_{_{QQL}}, C_{_{QQQ_1}}, C_{_{QQQ_2}}$,
are listed in Table \ref{ra6rbard5_values}.
On the other hand, if we impose that $C_{_{QQL}}$ takes the  value computed
by Foffa and Sturani, namely $C_{_{QQL}}=\frac{8}{15}$, the simplest solutions we found
all had  larger denominators. The simplest of them involved  $175=5^2\cdot7$ and was
\be \label{supsol}
\left[\frac{8}{15}, \frac{81}{175}, \frac{9}{25}\right]\,.
\ee
We added this possible solution in Table \ref{ra6rbard5_values}.

Let us also note that the expressions for the rational numbers $R_{a_6}$ and $R_{\bar d_5}$
can be reduced if we make use of the constraint \eq{constr_eq}. This can be done in many
ways (depending on which coefficient we wish to eliminate). For instance, if we
eliminate $C_{_{QQQ_1}}$ we get
\bea \label{R6R5new1}
R_{a_6}&=&  -\frac{141907229}{132825}- \frac{256}{11} C_{_{QQL}}-\frac{21760}{759} C_{_{QQQ_2}}\,,\nonumber\\ 
R_{\bar d_5}&=& -\frac{108672257}{17325} - \frac{1216}{11}C_{_{QQL}} - \frac{11584}{99}C_{_{QQQ_2}}
\,. \nonumber\\
\eea
If we  assume the value $C_{_{QQL}}=\frac{8}{15} \equiv C_{_{QQL}}^*$,
Eq. \eq{R6R5new1} yields
\bea \label{R6R5new2}
R_{a_6}^*&=&  -\frac{143555869}{132825}-\frac{21760}{759} C_{_{QQQ_2}}\,,\nonumber\\ 
R_{\bar d_5}^*&=& -\frac{109693697}{17325}  -\frac{11584}{99} C_{_{QQQ_2}}
\,. 
\eea 
In absence of computed values of $ C_{_{QQL}}, C_{_{QQQ_1}}, C_{_{QQQ_2}}$  consistent with 
our constraint \eq{constr_eq}, we cannot derive any precise values for the two
5PN TF parameters ${a}_6^{\nu^2}$ and ${\bar d}_5^{\nu^2}$. However, in view
of the fact that all the currently published values of $ C_{_{QQL}}, C_{_{QQQ_1}}, C_{_{QQQ_2}}$,
and notably the latest ones, Eqs. \eq{CFS}, have an absolute magnitude which is smaller than 1,
it seems reasonable to assume that $ C_{_{QQL}}, C_{_{QQQ_1}}, C_{_{QQQ_2}}$ are
all contained in the interval $[-1,1]$. Making this assumption (say, ``assumption ${\cal A}_1$'') then yields constraints on the
possible values of  $R_{a_6}$, $R_{\bar d_5}$, and, therefore on  ${a}_6^{\nu^2}$ and ${\bar d}_5^{\nu^2}$. More precisely, we find that, under the assumption ${\cal A}_1$ (constrained by
\eq{constr_eq})\footnote{Note that this yields a linear programming problem.}, the ranges of possible values of 
${a}_6^{\nu^2}$ and ${\bar d}_5^{\nu^2}$ are
\bea \label{range1}
&&-119.68  \leq a_6^{\nu^2}\leq -30.64\,, \nonumber\\
&&-582.96  \leq {\bar d_5}^{\nu^2} \leq -198.34\,.
\eea
Let us finally consider the less constraining assumption ${\cal A}'_1$
that only $ C_{_{QQL}}$ and $ C_{_{QQQ_2}}$ are within the interval $[-1,1]$. Under this assumption,
 we find (using Eq. \eq{R6R5new1}) that the possible ranges for ${a}_6^{\nu^2}$ and ${\bar d}_5^{\nu^2}$ become
 \bea \label{range1'}
&&-121.37  \leq a_6^{\nu^2}\leq -17.48\,, \nonumber\\
&&-590.99  \leq {\bar d_5}^{\nu^2} \leq -135.88\,.
\eea
Inserting the estimates \eq{range1} (or \eq{range1'}) in Eqs. (8.26)-(8.29) of Ref. \cite{Bini:2020hmy}
 will yield 5PN-level numerical estimates for both the binding energy and the periastron advance along
 circular orbits which might eventually be compared with forthcoming second order self-force results
 for these quantities.

To illustrate these estimated ranges,  we list in Table \ref{ra6rbard5_values} the  values of $a_6^{\nu^2}$ and ${\bar d_5}^{\nu^2}$ corresponding to a sample of  smallish-denominator solutions of \eq{constr_eq},
respecting the assumption ${\cal A}_1$, together with the solution \eq{supsol}.
The listed values are compatible (as they should) with the
range  \eq{range1}.


\begin{table}[ht]
\caption{ \label{ra6rbard5_values}
Values of the 5PN-level EOB  coefficients $a_6^{\nu^2}$, ${\bar d_5}^{\nu^2}$ corresponding to a
sample of rational solutions of the constraint  Eq. \eqref{constr_eq} on the coefficients 
$C_{_{QQL}},C_{_{QQQ_1}},C_{_{QQQ_2}}$ of the 5PN-level EFT action contribution \eq{SQQ}.
We have selected solutions having denominators involving only small powers of 2, 3, 5 and 7.
The last solution (labelled by a ${}^\ast$) is the simplest solution satisfying $C_{_{QQL}}= \frac{8}{15}$.
}
\begin{ruledtabular}
\begin{tabular}{lll}
$[C_{_{QQL}},C_{_{QQQ_1}},C_{_{QQQ_2}}]$ & $a_6^{\nu^2}$ & $\bar d_5^{\nu^2}$ \\
\hline
$\left[\frac{2}{25}, -\frac{3}{25}, -\frac{3}{7}\right]$&$-59.0019$&$-322.1289$\\ 
$\left[\frac{13}{25}, \frac{24}{25}, -\frac{3}{7}\right]$&$-69.2419$&$-370.7689$\\
$\left[-\frac{1}{15}, -\frac{12}{25}, -\frac{3}{7}\right]$ &$-55.5886$&$-305.9156$\\
$\left[-\frac{4}{27}, -\frac{17}{25}, -\frac{3}{7}\right]$&$-53.6923$&$-296.9082$\\
$\left[\frac{7}{27}, \frac{8}{25}, -\frac{3}{7}\right]$&$-63.1738$&$-341.9452$\\
$\left[\frac{16}{25}, \frac{13}{35}, \frac{31}{35}\right]$&$-109.7143$&$-537.8191$\\
$\left[\frac{13}{21}, \frac{8}{25}, \frac{31}{35}\right]$&$-109.2267$&$-535.5029$\\
$\left[\frac{11}{25}, \frac{19}{35}, -\frac{1}{10}\right]$&$-76.8000$&$-400.3714$\\
$\left[\frac{6}{25}, -\frac{9}{35}, \frac{9}{25}\right]$&$-85.3333$&$-432.0870$\\
\hline
$\left[\frac{8}{15}, \frac{81}{175}, \frac{9}{25}\right]^*$&   $ -92.1600 $&$ -464.5137 $\\
\end{tabular}
\end{ruledtabular}
\end{table}

\section{Nonlinear radiation-reaction contributions  to scattering}
\label{nonlinearrr}

The derivation we gave above of radiation-reaction contributions to scattering was systematically
based on a first-order treatment, {\it linear in the radiation-reaction force} ${\cal F}_{rr}$.
In this final Section, we wish to point out some of the subtleties that arise when
trying to go beyond  such linear-in-${\cal F}_{rr}$ effects.

Let us start by recalling that the PM-PN order of magnitude of ${\cal F}_{rr}$ is $O(\frac{G^2}{c^5})$,
i.e. ${\cal F}_{rr}$ is at the 2PM level and the 2.5PN level. Indeed, 
the leading-order value of the radiation-reaction force 
${\mathbf  F}_1^{\rm rad}$ (to be added to the conservative equations of motion of
 the first body in a 2-body system)  is given, in harmonic
 coordinates, by  (see Eq. (5a) in \cite{Damour:1981bh}) 
 \bea \label{LOFrr}
 \frac{{\mathbf  F}_1^{\rm rad}}{m_1}&=& - \frac45 \frac{G^2}{c^5} m_1 m_2 \frac{v_{12}^2}{r_{12}^3} \left[ {\mathbf v}_{12} - 3 ({\mathbf v}_{12}\cdot {\mathbf n}_{12}) {\mathbf n}_{12}\right]\nonumber\\
&&
 + O\left( \frac{G^3}{c^5} \right)+ O\left( \frac{G^2}{c^7}\right)\,,
 \eea
where ${\mathbf v}_{12}= {\mathbf v}_{1} - {\mathbf v}_{2}$, and ${\mathbf n}_{12}= ({\mathbf x}_{1} - {\mathbf x}_{2})/r_{12}$.

In view of the expression \eq{LOFrr}, one would a priori expect that the contributions that are
of second-order in radiation-reaction will be of order  $ \frac{G^4}{c^{10}}$, i.e. at the 4PM level and the 5PN level. In addition, these contributions will necessarily contains (at least) a factor $(m_1 m_2)^2$,
i.e. a factor $\nu^2$. In particular, they are expected to contribute to the scattering angle (of each
body in the c.m. frame) a term of order (using dimensional analysis)
\be \label{chiFrr2}
\delta^{{\cal F}_{rr}^2}\chi \sim \frac{G^4 m_1^2 m_2^2}{c^8 b^4} \frac{v_{12}^2}{c^2}\,.
\ee
We wish to emphasize that such a contribution is at the same (PM, PN and $\nu$) levels as the terms
on the second line of Eqs. \eq{tildechi4consminschwTF} and \eq{tildechi4consminschwEFT}
for $\tilde \chi_4$. Note that these terms are, in particular, comparable to the scattering
effects generated by $C_{QQL}$, $ C_{QQQ_1}$, and $ C_{QQQ_2}$.
In addition, one expects (when considering the higher-order-in-$G$ corrections
in Eq. \eq{LOFrr}) that  quadratic-in-${\cal F}_{rr}$ effects will also yield contributions
to $\tilde \chi_5$ and $\tilde \chi_6$ that will be again similar to those generated
by $C_{QQL}$, $ C_{QQQ_1}$, and $ C_{QQQ_2}$ in Eqs. \eq{tildechi5consminschwEFT},
\eq{tildechi6consminschwEFT}. Such terms are also similar to those generated by the conservative
TF parameters ${a}_6^{\nu^2}$ and ${\bar d}_5^{\nu^2}$. 

Summarizing so far: scattering effects quadratic-in-${\cal F}_{rr}$ look like
{\it conservative 5PN-level}  effects quadratic in $\nu$ (second-order self-force level), and of
PM orders 4, 5 and 6.

Before further discussing this issue, let us explicitly check that
a second-order treatment of ${\cal F}_{rr}$ does indeed induce (nonvanishing) contributions to the scattering 
of the expected order \eq{chiFrr2}. We consider the (PN-expanded) radiation-reaction
corrected equation of motion of body 1 (in harmonic coordinates\footnote{The second-order 
contribution \eq{v2}, considered separately, is gauge-dependent.}), namely
\be \label{x1rr}
\frac{d^2 {\mathbf x}_{1}}{dt^2}= - G m_2  \frac{{\mathbf x}_{1} - {\mathbf x}_{2}}{r_{12}^3}
- \epsilon_{rr} \frac{v_{12}^2}{r_{12}^3} \left[ {\mathbf v}_{12} - 3 ({\mathbf v}_{12}\cdot {\mathbf n}_{12}) {\mathbf n}_{12}\right]\,, 
\ee
where
\be
\epsilon_{rr} \equiv + \frac45 \frac{G^2 m_1 m_2}{c^5}\,.
\ee
We can then  integrate Eq. \eq{x1rr} by using second-order perturbation theory in $\epsilon_{rr}$, i.e.
\be
{\mathbf x}_{1}(t)= {\mathbf x}_{1}^{(0)}(t)+ \epsilon_{rr} {\mathbf x}_{1}^{(1)}(t)+ \epsilon_{rr}^2{\mathbf x}_{1}^{(2)}(t)+ O(\epsilon_{rr}^3)\,.
\ee
As we are primarily interested in checking the presence of the leading-order $G^4$ contribution \eq{chiFrr2},
we can even neglect the first, $O(G^1)$, Newtonian acceleration, and therefore use as 
zeroth-order motion ${\mathbf x}_{1}^{(0)}(t)$ a straight-line uniform motion. 
The scattering information is then obtained by considering the $t\to +\infty$ limit of 
$\dot {\mathbf x}_{1}(t)= {\mathbf v}_{1}^{(0)}(t)+ \epsilon_{rr} {\mathbf v}_{1}^{(1)}(t)+ \epsilon_{rr}^2{\mathbf  v}_{1}^{(2)}(t)$. One finds, $\lim_{t\to +\infty}\epsilon_{rr} {\mathbf v}_{1}^{(1)}(t)=0$,
and
\be \label{v2}
\lim_{t\to +\infty}  \epsilon_{rr}^2 {\mathbf v}_{1}^{(2)}(t)= - \frac{3 \pi}{8} \epsilon_{rr}^2 \frac{v_0^3}{b^4} \hat {\mathbf  b}\,,
\ee
where $ \hat  {\mathbf b}$ denotes the unit vector in the direction of the impact parameter (directed from body
2 towards body 1). This result does correspond to a (positive) second-order contribution of the type
indicated in Eq. \eq{chiFrr2}.

The existence of $O({\cal F}_{rr}^2)$ scattering effects has nothing paradoxical in itself, but 
raises  delicate issues concerning the inclusion of radiative effects in
computations of PM scattering at order $G^4$ and beyond. So far, radiation-reaction effects
in PM scattering (considered to all PN orders)
have been explicitly included (both in quantum-based and in classical-based computations) 
only at order $G^3$. We have explicitly shown above how $O(G^3)$ radiation-reaction effects
can be completely derived from linear-in-${\cal F}_{rr}$ computations.
Let us now recall that a crucial ingredient in any PM computation is the
choice of Green's functions when iteratively computing the gravitational interaction between
two massive objects. When doing a classical PM scattering computation one can proceed
along two different ways. 

A first way consists of 
deriving the  Fokker-Wheeler-Feynman-type \cite{Fokker1929,Wheeler:1949hn} reduced
action, defined by iteratively using the {\it time-symmetric}
 graviton propagator, whose Fourier-space kernel is (P denoting the principal value)
 \be \label {Gsym}
 G_{\alpha \beta ;\alpha' \beta'}(k)= (\eta_{\alpha \alpha'} \eta_{\beta \beta'} -\frac12 \eta_{\alpha \beta} \eta_{\alpha' \beta'}) {\rm P}\frac1{k^2},
 \ee
and by then completing the so-defined conservative dynamics by introducing an additional
radiation-reaction force ${\cal F}_{rr}$ (defined so as to ensure balance with radiative
losses of energy, momentum and angular momentum). 
Consistently with the TF approach (based on Fokker-Wheeler-Feynman dynamics) our present treatment follows this first path
(in particular we {\it define} the conservative dynamics in the TF way\footnote{As a consequence $\chi_4^{\rm cons}$} is fully determined by  1SF information, see Fig. 1 in Ref. \cite{Bini:2020nsb}).

A second way consists of bypassing the
use of any conservative dynamics to derive from the start the PM equations of motion of
the two bodies interacting via the {\it retarded} propagator\footnote{This was the route
followed in Refs. \cite{Damour:1981bh,Damour:1983tz,Damour:1982wm}.}. In the latter, the term 
${\rm P}\frac1{k^2}$ in Eq. \eq{Gsym} would be replaced by the kernel of the
retarded Green's function, i.e.
\be \label{Gret}
 \frac1{k^2 - {\rm sign}(k^0) i 0}= {\rm P}\frac1{k^2} + i \pi {\rm sign}(k^0) \delta(k^2).
\ee
When considering a source (say $T_{\mu \nu}$)
which is {\it linearly} coupled to a (linearized) gravitational field $h_{\mu \nu}$,
it is well known (since the classic work of Dirac \cite{Dirac:1938nz})
 that the radiation-reaction force needed to ensure balance is derived
from the {\it reaction} field $h^{\rm reac}_{\mu \nu}$, generated
by the source via the {\it reaction} Green's function
\be
G_{\rm reac} \equiv \frac12 \left[ G_{\rm ret}- G_{\rm ad}\right],
\ee
whose Fourier kernel involves the second term on the r.h.s. of Eq. \eq{Gret}, i.e.
\be
G_{\rm reac}(k) =  i \pi {\rm sign}(k^0) \delta(k^2).
\ee
[See, e.g., Section IV F of  \cite{Buonanno:1998is} for a detailed discussion.]
However, when considering the 3-loop diagrams representing the $G^4$ retarded
gravitational interaction of two worldlines, it becomes technically unclear how 
to derive the needed radiation-reaction force by separating out the on-shell
term $G_{\rm reac}(k)$ from the many intermediate interactions involving
$G_{\rm ret}(k) =G_{\rm sym}(k) + G_{\rm reac}(k)$ so as to relate the
retarded interaction to the time-symmetric one.

The situation is even more involved when considering (beyond the linear interaction
of sources) a quantum scattering computation,
involving the Feynman Green's function, namely
\be \label{GF}
G_F(k)= \frac1{k^2 - i 0}= {\rm P}\frac1{k^2} + i \pi \delta(k^2)\,.
 \ee
Observables can be computed by using the formalism of  Ref. \cite{Kosower:2018adc}, but 
the resulting expressions for the impulses are rather complex and make it unclear how to 
read off nonlinear radiation-reaction effects by separating out the on-shell term in
$G_F(k)= G_{\rm sym}(k) + i \pi \delta(k^2)$.

Let us also note that these technical difficulties in trying to directly relate a retarded (or Feynman)
scattering computation to a time-symmetric-plus-radiation-reaction one are compounded when
the $k$-space integrals are computed by the method of expansion by regions \cite{Beneke:1997zp}.
The latter method has been quite effective in relating conservative and radiation-reaction
effects up to order $G^3$, but it might encounter difficulties at the $G^4$ level because of
the $O({\cal F}_{rr}^2)$ scattering effects discussed above.

Let us end this discussion by commenting on the discrepancy we found above between 
the constraint Eq. \eq{constr_eq} derived from the TF approach, and the explicit values \eq{CFS}
derived in Refs.  \cite{Foffa:2019eeb,Foffa:2021pkg}. This discrepancy concerns conservative-like 5PN terms
that are of order $\nu^2$, and $G^4$, or beyond. The root of this discrepancy might be due to the
choice made in Refs.  \cite{Foffa:2019eeb,Foffa:2021pkg} of Green's functions for the nonlinear
interactions giving rise to the action contributions Eq. \eq{SQQ}. Indeed, Refs.  \cite{Foffa:2019eeb,Foffa:2021pkg} argued that the computation of these (initially nonlocal-in-time) interaction terms
should involve intermediate causal Green's functions (advanced or retarded), combined in a way to end up
with a final time-symmetric action. In the case of the first, $S_{_{QQL}}$, action contribution, it seems
that their computation would agree with a computation involving (as assumed in the TF approach) a
time-symmetric propagator in intermediate interactions. [This is why, we are ready to assume
that their estimate of $ C_{_{QQL}}$ should apply when comparing  TF to EFT.] By contrast,
we do not understand the meaning of the a  posteriori time-averaged action contributions
$S_{_{QQQ_1}}$ and $S_{_{QQQ_2}}$ that they compute. We leave to future work
a computation of the analogs of these terms when using time-symmetric propagators 
in intermediate interactions.

\section{Summary of results and concluding remarks}

We have presented several new results on radiative contributions to the classical two-body scattering
in General Relativity. 

We gave general formulas,  valid only to {\it linear order} in radiation reaction, expressing  the effect
of radiative losses (of energy, linear momentum and angular momentum)
on  the 4-momentum changes (a.k.a, impulses), $\Delta p_{a \mu} \equiv p_{a \mu}^+- p_{a \mu}^-$,
i.e. the radiation-reacted contribution $\Delta p_{a \mu}^{\rm rr, tot}$ in
\be
\Delta p_{a \mu}= \Delta p_{a \mu}^{\rm cons}(u_1^{-}, u_2^{-}, b)+ \Delta p_{a \mu}^{\rm rr, tot}(u_1^{-}, u_2^{-}, b).
\ee
We obtained this contribution as the sum of a {\it relative} term and a {\it recoil} one:
\be  
\Delta p_{a \mu}^{\rm rr, tot}(u_1^{-}, u_2^{-}, b)= \Delta p_{a \mu}^{\rm rr, rel}+ \Delta p_{a \mu}^{\rm rr, rec} + O({\cal F}_{\rm rr}^2)\,.
\ee
Our results are summarized in Table \ref{tab:impulse_coeffs1}.

We emphasized how the polynomial dependence of $\Delta p_{a \mu}^{\rm rr, tot}(u_1^{-}, u_2^{-}, b)$
can be exploited to yield some identity relating the various radiative losses. See Eqs. \eq{J3identity}, \eq{P4nu}.

We showed how the application of our general formulas at the $O(G^3)$ level led to a streamlined
classical derivation of  the  $O(G^3)$  radiative contribution to $\Delta p_{a \mu}$, which
was recently derived within a quantum approach in Ref. \cite{Herrmann:2021tct}.

Our general formulas involve the radiative losses (in the c.m. frame) of energy, angular momentum
and linear momentum. These losses admit a double PM and PN expansion, 
which can be expressed as (with $j = J/(G m_1 m_2)$, $\pinf = \sqrt{\g^2-1}$
and ${\mathbf P}^{\rm rad}= {P}^{\rm rad}_y {\mathbf e}_y$)
\bea
\frac{E^{\rm rad}}{M} &=&  + \nu^2  \left[ \frac{{ E}_{3}(\pinf)}{j^3}+ \frac{{ E}_{4}(\pinf)}{j^4}+  \cdots\right]  \,, \nonumber\\
\frac{ J^{\rm rad}}{J_{\rm c.m.}} &=& + \nu^1 \left[ \frac{{ J}_{2}(\pinf)}{j^2}+ \frac{{ J}_{3}(\pinf)}{j^3} + \cdots\right]\,, \nonumber\\
\frac{P_y^{\rm rad}}{M}&=& + \frac{m_2-m_1}{M} \nu^2 \left[ \frac{{ P}_{3}(\pinf)}{j^3}+ \frac{{ P}_{4}(\pinf)}{j^4}+  \cdots\right]\,.\nonumber\\
\eea
Here the subscripts $n$ (e.g., in $E_n$) label  the $n$PM order, i.e., $O(G^n)$.
The subsequent expansion of the various PM coefficients, $E_n(\pinf)$, $J_n(\pinf)$, $P_n(\pinf)$
in powers of $\pinf$ then corresponds to the usual PN expansion. 
The only radiative losses that are known in a PM-exact way are $J_2$ \cite{Damour:2020tta}
and, $E_3$ and $P_3$ \cite{Herrmann:2021tct}. In the present paper, we have computed the
fractionally 2PN-accurate expansions of the higher PM radiative losses $E_n(p_\infty)$, $J_n(p_\infty)$ and $P_n(p_\infty)$ up to $n=7$, including the contribution of tails (see Table \ref{tab:table_EnJnPn}).
Our new results are given in Eqs. \eqref{Erad_N}, \eqref{Erad_1PN}, \eqref{Erad_2PN} for the  expressions of the (instantaneous) radiated energy, Eqs. \eqref{Jrad_N}, \eqref{Jrad_1PN}, \eqref{Jrad_2PN} for the (instantaneous) radiated angular momentum and 
Eqs. \eqref{Prad_N}, \eqref{Prad_1PN}, \eqref{Prad_2PN} for the (instantaneous) radiated linear momentum.
 [We have also confirmed
the values of $E_3$ and $P_3$  at the 15th order in $\pinf$ (see Eqs. \eqref{hat_E_def_exp} and \eqref{J2E3P3}).]
At the 2PN level of accuracy, we have also given the  general expressions for the radiative losses of energy, angular momentum and linear momentum along hyperboliclike orbits in terms of two independent orbital parameters (namely, $\bar a_r$ and $e_r$) which, once re-expressed in terms of energy and angular momentum, can allow one to reach any order in a large-$j$ expansion limit.
Tail terms have instead been computed in this limit only up to $O(G^7)$ (see Eqs. \eq{Eradtail}, \eq{Jradtailfin}, \eq{Pradtailfin}).

Inserting the latter 2PN-accurate results in our general formulas for radiative scattering effects
allowed us to derive the $O(G^4)$ and $O(G^5)$ contribution to  $\Delta p_{a \mu}^{\rm rr, tot}$
with absolute 4.5PN accuracy, i.e. with terms of PN order
\be \label{PNdprr}
\Delta p_{a \mu}^{\rm rr, tot 4,5 PM} \sim \frac1{c^5} + \frac1{c^7} +\frac1{c^8} +\frac1{c^9}.
\ee
This absolute 4.5PN accuracy is consistent with the limitation of our $O({\mathcal F}_{\rm rr}^1)$ treatment, because contributions nonlinear in ${\mathcal F}_{\rm rr}$ start at order $O(G^4/c^{10})$.

Our explicit results are contained in Section \ref{PN_impulse_coeffs}. Let us note that, in the latter expansions, the terms
of absolute PN order $\frac1{c^8}$ (i.e. 4PN order) come from taking into account the
leading-order {\it tail} contribution to the radiative multipole moments. Our computation
of the relevant time-symmetric projections of these tail contributions is discussed
in Appendices \ref{App_tail_E}, \ref{App_tail_J} and \ref{App_tail_P}.

We also completed the recently derived potential-graviton contribution to {\it conservative}
scattering \cite{Bern:2021dqo} (see also \cite{Dlapa:2021npj}) by extracting from our
recent TF results the (conservative)  radiation-graviton contribution to 6PN accuracy,
see Eq.\eqref{Mrad_finite}.

Finally, we went beyond the $G^4$ level and combined information from our TF results and
from 5PN EFT results of Refs. \cite{Foffa:2019eeb,Blumlein:2020pyo,Foffa:2021pkg}
to  derive explicit theoretical expressions for the two hitherto undetermined $O(G^5)$ and
$O(G^6)$ parameters, dubbed  ${\bar d}_5^{\nu^2}$, and ${a}_6^{\nu^2}$,
entering the 5PN dynamics (as described through the TF approach).
We expressed ${\bar d}_5^{\nu^2}$, and ${a}_6^{\nu^2}$ as explicit linear
combinations of $\pi^2$ and of the (rational) coefficients entering various (local-in-time) nonlinear
contributions to the conservative effective 5PN action induced by radiation-gravitons \cite{Foffa:2019eeb,Foffa:2021pkg}. Our results are given in Eqs. \eqref{a6d5} and \eqref{R6R5}.
These results confirm (for the $\pi^2$ contributions) and extend (by providing the
dependence on $C_{_{QQL}},  C_{_{QQQ_1}},  C_{_{QQQ_2}}$)
results of  Bl\"umlein et al. \cite{Blumlein:2020pyo}. On the other hand, our results
exhibit several disagreements with results of Foffa and Sturani  \cite{Foffa:2019eeb,Foffa:2021pkg}.
We point out that part of these disagreements might be rooted in subtleties linked to contributions
that are nonlinear in radiation-reaction effects (see Section \ref{nonlinearrr}).

We leave to future work a deeper study of the latter nonlinear radiation-reaction effects,
and emphasize here that all our PN-expanded results, Eq. \eq{PNdprr}, for the radiative-losses
related contribution to the scattering impulses only depend on linear-in-radiation-reaction effects,
because, as discussed in Section \ref{nonlinearrr}, the leading-order nonlinear-in-radiation-reaction effect
is of order $\frac{G^4}{c^{10}}$. We also recall that in the TF framework, and in the present work, the {\it conservative} 
scattering angle is defined as the one coming from the time-symmetric Fokker-Wheeler-Feynman dynamics.
Any  comparison with  forthcoming  $O(G^4)$ \lq\lq conservative" computations should check that the meaning of conservative is the same, and any comparison with forthcoming $O(G^4)$ \lq\lq physical, retarded" computations should take into account all the needed contributions of order ${\mathcal F}_{\rm rr}^2$.

\section*{Acknowledgments}

We thank  Zvi Bern, Johannes Bl\"umlein, Stefano Foffa, Peter Marquard, Julio Parra-Martinez, Rafael Porto, Riccardo Sturani
and Gabriele Veneziano for useful exchanges. DB thanks the International Center for Relativistic Astrophysics Network, ICRANet, for partial support.
DB also acknowledges the hospitality  and the highly stimulating environment  of the Institut des Hautes Etudes Scientifiques. 
DB and AG thank MaplesoftTM for providing a complimentary license of Maple 2020.

\appendix

\section{Notation and useful relations}

We collect here some definitions and relations which are often used. 

The masses of the two bodies are denoted as $m_1$ and $m_2$, with the convention $m_1 \leq m_2$.
The symmetric mass ratio $\nu$ is the ratio of the reduced mass $\mu\equiv m_1 m_2/(m_1+m_2)$ to the total mass $M=m_1+m_2$,
\beq 
\nu\equiv \frac{m_1m_2}{(m_1+m_2)^2}=\frac{\mu}{M}\,.
\eeq
We also define $\Delta \equiv\sqrt{1-4\nu}$ which enters the mass ratios 
\beq
X_1=\frac{m_1}{M}=\frac12(1-\Delta)\,,\quad X_2=\frac{m_2}{M}=\frac12(1+\Delta)\,,
\eeq
with the property $X_1+X_2=1$.

The asymptotic 4-momenta are denoted by $p_{a}^\pm = m_a u_{a}^\pm$, with $a=1,2$, and the asymptotic energies by $E_a^\pm$.
We often work within the incoming c.m. frame of the system, with time axis
\be
U^{-} \equiv \frac{ p_1^{ -} + p_2^{ -}}{ |p_1^{ -}+  p_2^{ -} |}
=\frac{m_1u_1^-+m_2u_2^-}{E_{\rm c.m.}^-}\,,
\ee
where $E_{\rm c.m.}^-=(E_1+E_2)^-$ is the incoming c.m. energy, which is related to the incoming value of the momentum by
\beq
P_{\rm c.m.}^-= \frac{m_1m_2}{E_{\rm c.m.}^-}\sqrt{\gamma^2-1}= \frac{m_1m_2 }{E_{\rm c.m.}^-}\pinf
\,,
\eeq
with
\beq
\g \equiv  -u_1^-\cdot u_2^- = -  \frac{p_1^-\cdot p_2^-}{m_1 m_2}\,,\qquad
\pinf \equiv \sqrt{\g^2-1}\,,
\eeq
so that $E_{\rm c.m.}^- P_{\rm c.m.}^- = m_1 m_2 \pinf$.
The total incoming energy can also be written as
\beq
\frac{E_{\rm c.m.}^-}{Mc^2} \equiv h(\g, \nu)=\sqrt{1+2\nu (\g-1)}
\,,
\eeq 
implying that $P_{\rm c.m.}^-=\mu {\pinf}/{h}$ and
\beq
\frac{G E_{\rm c.m.}^-}{b} \equiv \frac{GM h}{b} = \frac{\pinf}{j}\,,
\eeq
where $b$ is the impact parameter and
\beq 
 j\equiv \frac{c J_{\rm c.m}^-}{G m_1 m_2}=\frac{c J_{\rm c.m}^-}{G M \mu}\,,
 \eeq
is a dimensionless rescaled version of the total center-of-mass angular momentum $J_{\rm c.m}^-$.
The incoming energies 
\beq
\frac{E_1^-}{m_1}=\frac{m_2 \g + m_1}{E_{\rm c.m.}^-}\,,\qquad
\frac{E_2^-}{m_2}=\frac{m_1 \g + m_2}{E_{\rm c.m.}^-}\,,
\eeq
can also be cast in the form
\bea
E_1^-&=&\frac{E_{\rm c.m.}^-}{2}(1-\sqrt{1-4\xi}) \,,  \nonumber\\
E_2^-&=&\frac{E_{\rm c.m.}^-}{2}(1+\sqrt{1-4\xi})\,, 
\eea
where $\xi$ denotes the symmetric energy ratio (simply related to the symmetric mass ratio and the c.m. energy)
\beq
\xi = \frac{E_1^-E_2^-}{(E_{\rm c.m.}^-)^2}\,, \quad
1-4\xi=\frac{1-4\nu}{h^4}\,.
\eeq

In the following Appendices we mostly use units where $c$ and $G$ are set to unity for simplicity.

\section{PM expansion of main quantities}
\label{PMAppendix}

It is convenient to express the scattering angle as well as the radiative losses as power series expansions in the dimensionless variable $\frac1j= \frac{G m_1 m_2}{J}$, with coefficients depending on $\g$ and the symmetric mass-ratio $\nu$.
The latter exhibit a simple $\nu$-structure when suitably rescaled by powers of $h$.

The relative scattering angle is given by 
\be 
\chi^{\rm rel}= \chi^{\rm cons} + \delta^{\rm rr} \chi^{\rm rel}\,.
\ee
The PM expansion of the conservative part is
\beq 
\chi^{\rm cons} = \sum_{n=1}^\infty \frac{2\chi^{\rm cons}_n }{j^n}\,.
\eeq
The only coefficients which are currently exactly known are
\bea
\label{chi3PMcons}
\chi_1^{\rm cons}(\g)&=&\frac{2 \, \g^2-1 }{\sqrt{\g^2-1}}\,,\nonumber\\
\chi_2^{\rm cons}(\g, \nu)&=&  \frac38 \pi   \frac{5\gamma^2-1}{h}\,,\nonumber\\
\chi_3^{\rm cons}(\g, \nu)&=&\frac{64 \gamma^6+120 \gamma^4+60\gamma^2-5}{3(\gamma^2-1)^{3/2}}\nonumber\\
&-&\frac{2\nu \sqrt{\gamma^2-1}}{h^2}
\bar C^{\rm cons}(\gamma)\,, 
\eea
with
\bea
\bar C^{\rm cons}(\gamma)&=&\frac23 \gamma (14\gamma^2+25)\nonumber\\
&+&2 \frac{4\gamma^4-12\gamma^2-3}{\sqrt{\gamma^2-1}}{\rm arctanh}\left(\frac{\sqrt{\gamma^2-1}}{\gamma}\right)\,.\nonumber\\
\eea
Similarly, the expansion of the radiation-reaction part \eq{deltarrchirel} reads
\beq
\delta^{\rm rr} \chi^{\rm rel} = \sum_{n=3}^\infty  \frac{ 2\chi^{\rm rr}_n}{j^n}\,,
\eeq
with coefficients
\bea 
\label{PMexpchirad}
\nu^{-1}2\chi^{\rm rr}_3 &=& \chi^{\rm cons}_1 J_2 
\,, \nonumber\\
\nu^{-1}2\chi^{\rm rr}_4 &=& \chi^{\rm cons}_1J_3 +2 \chi^{\rm cons}_2 J_2 - h E_3  \frac{d\chi^{\rm cons}_1}{d\gamma} 
\,,\nonumber\\
\nu^{-1}2\chi^{\rm rr}_5 &=&  \chi^{\rm cons}_1 J_4+2\chi^{\rm cons}_2 J_3+3\chi^{\rm cons}_3J_2 \nonumber\\
&-& h  \left( E_4 \frac{d\chi^{\rm cons}_1}{d\gamma} + E_3  \frac{d\chi^{\rm cons}_2}{d\gamma} \right)  
\,,\nonumber\\
\nu^{-1}2\chi^{\rm rr}_6 &=&  \chi^{\rm cons}_1 J_5+2\chi^{\rm cons}_2 J_4+3\chi^{\rm cons}_3J_3 +4\chi^{\rm cons}_4J_2\nonumber\\
&-& h  \left( E_5 \frac{d\chi^{\rm cons}_1}{d\gamma} + E_4  \frac{d\chi^{\rm cons}_2}{d\gamma} + E_3  \frac{d\chi^{\rm cons}_3}{d\gamma} \right)  
\,,\nonumber\\
\nu^{-1}2\chi^{\rm rr}_7 &=&  \chi^{\rm cons}_1 J_6+2\chi^{\rm cons}_2 J_5+3\chi^{\rm cons}_3J_4 +4\chi^{\rm cons}_4J_3\nonumber\\
&+& 5\chi^{\rm cons}_5J_2
-h  \left( E_6 \frac{d\chi^{\rm cons}_1}{d\gamma} + E_5  \frac{d\chi^{\rm cons}_2}{d\gamma}\right.\nonumber\\
&+&\left.
E_4  \frac{d\chi^{\rm cons}_3}{d\gamma} + E_3  \frac{d\chi^{\rm cons}_4}{d\gamma} \right)  
\,,
\eea
etc. Here, $E_n$ and $J_n$ denote the coefficients of the PM expansion of the energy and angular momentum radiative losses \eq{ErrJrr}
\bea
\label{PMexpEJrad}
-\frac{ \delta^{\rm rr} J}{J_{\rm c.m.}}=\frac{ J^{\rm rad}}{J_{\rm c.m.}} &=& \nu  \sum_{n=2}^\infty \frac{{ J}_{n}}{j^n} \,, \nonumber\\
-\frac{ \delta^{\rm rr} E}{M}=\frac{E^{\rm rad}}{M} &=& \nu^2   \sum_{n=3}^\infty \frac{ E_{n}}{j^n} \,,
\eea
respectively.
The only coefficients of the above PM expansions which are currently exactly known are $J_2$ and $E_3$.
They are given by
\be
{\widehat J}_2(\gamma)=h^2 J_2= 2(2\gamma^2-1)\sqrt{\gamma^2-1}{\mathcal I}(v) \,,
\ee
with  ${\mathcal I}(v)$ defined in Eq. \eqref{cal_I_di_v_def}, and
\bea
\widehat E_3(\g)=h^4 E_3=\pi p_\infty^3 \widehat {\mathcal E}(\g)\,,
\eea
with $\widehat  {\mathcal E}(\g)$ defined in Eq. \eqref{hat_E_def}.
The remaining coefficients are known only in PN sense.
Their explicit expressions up to 2PN order are given in Appendix \ref{PNAppendix}.

The energy-rescaled coefficients
\beq
\tilde \chi^{\rm X}_n = h^{n-1} \chi^{\rm X}_n = P^{\g}_{[\frac{n-1}{2}]}(\nu)  \,,   
\eeq
with X = cons, rr, exhibit a simple $\nu$-structure, i.e., polynomial of degree $[\frac{n-1}{2}]$, with coefficients depending on $\gamma$.
Similarly, the $\nu$ dependence of $E_n$ satisfies the property
\be 
\widehat E_n(\g)=h^{n+1} E_n(\g, \nu) = P^\g_{[\frac{n-2}{2}]}(\nu)\,,
\ee
where $ P^\g_N(\nu)$ denotes a polynomial in $\nu$ of order $N=[\frac{n-2}{2}]$, with coefficients depending on $\gamma$. 
Concerning the angular momentum coefficients, we have shown that the quantity
\be
\widehat J_3 \equiv h^3J_3+h^2\nu E_3
= h^3 J_3 + \frac{\nu \pi p_\infty^3\widehat {\mathcal E}}{h^2}\,,
\ee
must be  independent of $\nu$, and be only a function of $\g$. 
We have checked this result by explicitly computing $J_3$ at the 2PN fractional accuracy.
This is also true for the quantity
\beq
\widehat J_4=h^4J_4+h^3\nu E_4\,.
\eeq

Finally, the PM expansion of the radiative linear momentum loss \eq{Prr} is 
\beq
\label{PMexpPrad}
-\frac{ \delta^{\rm rr} P_y}{M}=\frac{P_y^{\rm rad}}{M}= + \frac{m_2-m_1}{M} \nu^2 \sum_{n=3}^\infty \frac{P_{n}}{j^n}\,,
\eeq
the only PM-known coefficient being 
\beq
P_3=\sqrt{\frac{\gamma-1}{\gamma+1}}E_3\,.
\eeq
We found that the quantities
\bea
h^3 \left[P_4 -\sqrt{\frac{\gamma-1}{\gamma+1}}E_4 \right]&=&\left[P_4 -\sqrt{\frac{\gamma-1}{\gamma+1}}E_4 \right]_{\nu=0}
\,,\nonumber\\
h^4 \left[P_5 -\sqrt{\frac{\gamma-1}{\gamma+1}}E_5 \right]&=&\left[P_5 -\sqrt{\frac{\gamma-1}{\gamma+1}}E_5 \right]_{\nu=0}
\,,\nonumber\\
\eea
do not depend on $\nu$, by explicitly computing the coefficients at the 2PN fractional accuracy.

Concerning the impulse coefficients $c_b^{1\rm X}$,  $c_{u_1}^{1\rm X}$,  $c_{u_2}^{1\rm X}$, we have given their PM expansion in powers of $GM/b$ instead of $1/j$ by using the relation $\frac1j= \frac{G M h}{ b \pinf}$ between the dimensionless angular momentum $j$ and the impact parameter $b$, namely
 \bea
c_b^{1,\rm X} &=&  \sum_{n=3}^\infty \frac{c_{b,G^n}^{\rm X}}{b^n}\,, \nonumber\\
c_{u_1}^{1,\rm X} &=&  \sum_{n=3}^\infty \frac{c_{u_1,G^n}^{\rm X}}{b^n}\,, \nonumber\\
c_{u_2}^{1,\rm X} &=&  \sum_{n=3}^\infty \frac{c_{u_2,G^n}^{\rm X}}{b^n}\,, 
\eea
with X = cons, rr.
The expressions for the coefficients up to $n=5$ can be easily deduced from the 5PM result for $\Delta p_1$ given in Table \ref{tab:impulse_coeffs3}, fully known up to 3PM and only in PN sense up to 2PN at higher PM orders.
The fractionally 2PN-accurate results at orders $O(G^{\le 3})$ are listed in Appendix \ref{PNAppendix}, whereas the new $O(G^4)$ and $O(G^5)$ results are given in Section \ref{PN_impulse_coeffs}.

\section{Fractionally 2PN-accurate radiated energy along hyperboliclike orbits (without the 1.5PN tail contribution)}

Most of the literature on radiated fluxes of energy (or angular momentum, or linear momentum)
focusses on ellipticlike orbits.  Little attention has been given to radiative losses along hyperbolic motions. Exceptions are the fractionally 1PN-accurate computations of  radiative energy loss \cite{Blanchet:1989cu}
and angular momentum loss \cite{Junker:1992kle}\footnote{Note that in \cite{Junker:1992kle},
Eq. (42), second line there is a missing term of $+42120\nu(e_r^2-1)^2$ inside the bracketed expression proportional to the arccos.
}.  In addition, the latter
reference gave the leading-order radiative loss of linear momentum along an hyperbolic motion.

For our present purposes, we improved these results by computing with fractional 2PN accuracy the
energy and angular momentum losses along hyperbolic motions. This was done by inserting in the standard
energy-momentum and angular momentum fluxes \cite{Thorne:1980ru}, the knowledge
of the 2PN-accurate {\it source} multipole moments \cite{Blanchet:1995fg} with a 2PN-accurate quasi-Keplerian solution of hyperbolic motions \cite{Damour:1988mr}.
More precisely, the gravitational wave energy flux is given by ${\mathcal F}^{\rm GW}=\frac{dE_{\rm rad}}{dU}$ (where $U$ denotes the retarded time in radiative coordinates; see Eq. (68a) of Ref. \cite{Blanchet:2013haa})
\bea
\label{energy_flux}
{\mathcal F}^{\rm GW}&=&\sum_{l=2}^\infty \frac{G}{c^{2l+1}}
\left\{
\frac{(l+1)(l+2)}{(l-1)l l!(2l+1)!!}U_L^{(1)}U_L^{(1)}\right.\nonumber\\
&+&\left. \frac{4l(l+2)}{c^2(l-1)(l+1)!(2l+1)!!}V_L^{(1)}V_L^{(1)}
\right\}\,.
\eea
Here $U_L$ and $V_L$ denote the radiative multipole moments that parametrize the
asymptotic waveform. These moments are perturbatively computed in terms of the source
variables, and involve both instantaneous and past-hereditary contributions. For example, at the
2PN accuracy, one has (with some rational constants $\kappa_l$ and $\pi_l$; see  \cite{Blanchet:2013haa})
\bea
\label{U_LandV_L}
U_L(U)&=& U_L^{\rm inst}(U)+U_L^{\rm tail}(U)\nonumber\\
&=& I_L^{(l)}(U)\nonumber\\
&+&\frac{2G E_{\rm c.m.}}{c^3}\int_0^\infty d\tau I_L^{(l+2)}(U-\tau)\left[\ln \left(\frac{c\tau}{2r_0}\right)+\kappa_l  \right]
\,,\nonumber\\
V_L(U)&=& V_L^{\rm inst}(U)+V_L^{\rm tail}(U)\nonumber\\
&=& J_L^{(l)}(U)\nonumber\\
&+&\frac{2G E_{\rm c.m.}}{c^3}\int_0^\infty d\tau J_L^{(l+2)}(U-\tau)\left[\ln \left(\frac{c\tau}{2r_0}\right)+\pi_l  \right]
\,,\nonumber\\
\eea
the superscript in parenthesis  denoting  repeated time-derivatives.
When aiming at the 2PN fractional accuracy for ${\mathcal F}^{\rm GW}$ one can use
\bea
\mathcal{F}^\mathrm{GW}&=&{G\over c^5}\biggl\{ {1\over 5}
U^{(1)}_{ij} U^{(1)}_{ij}\nonumber\\
&+&
{1\over c^2} \left[ {1\over
189} U^{(1)}_{ijk} U^{(1)}_{ijk} +{16\over 45} V^{(1)}_{ij}
V^{(1)}_{ij}\right]\nonumber\\
&+&{1\over c^4} \left[ {1\over 9072} U^{(1)}_{ijkm}
U^{(1)}_{ijkm}+{1\over 84} V^{(1)}_{ijk}
V^{(1)}_{ijk}\right]\biggr\}
\,,\nonumber\\
\eea
in which one must insert Eqs. \eq{U_LandV_L} with the appropriately PN-accurate multipoles
(notably the 2PN-accurate quadrupole $U_{ij}$, for which $\kappa_2=\frac{11}{12}$).

To evaluate the total radiated energy\footnote{We use henceforth the fact that the (radiative)
retarded time $U$ only differs by an additive contribution from the (harmonic) coordinate time $t$.}
\beq
E^{\rm rad}=\int_{-\infty}^\infty dt {\mathcal F}^\mathrm{GW}\,,
\eeq
one needs to explicitly evaluate the multipole moments along the hyperbolic orbit. 
This is conveniently done by using the 2PN-accurate quasi-Keplerian parametrization 
of the hyperboliclike motion \cite{Damour:1990jh,Cho:2018upo}
\begin{eqnarray} \label{hypQK2PN}
r&=& \bar a_r (e_r \cosh v-1)\,,\nonumber\\
\bar n t&=&e_t \sinh v-v + f_t V+g_t \sin V\,,\nonumber\\
\phi &=&K[V+f_\phi \sin 2V+g_\phi \sin 3V]\,,
\end{eqnarray}
with
\beq
\label{Vdef}
V(v)=2\, {\rm arctan}\left[\sqrt{\frac{e_\phi+1}{e_\phi-1}}\tanh \frac{v}{2}  \right]\,.
\eeq
Here we use dimensionless variables $t= c^3 t^{\rm phys}/(GM)$ and $r= c^2 r^{\rm phys}/(GM)$ as well as dimensionless rescaled orbital parameters, such as a dimensionless semi-major axis $a_r \equiv c^2 a^{\rm phys}/( GM)$.
The expressions of the orbital parameters $\bar n$, $\bar a_r$, $K$, $e_t,e_r,e_\phi$, $f_t,g_t,f_\phi, g_\phi$ are given, e.g., in Table VIII of Ref. \cite{Bini:2020hmy} in harmonic coordinates as functions of the conserved energy and angular momentum of the system.

We computed $E^{\rm rad}$ with the 2PN (fractional) accuracy, including the 1.5PN tail
contribution coming from the hereditary integral terms in Eqs. \eq{U_LandV_L}. 
The tail contribution is discussed in the following Appendix. Let us display here our results
for the 2PN-accurate energy loss coming from the instantaneous multipole contributions, starting
with the quadrupolar one: $U_{ij}^{\rm inst}= I_{ij}^{(2)}$.

We first display $E^{\rm rad, inst}_{\rm 2PN}$ as an exact expression in terms of the harmonic-coordinate orbital parameters $e_r$ and $\bar a_r$ (instead of $e_r$ and $j$ as in our previous work \cite{Bini:2020hmy}, Eq. (D2), which generalizes the 1PN accurate results of Ref. \cite{Blanchet:1989cu}, Eq. (5.7))
\beq
 E^{\rm rad, inst}_{\rm 2PN}= \frac{\nu^2}{\bar a_r^{7/2}}\left(\Delta E^{\rm N}_{\rm resc}+\frac{\eta^2}{\bar a_r}\Delta E^{\rm 1PN}_{\rm resc}+\frac{\eta^4}{\bar a_r^2}\Delta E^{\rm 2PN}_{\rm resc}\right)\,,
\eeq
where we have set $G=M=c=1$ for simplicity, and
\beq
\Delta E^{\rm nPN}_{\rm resc}= \frac{A_E^{\rm nPN}}{\sqrt{e_r^2-1}}{\rm arccos}\left(-\frac{1}{e_r}\right)+B_E^{\rm nPN}\,,
\eeq
with coefficients
\begin{widetext}
\bea
(e_r^2-1)^3 A_E^{\rm N}&=&\frac{2}{15} (292 e_r^2+96+37 e_r^4)  
\,,\nonumber\\
(e_r^2-1)^3 B_E^{\rm N}&=&\frac{2}{45} (673 e_r^2+602)  
\,,\nonumber\\
(e_r^2-1)^{4}A_E^{\rm 1PN}&=&  -\frac{112}{5}\nu-\frac{1132}{21}
+\left(-\frac{37318}{105}-\frac{308}{3}\nu\right) e_r^2
+\left(-\frac{233}{2}-\frac{249}{5}\nu\right) e_r^4
+\left(\frac{1143}{140}-\frac{37}{15}\nu\right) e_r^6 
\,,\nonumber\\
(e_r^2-1)^4  B_E^{\rm 1PN}&=& -\frac{446}{9}\nu-\frac{210811}{1575}
+\left(-105\nu-\frac{592573}{1575}\right) e_r^2
+\left(-\frac{205}{9}\nu-\frac{47659}{6300}\right) e_r^4 
\,,\nonumber\\
(e_r^2-1)^{5}A_E^{\rm 2PN}&=& \frac{105146}{315}\nu+\frac{32}{5}\nu^2+\frac{44134}{405}
+\left(\frac{387220}{567}+\frac{1089412}{315}\nu+\frac{623}{15}\nu^2\right) e_r^2\nonumber\\
&&
+\left(-\frac{108326}{945}+\frac{622831}{210}\nu+\frac{587}{12}\nu^2\right) e_r^4
+\left(-\frac{424337}{2520}+\frac{273}{10}\nu^2+\frac{27875}{168}\nu\right) e_r^6\nonumber\\
&&
+\left(\frac{37}{20}\nu^2+\frac{114101}{5040}-\frac{1411}{168}\nu\right) e_r^8 
\,,\nonumber\\
(e_r^2-1)^5 B_E^{\rm 2PN}&=& \frac{607888}{675}\nu+\frac{151}{10}\nu^2+\frac{78464696}{297675}
+\left(\frac{9047}{180}\nu^2+\frac{84265357}{18900}\nu+\frac{164159833}{238140}\right) e_r^2\nonumber\\
&&
+\left(\frac{59688863}{37800}\nu-\frac{520075147}{1190700}+\frac{2048}{45}\nu^2\right) e_r^4
+\left(-\frac{6299}{280}\nu+\frac{2711041}{176400}+\frac{2723}{180}\nu^2\right) e_r^6\,.
\eea

\end{widetext}

Expanding then the above results in inverse powers of $j$ (once $(\bar a_r, e_r)$ have been reexpressed in terms of $(p_\infty,j)$), yields the following $G^7$-accurate (2PN) results
\be
E^{\rm rad, inst}_{\rm 2PN}=\nu^2\left(E_{\rm N}+\eta^2 E_{\rm 1PN}+\eta^4 E_{\rm 2PN}\right)\,,
\ee
where
\bea
\label{Erad_N}
E_{\rm N}&=&  \frac{37}{15} \pi\frac{p_\infty^4}{j^3}+\frac{1568}{45} \frac{p_\infty^3}{j^4}+\frac{122}{5}\pi \frac{p_\infty^2}{j^5}\nonumber\\
&+&\frac{4672}{45} \frac{p_\infty}{j^6}+\frac{85}{3} \pi\frac{1}{j^7}+O\left(\frac{1}{j^8}\right)\,,
\eea
\bea 
\label{Erad_1PN}
E_{\rm 1PN}&=&  \left(\frac{1357}{840}   -\frac{74}{15}\nu \right)\pi \frac{p_\infty^6}{j^3}\nonumber\\
&+&\left( \frac{18608}{525} -\frac{1424}{15}\nu \right)\frac{p_\infty^5}{j^4}\nonumber\\
&+&\left(\frac{13831}{280}-\frac{933}{10}\nu    \right)\pi\frac{p_\infty^4}{j^5}\nonumber\\
&+&\left(\frac{142112}{315}-\frac{26464}{45}\nu   \right)\frac{p_\infty^3}{j^6}\nonumber\\
&+&\left(\frac{2259}{8}-265\nu     \right)\pi \frac{p_\infty^2}{j^7}+O\left(\frac{1}{j^8}\right)\,,\nonumber\\ 
\eea
and
\bea
\label{Erad_2PN}
E_{\rm 2PN}&=&  \left(\frac{27953}{10080}-\frac{839}{420}\nu  +\frac{37}{5}\nu^2     \right)\pi \frac{p_\infty^8}{j^3}\nonumber\\
&+&\left(\frac{220348}{11025}-\frac{31036}{525}\nu     +172\nu^2\right)\frac{p_\infty^7}{j^4}\nonumber\\
&+&\left(-\frac{64579}{5040}    -\frac{187559}{1680}\nu   +\frac{2067}{10}\nu^2 \right)\pi \frac{p_\infty^6}{j^5}\nonumber\\
&+&\left(-\frac{293992}{1701}-\frac{6732728}{4725}\nu  +\frac{24424}{15}\nu^2  \right)\frac{p_\infty^5}{j^6}\nonumber\\
&+&\left(\frac{19319}{378}   -\frac{432805}{336}\nu   +\frac{7605}{8}\nu^2  \right)\pi \frac{p_\infty^4}{j^7}\nonumber\\
&+&O\left(\frac{1}{j^8}\right)\,.
\eea

From the above expressions one can easily get the PN expansion of the PM coefficients $E_n$ (see Eq. \eq{PMexpEJrad}).
For example
\bea
E_3&=&\frac{37}{15} \pi p_\infty^4
+\left(\frac{1357}{840}   -\frac{74}{15}\nu \right)\pi p_\infty^6\nonumber\\
&&
+\left(\frac{27953}{10080}-\frac{839}{420}\nu  +\frac{37}{5}\nu^2\right)\pi p_\infty^8\nonumber\\
&&
+O(p_\infty^{10})\,.
\eea
Note that no tails appear at the level of $E_3$ (see the next section).

\section{Fractionally 1.5PN tail contribution to the radiated energy along hyperboliclike orbits}
\label{App_tail_E}

It was  shown in Ref. \cite{Damour:2014jta} that the 4PN-level tail contribution
to the radiation-reaction force is given by the following time-antisymmetric force
\be \label{Frrtail}
{\cal F}^i_{a \, \rm tail}= - \frac{4 \pi}{5} \frac{G^2 M}{c^8} m_a x_a^j H[I_{ij}^{(6)}](t)\,.
\ee
Here, the superscript on the quadrupole moment $I_{ij}$ denoting the sixth time-derivative,
while $H$ denotes the Hilbert transform, defined as (with P denoting the principal value)
\be
\label{HT_def}
H[f](t) \equiv \frac1{\pi} {\rm P} \int_{- \infty}^{+ \infty} \frac{dt'}{t-t'} f(t')\,.
\ee
The time-integrated, tail-related loss of mechanical energy of the two-body system
 induced by the radiation-reaction force 
\eq{Frrtail} is $ - \delta^{\rm rr}E_{\rm system}=- \int dt \sum_a \dot x^i_a {\cal F}^i_{a \, \rm tail}$.
Replacing the symmetric-trace-free projection of  $\sum_a m_a \dot x^i_a  x^j_a$ by
(half) the time derivative of the quadrupole moment $I_{ij}$, and integrating by parts, one
easily finds that
\be \label{Eloss}
 - \delta^{\rm rr}E_{\rm system}=  \frac{2 \pi}{5} \frac{G^2 M}{c^8}\int dt I_{ij}^{(3)}(t) H[I_{ij}^{(4)}](t)\,.
\ee
Let us compare this tail-related loss of mechanical energy to the usually considered expression for
 the leading-order (LO) tail contribution to the {\it radiated} energy, coming
 from including in Eq. \eq{energy_flux} the $U_{ij}^{\rm tail}$ hereditary contribution of Eq. \eq{U_LandV_L}, namely 
\bea \label{Epast}
E^{\rm rad}_{\rm  past\,   tail}&=& \frac{4}{5}\frac{G^2 M}{c^8} \int dt I_{ij}^{(3)}(t){\mathcal I}^{(5)}_{ij \, \rm past}(t).
\eea
Here we used the notation of Ref. \cite{Arun:2007rg}
(applicable to a generic multipole moment of electric or magnetic type)
\bea
{\mathcal I}^{(n)}_{L \, \rm past}(t)& \equiv &\int_0^{+\infty} d\tau   \ln \left(\frac{\tau}{C_{I_L}}\right)I^{(n)}_{L}(t-\tau)\,, \qquad
\eea
where the constant $C_{I_L}$ depends on the considered multipole. Namely
\beq
C_{I_L}=2r_0e^{-\kappa_l}\,,\qquad C_{J_L}=2r_0e^{-\pi_l}\,,
\eeq
where $\kappa_l$ and $\pi_l$ are those of Eq. \eq{U_LandV_L},
and where  the length scale $r_0$ coincides with the one introduced in the multipolar post-Minkowskian formalism \cite{Blanchet:1985sp}; explicitly
\bea
\label{C_is}
[C_{I_2},C_{I_3},C_{I_4}]&=& 2r_0[e^{-11/12},e^{-97/60},e^{-59/30}]\,,\nonumber\\
{}[C_{J_2},C_{J_3}]&=& 2r_0[e^{-7/6},e^{-5/3}]\,.
\eea

The radiated energy Eq. \eq{Epast} a priori looks different from the energy loss \eq{Eloss},
notably because the tail-induced quadrupole moment ${\mathcal I}^{(5)}_{ij \, \rm past}(t)$
entering Eq. \eq{Epast} is {\it past-hereditary} (and therefore asymmetric under time-reversal),
 while the integrand of \eq{Eloss} is time-symmetric. [The Hilbert transform being time-odd.]
 However, it is easily seen that if one decomposes the integral transform entering  ${\mathcal I}^{(5)}_{ij \, \rm past}(t)$ into its time-even and time-odd parts, by replacing 
 $ I^{(5)}_{ij}(t-\tau)$ by $\frac12( I^{(5)}_{ij}(t-\tau)+  I^{(5)}_{ij}(t+\tau)) + \frac12( I^{(5)}_{ij}(t-\tau)-  I^{(5)}_{ij}(t+\tau))$, the time-odd part will yield (after integrating by parts)
 a vanishing contribution to the double integral
 \be
 \int_{- \infty}^{+ \infty} dt \int_0^{+\infty} d\tau I^{(3)}_{ij}(t) [ I^{(5)}_{ij}(t-\tau)-  I^{(5)}_{ij}(t+\tau)] \ln \tau =0\,.
 \ee
After a last integration by parts, one finds  that $E^{\rm rad}_{\rm  past\,   tail}$ is simply equal to
the mechanical energy loss $ - \delta^{\rm rr}E_{\rm system}$ given by the time-even expression \eq{Eloss}
involving the Hilbert transform $H[I_{ij}^{(4)}]$. 

This time-symmetric evaluation of tail effects is the concrete realization of the time-symmetric evaluation of GW radiation  mentioned in footnote \ref{foot_bisec}, and valid when treating radiation-reaction  effects to  linear order. Let us recall that the treatment used in this paper is only intended to be valid to first order in ${\mathcal F}^{\rm rr}$.

A more transparent way to understand the equality
between $E^{\rm rad}_{\rm  past\,   tail}$ and $ - \delta^{\rm rr}E_{\rm system}$ is to work in the
frequency domain. The time-domain convolutions entering both $H[I_{ij}^{(4)}]$
and  ${\mathcal I}^{(5)}_{ij \, \rm past}(t)$ become multiplications by the Fourier
transforms of their kernels, say $K(\omega) = \int dt e^{i \omega t} K(t)$. It is convenient to factor out 
some of the factors $ (- i \omega)^n$ corresponding to a $n$th time derivative so as to
write both integrals as bilinear forms in the Fourier transform of $I_{ij}^{(3)}(t)$. More
precisely, we shall write the various tail integrals in the form
\be
E^{\rm rad}_{\rm X}=\frac{2 \pi}{5} \frac{G^2 M}{c^8}\int_{- \infty}^{+ \infty} \frac{d \omega}{2 \pi } I_{ij}^{(3)}(- \omega) K^{\rm X}(\omega)  I_{ij}^{(3)}(\omega)\,,
\ee
where the label $X$ is either ${\rm rr}$ or  past tail, or sym tail, to be defined below. 

Using the fact that the kernel of the Hilbert transform is $- i \, {\rm sign}(\omega)$, the 
kernel $K^{\rm rr}(\omega)$ entering the mechanical energy loss $ - \delta^{\rm rr}E_{\rm system}$ 
is obtained as  
\be
K^{\rm rr}(\omega)= - (- i {\rm sign}(\omega)) (- i \omega) = + |\omega|\,.
\ee
On the other hand, using the integral (See, e.g., Refs. \cite{Blanchet:1993ec,Arun:2007rg})
\be \label{Krr}
\int_0^\infty du\, e^{i\omega u}\ln \left(\frac{u}{C}\right)
= -\frac{\pi}{2|\omega|}-\frac{i}{ \omega }\ln (C|\omega|e^{\gamma_E}),
\ee
the kernel entering the past-tail radiated energy reads
\be \label{Kpast}
K^{\rm  past\,   tail}(\omega)=  | \omega| + \frac{2  i}{\pi} \omega \ln (C_{I_2} |\omega| e^{\gamma_E}).
\ee
The crucial point here is that the second contribution on the r.h.s. of the past-tail kernel, Eq. \eq{Kpast},
 is {\it odd under frequency reversal}, $\omega \to - \omega$ (corresponding to time-reversal). The corresponding 
integrated term (sandwiched between $ I_{ij}^{(3)}(- \omega)$ and $  I_{ij}^{(3)}(\omega)$ then
{\it vanishes}.

In other words, if we define the {\it symmetric tail} as being the time-symmetric projection
of the hereditatry past tail, so that its frequency kernel is  the frequency-symmetric projection
of $K^{\rm  past\,   tail}(\omega)$, 
\bea
K^{\rm  sym\,   tail}(\omega) &=&  \frac12  \left [K^{\rm  past\,   tail}(\omega)+ K^{\rm  past\,   tail}(-\omega)\right] \nonumber\\
&=& | \omega| = K^{\rm rr}(\omega)\,,
\eea
we have the double result that the tail contribution to the total  radiated energy is
automatically equal to its time-symmetric projection, and  that it balances the
radiation-reaction energy loss
\be
E^{\rm rad}_{\rm  past\,   tail} = E^{\rm rad}_{\rm  sym\,   tail} =  - \delta^{\rm rr}E_{\rm system}\,,
\ee
where
\bea \label{Eradtail}
E^{\rm rad}_{\rm  sym\,   tail} 
 &=&\frac{2 \pi}{5} \frac{G^2 M}{c^8}\int_{- \infty}^{+ \infty} \frac{d \omega}{2 \pi } I_{ij}^{(3)}(- \omega) |\omega|  I_{ij}^{(3)}(\omega) \nonumber\\
  &=&\frac{2 \pi}{5} \frac{G^2 M}{c^8}\int_{- \infty}^{+ \infty} \frac{d \omega}{2 \pi } I_{ij}(- \omega) 
  |\omega| ^7 I_{ij}(\omega)\nonumber\\
   &=&\frac{2 }{5} \frac{G^2 M}{c^8}  \int_{0}^{+ \infty} d \omega \, \omega^7  I_{ij}(- \omega) I_{ij}(\omega)\,.
 \eea
 Besides being conceptually useful for clarifying the properties of tail integrals, the frequency-domain
 representation is very useful for  computing the explicit values of the various needed tail
 integrals along hyperbolic motions. Similarly to the various hyperbolic-motion integrals we
 encountered in our previous works \cite{Bini:2019nra,Bini:2020wpo,Bini:2020nsb,Bini:2020hmy,Bini:2020rzn}, we could evaluate the large-impact-parameter
 expansion of the needed tail integrals of the type \eq{Eradtail} by starting from the 
 Keplerian parametrization  of the hyperbolic motion. 
The appropriate Newtonian-level limit of Eqs. \eq{hypQK2PN} reads
\begin{eqnarray} \label{hypKN}
r&=& \bar a_r (e_r  \cosh v-1)\,,\nonumber\\
\bar n t&=&e_r  \sinh v-v\,,\nonumber\\  
\phi&=&2\, {\rm arctan}\left[\sqrt{\frac{e_r +1}{e_r -1}}\tanh \frac{v}{2}  \right]\,,
\end{eqnarray}
with orbital parameters $\bar n= p_\infty^3$, $\bar a_r=1/(2\bar E)=1/p_\infty^2$, $e_r=\sqrt{1+p_\infty^2j^2}$.
 
The first step consists in Fourier transforming the multipole moments entering Eq. \eq{Eradtail}, i.e.,
\beq
\label{I_ab_omega2}
\hat I_{ab}(\omega)=\int \frac{dt}{dv}e^{i\omega t(v)}I_{ab}(t)|_{t=t(v)} \, dv \,,
\eeq
and similarly for the other moments.
This is done by using the integral representation of the Hankel functions of the first kind of order $p \equiv \frac{q}{e_r}$ and argument $q \equiv i \, u$, where
\be 
u\equiv \omega e_r \bar a_r^{3/2}, 
\ee
namely
\beq
\label{Hankel_rep}
H_p^{(1)}(q)=\frac{1}{ i\pi }\int_{-\infty}^\infty e^{q\sinh v -p v}dv\,.
\eeq
As the argument $q=iu$  of the Hankel function is purely imaginary, the Hankel function becomes converted into a 
Bessel K function, according to the  relation
\beq
H_p^{(1)}(iu)=\frac{2}{\pi}e^{-i \frac{\pi}{2}(p+1)}K_p(u)\,.
\eeq
Using standard identities valid for Bessel functions we could express
the results in terms of only two orders: $p= \frac{ i u}{e_r}$ and $p+1= 1+\frac{ i u}{e_r} $. 
For instance, the averaged energy tail turns out to read (in scaled variables, provisionally setting $G=M=c=1$)
\beq \label{EGWu}
E^{\rm rad}
_{\rm   tail}=\frac{1}{e_r^2 \bar a_r^3}  \int_{0}^\infty  du  {\mathcal K}_{\rm tail}(u)\,,
\eeq
where we suppressed the qualification (past or sym) of the tail.

\begin{widetext}

\bea
\label{calKtail}
\mathcal K_{\rm  tail}(u)&=&
-\frac{64}{5}\frac{\nu^2}{\bar a_r^2}\frac{p^2}{u^3}e^{-i\pi p}\left\{
u^2(p^2+u^2+1)(p^2+u^2)K_{p+1}^2(u)\right.\nonumber\\
&&
-2u\left[\left(p-\frac32\right)u^2+p(p-1)^2\right](p^2+u^2)K_{p}(u)K_{p+1}(u)\nonumber\\
&&\left.
+2\left[\frac12u^6+\left(2p^2-\frac32p+\frac16\right)u^4+\left(\frac52p^4-\frac72p^3+p^2\right)u^2+p^4(p-1)^2\right]K_{p}^2(u)
\right\}\,.
\eea

\end{widetext}

The integral \eq{EGWu} cannot be performed in closed analytical form because the orders $p$ or $p+1$
 of the Bessel K functions depend on the $u$ integration variable.
 However, as $p= \frac{i u}{e_r}$ tends to zero when $e_r \to \infty$, the use of a large-eccentricity expansion allows
 one to express the integral \eq{EGWu} in terms of computable Bessel-function integrals.
 Indeed, Taylor-expanding to second order in $p$,
\begin{eqnarray}
\label{K_exp}
K_p(u)&=& K_0(u)+\frac12 p^2 \frac{\partial^2 K_\nu(u)}{\partial \nu^2}\Bigg|_{\nu=0}+O(p^3)
\,,\nonumber\\
K_{p+1}(u)&=& K_1(u)+\frac{p}{u} K_0(u) \nonumber\\
&&+\frac12 p^2 \frac{\partial^2 K_\nu(u)}{\partial \nu^2 }\Bigg|_{\nu=1}+O(p^3)\,,
\end{eqnarray}
yields the following  N$^3$LO  large-eccentricity expansion for the integrand \eqref{calKtail}:

\bea
\mathcal K_{\rm tail}(u)&=&\frac{\nu^2}{e_r^2\bar a_r^2}\left[\tilde {\mathcal K}_{\rm tail}^{\rm LO}(u)+\frac{\pi}{e_r}\tilde {\mathcal K}_{\rm  tail}^{\rm NLO}(u)\right.\nonumber\\
&+&\left.
\frac{1}{e_r^2}\tilde {\mathcal K}_{\rm  tail}^{\rm NNLO}(u)+\frac{\pi}{e_r^3}\tilde {\mathcal K}_{\rm  tail}^{\rm N^3LO}(u)\right]\,,
\eea
where
\begin{widetext}

\bea
\tilde {\mathcal K}_{\rm tail}^{\rm LO}(u)&=&\frac{64}{5}u^3\left[
\left(\frac{1}{3}+u^2\right)K_0^2(u)+3u K_0(u)K_1(u)+(1+u^2)K_1^2(u)
\right]
\,,\nonumber\\
\tilde {\mathcal K}_{\rm  tail}^{\rm NLO}(u)&=&u\tilde {\mathcal K}_{\rm  tail}^{\rm LO}(u)
\,,\nonumber\\
\tilde {\mathcal K}_{\rm  tail}^{\rm NNLO}(u)&=&\frac{\pi^2}{2}u^2\tilde {\mathcal K}_{\rm  tail}^{\rm LO}(u)
-\frac{64}{5}u^3\left\{
(1+3u^2)K_0^2(u)+7u K_0(u)K_1(u)+(1+2u^2)K_1^2(u)\right.\nonumber\\
&&\left.
+u^2\left[\left(u^2+\frac13\right)K_0(u)+\frac32uK_1(u)\right]\frac{\partial^2 K_\nu(u)}{\partial \nu^2}\Bigg|_{\nu=0}
+u^2\left[\frac32uK_0(u)+(u^2+1)K_1(u)\right]\frac{\partial^2 K_\nu(u)}{\partial \nu^2}\Bigg|_{\nu=1}
\right\}
\,,\nonumber\\
\mathcal K_{\rm  tail}^{\rm N^3LO}(u)&=&u\mathcal K_{\rm  tail}^{\rm NNLO}(u)-\frac{\pi^2}{3}u^3\mathcal K_{\rm  tail}^{\rm LO}(u)
\,.
\eea

\end{widetext}
We could then evaluate all the needed integrals. Converting the large-$e_r$ expansion into
a large-$j$ expansion finally leads to the following explicit result
\bea
\label{Eradtailfin}
 E^{\rm rad}_{\rm tail}&=&\nu^2\left[\frac{3136}{45} \frac{p_\infty^6}{j^4} 
+ \pi  \frac{297\pi^2}{20}\frac{p_\infty^5}{j^5}\right.\nonumber\\
&+&  \left(\frac{9344}{45}  + \frac{88576}{675} \pi^2\right)\frac{p_\infty^4}{j^6} \nonumber\\
&+&\left. \pi\left(-\frac{2755}{64} \pi^4 + \frac{1579}{3} \pi^2\right)\frac{p_\infty^3}{j^7} \right]\nonumber\\
&+&O\left(\frac{1}{j^8}\right)\,.
\eea

From the above expression, including the results of the previous section, one easily gets the PN expansion of the PM coefficients $E_n$ (see Eq. \eq{PMexpEJrad}).
For example
\bea
E_4&=&\frac{1568}{45} p_\infty^3
+\left( \frac{18608}{525} -\frac{1424}{15}\nu \right)p_\infty^5\nonumber\\
&&
+\frac{3136}{45} p_\infty^6\nonumber\\
&&
+\left(\frac{220348}{11025}-\frac{31036}{525}\nu     +172\nu^2\right)p_\infty^7\nonumber\\
&&
+O(p_\infty^8)\,,
\eea
which incorporates the 1.5PN tail contribution.

\section{Fractionally 2PN-accurate radiated angular momentum along hyperboliclike orbits (without the 1.5PN tail contribution)}
\label{App_J}

The radiated angular momentum flux ${\mathcal G}_i=\frac{dJ^{\rm rad}_i}{dU}$ in terms of radiative multipole moments reads
\bea
\label{angmom_flux}
{\mathcal G}_i&=&\epsilon_{iab}\sum_{l=2}^\infty \frac{G}{c^{2l+1}}
\left\{
\frac{(l+1)(l+2)}{(l-1) l!(2l+1)!!}U_{aL-1}U_{bL-1}^{(1)}\right.\nonumber\\
&&\left.+\frac{4l^2(l+2)}{c^2(l-1)(l+1)!(2l+1)!!}V_{aL-1}V_{bL-1}^{(1)}
\right\}\,.
\eea
By replacing both $U_L$ and $V_L$ by their corresponding expressions \eqref{U_LandV_L} in Eq. \eqref{angmom_flux}, one finds for the instantaneous contribution
to the radiated angular momentum
\beq
J_i^{\rm rad, inst}=\int_{-\infty}^\infty dt {\mathcal G}_i^\mathrm{rad, inst}\,,
\eeq
with
\bea
\mathcal{G}_i^\mathrm{rad, inst}&=&{G\over c^5}\epsilon_{iab}\biggl\{ {2\over 5}
I^{(2)}_{aj} I^{(3)}_{bj}\nonumber\\
&+&
{1\over c^2} \left[ {1\over
63} I^{(3)}_{ajk} I^{(4)}_{bjk} +{32\over 45} J^{(2)}_{aj}
J^{(3)}_{bj}\right]\nonumber\\
&+&{1\over c^4} \left[ {1\over 2268} I^{(4)}_{ajkm}
I^{(5)}_{bjkm}+{1\over 28} J^{(3)}_{ajk}
J^{(4)}_{bjk}\right]\biggr\}\,,\nonumber\\
\eea
at the 2PN level of accuracy.
The only nonvanishing component is the $z$ one. 
Suppressing the $z$ label we find
\beq
J^{\rm rad, inst}_{\rm 2PN}= \frac{\nu^2}{\bar a_r^{2}}\left(\Delta J^{\rm N}_{\rm resc}+\frac{\eta^2}{\bar a_r}\Delta J^{\rm 1PN}_{\rm resc}+\frac{\eta^4}{\bar a_r^2}\Delta J^{\rm 2PN}_{\rm resc}\right)\,,
\eeq
where
\beq
\Delta J^{\rm nPN}_{\rm resc}= \frac{A_J^{\rm nPN}}{\sqrt{e_r^2-1}}{\rm arccos}\left(-\frac{1}{e_r}\right)+B_J^{\rm nPN}\,,
\eeq
with coefficients
\begin{widetext}
\bea
(e_r^2-1)^{3/2} A_J^{\rm N}&=& \frac{64}{5} +\frac{56}{5} e_r^2 
\,,\nonumber\\
(e_r^2-1)^{3/2} B_J^{\rm N}&=&\frac{104}{5}  +\frac{16}{5} e_r^2 
\,,\nonumber\\
(e_r^2-1)^{5/2} A_J^{\rm 1PN}&=& -\frac{144}{5}\nu-\frac{3644}{105} 
+\left(-\frac{208}{3}\nu-\frac{9596}{105}\right)e_r^2
+\left(-\frac{158}{15}\nu+\frac{739}{42}\right) e_r^4
\,,\nonumber\\
(e_r^2-1)^{5/2} B_J^{\rm 1PN}&=&-\frac{2524}{45}\nu-\frac{8089}{105}
+\left(-\frac{2294}{45}\nu-\frac{9007}{210}\right)e_r^2
+  \left(-\frac{8}{5}\nu+\frac{80}{7}\right)e_r^4
\,,\nonumber\\
(e_r^2-1)^{7/2} A_J^{\rm 2PN}&=& \frac{93316}{2835}+\frac{73316}{315}\nu+16\nu^2
+\left(\frac{32636}{21}\nu-\frac{255182}{945}+\frac{488}{5}\nu^2\right) e_r^2\nonumber\\
&&
+\left(\frac{76831}{210}\nu-\frac{743291}{1890}+\frac{1462}{15}\nu^2\right) e_r^4
+\left(\frac{58}{5}\nu^2-\frac{3328}{105}\nu+\frac{5509}{180}\right) e_r^6
 \nonumber\\
(e_r^2-1)^{7/2} B_J^{\rm 2PN}&=& \frac{1628}{45}\nu^2+\frac{2947753}{4725}\nu+\frac{1297277}{42525} 
+\left(-\frac{2823452}{6075}+\frac{1816}{15}\nu^2+\frac{14194393}{9450}\nu\right) e_r^2\nonumber\\
&&
+\left(\frac{5686}{1575}\nu-\frac{137559}{700}+\frac{578}{9}\nu^2\right) e_r^4
+\left(\frac{6}{5}\nu^2-\frac{298}{35}\nu+\frac{1952}{63}\right) e_r^6
\,.
\eea
\end{widetext}

In the large-$j$ expansion limit one has
\be
J^{\rm rad, inst}_{\rm 2PN}= \nu^2\left(J_{\rm N}+\eta^2 J_{\rm 1PN}+\eta^4 J_{\rm 2PN}\right)\,,
\ee
where
\bea 
\label{Jrad_N}
J_{\rm N}&=&  \frac{16}{5} \frac{p_\infty^3}{j}+\frac{28}{5} \pi \frac{p_\infty^2}{j^2}+\frac{176}{5} \frac{p_\infty}{j^3}\nonumber\\
&+&12 \pi\frac{1}{j^4}+\frac{304}{15 p_\infty  j^5}-\frac{144}{ 25 p_\infty^3  j^7}+O\left(\frac{1}{j^8}\right)\,,  
\eea
\bea
\label{Jrad_1PN}
J_{\rm 1PN}&=&  \left(-\frac{16}{5}\nu  +\frac{176}{35} \right)\frac{p_\infty^5}{j}\nonumber\\
&+&\left(  \frac{739}{84}-\frac{163}{15}\nu \right)\pi \frac{p_\infty^4}{j^2}\nonumber\\
&+&\left(\frac{8144}{105} -\frac{1072}{9}\nu  \right)\frac{p_\infty^3}{j^3}\nonumber\\
&+&\left( \frac{107}{2} -\frac{374}{5}\nu \right)\pi \frac{p_\infty^2}{j^4}\nonumber\\
&+&\left(\frac{84368}{315}-\frac{13456}{45}  \nu \right)\frac{p_\infty}{j^5}\nonumber\\
&+&\left( \frac{359}{4} -\frac{235}{3}\nu \right)\pi\frac{1}{j^6}\nonumber\\
&+&\left(\frac{16816}{105}  -\frac{8336}{75} \nu \right)\frac{1}{p_\infty j^7}+O\left(\frac{1}{j^8}\right)\,,
\eea
and
\bea  
\label{Jrad_2PN}
J_{\rm 2PN}&=& \left(-\frac{608}{315}   -\frac{148}{35}\nu  +\frac{16}{5}\nu^2 \right)\frac{p_\infty^7}{j}\nonumber\\
&+&\left(-\frac{5777}{2520}-\frac{5339}{420}\nu  +\frac{50}{3}\nu^2  \right)\pi \frac{p_\infty^6}{j^2}\nonumber\\
&+&\left(-\frac{93664}{1575}-\frac{247724}{1575}\nu     +\frac{11056}{45}\nu^2 \right)\frac{p_\infty^5}{j^3}\nonumber\\
&+&\left( -\frac{101219}{1512}  -\frac{15856}{105}\nu+209  \nu^2   \right)\pi\frac{p_\infty^4}{j^4}\nonumber\\
&+&\left(-\frac{552320}{1701}-\frac{31828}{27}\nu     +\frac{3568}{3} \nu^2\right)\frac{p_\infty^3}{j^5}\nonumber\\
&+&\left(-\frac{2999}{216}-\frac{8497}{12}\nu   +\frac{1484}{3}  \nu^2 \right)\pi\frac{p_\infty^2}{j^6}\nonumber\\
&+&\left(\frac{10005568}{14175}  -\frac{4719436}{1575}  \nu +\frac{4432}{3} \nu^2\right)\frac{p_\infty}{j^7}\nonumber\\
&+& O\left(\frac{1}{j^8}\right)\,.
\eea

\section{Fractionally 1.5PN tail contribution to the radiated angular momentum  along hyperboliclike orbits}
\label{App_tail_J}

The tail contribution to the radiated angular momentum  is obtained as in the case of the radiated energy. A direct calculation shows again that the integrated past-tail is equal to its time-symmetric projection:
\bea
\label{tail_J}
(\Delta J_i)_{\rm past\, tail}&=& (\Delta J_i)_{\rm sym\, tail} \nonumber\\
&=& \frac{G^2M}{c^8} \frac{4}{5}i  \epsilon_{ijk}    \int_0^\infty  d\omega \omega^{6}   I_{jl}(\omega)I_{kl}(-\omega)\,,
\nonumber\\
\eea
with only surviving component along the $z$-axis, with value
\bea
\label{Jradtailfin}
\Delta J{}_{\rm sym\, tail}&=&\nu^2 \left[\frac{448}{5}\frac{p_\infty^4}{j^3}  
+\pi \frac{69\pi^2}{5} \frac{p_\infty^3}{j^4}\right.\nonumber\\
&+&  \left(\frac{128}{15}+\frac{4352}{45}  \pi^2  \right)\frac{p_\infty^2}{j^5}\nonumber\\
&+&\pi \left(-\frac{423}{16}  \pi^4 + 303 \pi^2 \right)\frac{p_\infty}{j^6}\nonumber\\
&+&\left.O\left(\frac{1}{j^7}\right)\right]\,,
\eea
where we have suppressed the $z$ label and the large-$j$ expansion limit has been considered as usual.

\section{Fractionally 2PN-accurate radiated linear momentum along hyperboliclike orbits (without the 1.5PN tail contribution)}

The general expression for the linear momentum flux at the 2PN level of accuracy in terms of the 
radiative multipole moments \cite{Blanchet:1985sp,Blanchet:1987wq,Blanchet:1989ki,Damour:1990ji,Blanchet:1998in,Poujade:2001ie} is
\begin{widetext}
\bea
\frac{dP^{\rm rad}_j}{dU}&=&\sum_{l=2}^\infty \left[\frac{G}{c^{2l+3}}\frac{2(l+2)(l+3)}{l(l+1)!(2l+3)!!}U_{jL}^{(2)}U_L^{(1)}+\frac{G}{c^{2l+5}}\frac{8(l+3)}{(l+1)!(2l+3)!!}V_{jL}^{(2)}V_{L}^{(1)}\right.\nonumber\\
&&\left. 
+\frac{G}{c^{2l+3}}\frac{8(l+2)}{(l-1)(l+1)!(2l+1)!!}\epsilon_{jab}U_{aL}^{(1)}V_{bL}^{(1)}
\right]\,.
\eea
The 2PN-accurate instantaneous contribution to the above expression reads
\bea
\label{dP_dt_eq}
\frac{dP^{\rm rad, inst}_j}{dU}&=&\frac{G}{c^7}\left[ \frac{2}{63}I_{jpq}^{(4)}I_{pq}^{(3)}+\frac{16}{45}\epsilon_{jpq}I_{pr}^{(3)}J_{qr}^{(3)}\right. \nonumber\\
&&+\frac1{c^2}\left(\frac{4}{63}J^{(4)}_{jpq}J^{(3)}_{pq}+\frac{1}{1134}I^{(5)}_{jpqr}I^{(4)}_{pqr}+\frac{1}{126}\epsilon_{j p q}I^{(4)}_{prs}J^{(4)}_{qrs} \right)\nonumber\\
&&\left.
+\frac1{c^4}\left(
 \frac{2}{945}J^{(5)}_{jpqr}J^{(4)}_{pqr}+\frac{1}{59400}I^{(6)}_{jpqrs}I^{(5)}_{pqrs}+\frac{2}{14175}\epsilon_{j p q}
I^{(5)}_{pabc}J^{(5)}_{qabc}\right)
 \right]\,.
\eea
Integrating Eq. \eqref{dP_dt_eq} along the hyperbolic orbit we find $\Delta P^{\rm rad, inst}_x=0=\Delta P^{\rm rad, inst}_z$, and
\bea
\Delta P^{\rm rad, inst}_y &=&\frac{\nu^2\sqrt{1-4\nu}}{(e_r^2-1)^4\bar a_r^4} \left(Q_0+\frac{\eta^2}{(e_r^2-1)e_r\bar a_r} Q_2
+\frac{e_r\eta^4}{(e_r^2-1)^2\bar a_r^2} Q_4\right)\,,
\eea
where
\bea
Q_0&=&A_1e_r\arccos\left(-\frac1{e_r}\right)+\frac{(e_r^2-1)^{1/2}}{e_r}A_0\,,\nonumber\\
Q_2&=&B_1\arccos\left(-\frac1{e_r}\right)+(e_r^2-1)^{1/2}B_0\,,\nonumber\\
Q_4&=&C_3\arccos^3\left(-\frac1{e_r}\right)+\frac{C_2(e_r^2-1)^{1/2}}{e_r^2}\arccos^2\left(-\frac1{e_r}\right)
+\frac{C_1}{e_r^2}\arccos\left(-\frac1{e_r}\right)+\frac{C_0(e_r^2-1)^{1/2}}{e_r^2}\,,
\eea
with coefficients
\bea
A_0&=&\frac{64}{45}+\frac{1502}{45}e_r^2+\frac{283}{15}e_r^4
\,,\nonumber\\
A_1&=&\frac{104}{5}+\frac{152}{5}e_r^2+\frac{37}{15}e_r^4
\,,\nonumber\\
B_0&=& -\frac{77986}{4725}-\frac{8}{5}\nu+\left(-\frac{1778}{45}\nu-\frac{341864}{945}\right) e_r^2+\left(-\frac{1768}{45}\nu-\frac{7837243}{18900}\right) e_r^4+\left(-\frac{283}{30}\nu+\frac{5007}{1400}\right) e_r^6
\,,\nonumber\\
B_1&=&  -\frac{64}{15}+\left(-\frac{12692}{63}-24\nu\right) e_r^2+\left(-\frac{232}{5}\nu-\frac{7544}{15}\right) e_r^4+\left(-\frac{54577}{630}-\frac{91}{5}\nu\right) e_r^6+\left(\frac{1661}{280}-\frac{37}{30}\nu\right) e_r^8
\,,\nonumber\\
C_0&=&  \frac{443180371}{5953500}+\frac{3194}{63}\nu
+\left(\frac{4}{45}\nu^2+\frac{17629369}{9450}\nu+\frac{803259908}{496125}\right)e_r^2
+\left(\frac{146}{45}\nu^2+\frac{70624283}{18900}\nu+\frac{6582331379}{3175200}\right)e_r^4\nonumber\\
&&
+\left(-\frac{4695130141}{13608000}+\frac{863}{60}\nu^2+\frac{1104757}{1512}\nu\right)e_r^6
+\left(\frac{283}{40}\nu^2-\frac{153691}{8400}\nu+\frac{4955593}{529200}\right)e_r^8
\,,\nonumber\\
C_1&=&  \frac{54466}{1575}+\frac{40}{3}\nu
+\left(\frac{42026}{45}\nu+\frac{4103647}{4050}\right)e_r^2
+\left(\frac{229891}{63}\nu+\frac{8}{5}\nu^2+\frac{6491512}{2835}\right)e_r^4\nonumber\\
&&
+\left(\frac{10538077}{45360}+\frac{34469}{20}\nu+\frac{53}{5}\nu^2\right)e_r^6
+\left(\frac{35}{3}\nu^2-\frac{3136361}{20160}+\frac{63887}{1260}\nu\right)e_r^8
+\left(-\frac{8609}{1680}\nu+\frac{83359}{5040}+\frac{37}{40}\nu^2\right) e_r^{10}
\,,\nonumber\\
C_2&=&  -\frac{32}{5}-\frac{751}{5} e_r^2-\frac{849}{10} e_r^4
\,,\nonumber\\
C_3&=&  -\frac{156}{5}-\frac{228}{5} e_r^2-\frac{37}{10} e_r^4
\,.
\eea

\end{widetext}  

The  large-$j$ expansion of this result yields
\be
{\bf P}^{\rm rad, inst}=\nu^2 (m_2-m_1)[{\mathcal P}_{\rm rad, N}^y+\eta^2 {\mathcal P}_{\rm rad, 1PN}^y
+\eta^4 {\mathcal P}_{\rm rad, 2PN}^y
]  {\mathbf e}_y\,,
\ee
where
\bea
\label{Prad_N}
{\mathcal P}_{\rm rad, N}^y&=&\frac{37}{30}\pi \frac{p_\infty^5}{j^3}+\frac{64}{3} \frac{p_\infty^4}{j^4}+\frac{1097}{60}\pi \frac{p_\infty^3}{j^5}\nonumber\\
&+&\frac{4384}{45} \frac{p_\infty^2}{j^6}+\frac{2841}{80}\pi \frac{p_\infty}{j^7} +O\left(\frac{1}{j^8}\right)
\,,\nonumber\\
\eea
\bea
\label{Prad_1PN}
{\mathcal P}_{\rm rad, 1PN}^y&=&
\left(\frac{839}{1680} -\frac{37}{15}\nu \right)\pi \frac{p_\infty^7}{j^3}\nonumber\\
&+&\left(\frac{1664}{175} -\frac{160}{3}\nu \right)\frac{p_\infty^6}{j^4}\nonumber\\
&+&\left(\frac{148507}{10080}-\frac{3529}{60}\nu  \right)\pi\frac{p_\infty^5}{j^5}\nonumber\\
&+&\left(\frac{813248}{4725} -\frac{18896}{45}\nu \right)\frac{p_\infty^4}{j^6}\nonumber\\
&+& \left(\frac{5666863}{40320}-\frac{4373}{20}\nu  \right)\pi \frac{p_\infty^3}{j^7} \nonumber\\
&+&O\left(\frac{1}{j^8}\right)\,,
\eea
and
\bea
\label{Prad_2PN}
{\mathcal P}_{\rm rad, 2PN}^y&=&
\left(\frac{2699}{2016}  -\frac{107}{280}\nu  +\frac{37}{10}\nu^2    \right)\frac{\pi p_\infty^9}{j^3} \nonumber\\
&+&\left(\frac{227776}{33075}-\frac{1096}{105}\nu +\frac{280}{3}\nu^2  \right)\frac{p_\infty^8}{j^4}\nonumber\\
&+&\left(-\frac{1131443}{40320}  -\frac{55009}{2016}\nu  +\frac{608}{5}\nu^2 \right)  \frac{\pi p_\infty^7}{j^5}\nonumber\\
&+&\left( -\frac{115582624}{212625}-\frac{2269324}{4725}\nu  +\frac{15652}{15}\nu^2 \right) \frac{p_\infty^6}{j^6}\nonumber\\
&+&\left( -\frac{178354019}{362880}-\frac{37}{80}\pi^2\right.\nonumber\\
&&\left.
-\frac{5607509}{10080}\nu  +\frac{26789}{40}\nu^2 \right) \frac{\pi p_\infty^5}{j^7}\nonumber\\
&+&O\left(\frac{1}{j^8}\right) 
\,.
\eea

\section{Fractionally 1.5PN tail contribution to the radiated linear momentum  along hyperboliclike orbits}
\label{App_tail_P}

Contrary to what happened for the tail contributions to the energy and the angular momentum
(for which the integration automatically projected the hereditary past tail into its time-symmetric
projection), a subtlety arises for the tail contribution to the radiated linear momentum.

The time-symmetric tail contribution to the radiated linear momentum (which, in view of the
radiation-reaction derivation of energy loss presented above, should be the only relevant
quantity)  has the following Fourier-domain expression
\begin{widetext}
\bea
P_{i \, \rm sym\, tail}^{\rm rad}&=& \frac{G^2M}{c^{10}} \left[\frac{2i}{63}\int_0^\infty d\omega \omega^8 [I_{ijk}(-\omega)I_{jk}(\omega)-I_{ijk}(\omega)I_{jk}(-\omega) ]\right. \nonumber\\
&&\left.+\frac{16}{45}\epsilon_{ijk}\int_0^\infty d\omega \omega^7 [I_{jl}(\omega)J_{kl}(-\omega)+I_{jl}(-\omega)J_{kl}(\omega)]  \right]\,.
\eea
\end{widetext}
Performing the integration in the large-$j$ expansion limit we find that the only nonvanishing
component is along the $y$ axis ($P_x{}_{\rm sym\, tail}^{\rm rad}=0=P_z{}_{\rm sym\, tail}^{\rm rad}$):
\bea
\label{Pradtailfin}
 P_y{}_{\rm sym\, tail}^{\rm rad}&=& -\nu^2\sqrt{1-4\nu}  \left\{
 -\frac{128}{3}\frac{p_\infty^7}{j^4}
 - \frac{1509\pi^3}{140}\frac{p_\infty^6}{j^5}\right.\nonumber\\
&&
 + \left(-\frac{8768}{45} - \frac{521216\pi^2}{4725}\right) \frac{p_\infty^5}{j^6}\nonumber\\
&&
+ \left(\frac{36885}{896}\pi^5 - \frac{142391}{280}\pi^3\right) \frac{p_\infty^4}{j^7}\nonumber\\
&&\left.
+O\left(\frac{1}{j^8}\right)
\right\}\,. 
\eea
When computing the past-tail version of $P_{i \, \rm tail}^{\rm rad}$, one finds that its $y$ and $z$ components
are equal to their just-written sym-tail counterparts. By contrast, the $x$-component of
the past-tail contribution to the radiated linear momentum no longer vanishes  (because
of the dissymmetry between the scales $C_{I_3}$ and $C_{J_2}$ entering the tail logarithms). 
Its value is equal to
\bea
 P_x{}_{\rm past\, tail}^{\rm  rad}&=&-\nu^2\sqrt{1-4\nu}\left[\pi \frac{1491}{400}\frac{p_\infty^7}{j^4}
 +\frac{20608}{225}\frac{p_\infty^6}{j^5}\right.\nonumber\\
&+&\left.\pi  \frac{267583}{2400} \frac{p_\infty^5}{j^6}
+ \frac{64576}{75} \frac{p_\infty^4}{j^7}
\right]   \nonumber\\
&+&O\left(\frac{1}{j^8}\right)\,.\nonumber\\
\eea
This past-tail contribution is along the same direction as the LO impulse, and therefore conservative-like.
The reasoning (coming from \cite{Damour:2014jta}) recalled in Appendix \ref{App_tail_E} indicates
that such a contribution (which is at the absolute 5PN level) 
is already included in the time-symmetric conservative dynamics
and should not be considered among the radiation-reaction effects.

\section{PN evaluation of the impulse coefficients and the radiative losses and reminders on the PN-expanded scattering angle (conservative and radiative)}
\label{PNAppendix}

We give below, in the form of various tables, the fractionally 2PN-expanded form of the various impulse coefficients $c_b^{1,\rm X}$, $c_{u_1}^{1, \rm X} $, $ c_{u_2}^{1, \rm X}$ and $\widehat c_{u_1}^{1, \rm X} $, with X = cons, rr rel, rr rec, entering the decomposition of $\Delta p_1^\mu$, up to 3PM order (see Table \ref{tab:table5}, \ref{tab:table6}, \ref{tab:table7} and \ref{tab:table8}, respectively). This is
completed by writing down some useful PN relations concerning the conservative 
and radiative parts of the scattering angle (see Table \ref{tab:table_chi_cons} and \ref{tab:table_chi_rr}, respectively) as well as the coefficients of the PM expansion of the energy and momentum radiative losses (see Table \ref{tab:table_EnJnPn}). 

The PN expansion of $\chi^{\rm cons}(p_\infty,j)$ is known at present up to the 6PN level \cite{Bini:2019nra,Bini:2020wpo,Bini:2020nsb,Bini:2020hmy,Bini:2020rzn}. 
However, since the PN knowledge of the radiated energy and angular momentum is only at the
fractional 2PN level, it is enough to use $\chi^{\rm cons}$ (which starts at
the Newtonian level) with an absolute 2PN accuracy. Below $\eta=\frac1c$  is a place-holder for keeping
track of (absolute or relative) PN expansions. Beware that most of the time we use the phrase 
{\it 2PN accuracy} in a fractional sense. One should keep in mind that radiation-reaction effects start
at the 2.5PN level, i.e., $\eta^5$.

\begin{widetext}


\begin{table*} 
\caption{\label{tab:table5} PN expansion of the various coefficients of $c_b^1=\mu \sum_{n=1}^\infty [c_{b,G^n}^{\rm cons}+c_{b,G^n}^{\rm rr, rel}+c_{b,G^n}^{\rm rr, rec}]\left( \frac{GM}{b}\right)^n$, with $n\leq3$.}
\begin{ruledtabular}
\begin{tabular}{cclll}
  &cons&  $-\frac{2}{p_\infty}-4p_\infty$ \\
$G^1$ &rr, rel & $-$\\
  &rr, rec & $-$\\
\hline
  &cons & $\pi \left(-\frac{3}{p_\infty}-\frac{15}{4} p_\infty\right)$ \\
$G^2$ &rr, rel  & $-$\\
  &rr, rec & $-$\\
\hline
  &cons & $\frac{2}{p_\infty^5}+\frac{2\nu}{p_\infty^3}+\left(\frac{31}{2}\nu-32\right)\frac{1}{p_\infty}+O(p_\infty)$ \\
$G^3$ &rr, rel & $\nu\left(-\frac{16}{5} -\frac{80}{7}  p_\infty^2-\frac{512}{63}  p_\infty^4+O(p_\infty^6)\right)$\\
  &rr, rec & $-$\\
\end{tabular}
\end{ruledtabular}
\end{table*}


\begin{table*}  
\caption{\label{tab:table6} PN expansion of the various coefficients of $c_{u_1}^1=\mu \sum_{n=2}^\infty [c_{u_1,G^n}^{\rm cons}+c_{u_1,G^n}^{\rm rr, rel}+c_{u_1,G^n}^{\rm rr, rec}]\left( \frac{GM}{b}\right)^n$, with $n\leq3$.}
\begin{ruledtabular}
\begin{tabular}{cclll}
  &cons & $-\frac{2}{p_\infty^4}+\frac{\left(-\frac{17}{2}+\frac12 \Delta \right)}{p_\infty^2}-\frac{79}{8}+\frac{15}{8}\Delta+O(p_\infty^2)$ \\
$G^2$  &rr, rel  & $-$\\
  &rr, rec & $-$\\
\hline
  &cons & $\pi\left[-\frac{6}{p_\infty^4}+\frac{\left(\frac32\Delta-21\right)}{p_\infty^2}
-\frac{39}{2}+\frac92\Delta
+O(p_\infty^2)\right]$ \\
$G^3$  &rr, rel & $\pi\nu \left[-\frac{37}{15p_\infty}+\left(-\frac{2911}{840} +\frac{37}{60} \Delta+\frac{37}{15}\nu \right) p_\infty+\left(\frac{2393}{840}\nu -\frac{37}{15}\nu^2 +\frac{125}{224} \Delta-\frac{37}{60}\Delta\nu -\frac{1379}{360} \right) p_\infty^3
+O(p_\infty^5)\right]$\\
  &rr, rec & $-\pi \nu \left[ \left(\frac{37}{15}\nu -\frac{37}{60} +\frac{37}{60}\Delta\right) p_\infty
+\left(-\frac{37}{60} \Delta\nu +\frac{2393}{840} \nu -\frac{37}{15}\nu^2 -\frac{125}{224}  +\frac{125}{224}\Delta \right) p_\infty^3
+O(p_\infty^5)\right]$\\
& rr, tot & 
$\pi \nu \left[-\frac{37}{15 p_\infty}-\frac{2393}{840} p_\infty-\frac{32987}{10080} p_\infty^3+O(p_\infty^5)\right]$
\\
\end{tabular}
\end{ruledtabular}
\end{table*}


\begin{table*}  
\caption{\label{tab:table7} PN expansion of the various coefficients of  $c_{u_2}^1=\mu \sum_{n=2}^\infty [c_{u_2,G^n}^{\rm cons}+c_{u_2,G^n}^{\rm rr, rel}+c_{u_2,G^n}^{\rm rr, rec}]\left( \frac{GM}{b}\right)^n$, with $n\leq3$.}
\begin{ruledtabular}
\begin{tabular}{cclll}
  &cons & $\frac{2}{p_\infty^4}+\frac{\left(\frac{17}{2}+\frac{1}{2}\Delta \right)}{p_\infty^2}
+\frac{79}{8}+\frac{15}{8}\Delta+O(p_\infty^2) $ \\
$G^2$  &rr, rel  & $-$\\
  &rr, rec & $-$\\
\hline
  &cons & $\pi \left[\frac{6}{p_\infty^4}+\frac{\left(\frac32\Delta+21\right)}{p_\infty^2}
+\frac{39}{2}+\frac{9}{2}\Delta
+O(p_\infty^2)\right]$ \\
$G^3$  &rr, rel & $\pi\nu \left[\frac{37}{15 p_\infty}+\left(-\frac{37}{60}\Delta+\frac{125}{56}-\frac{37}{15}\nu\right) p_\infty
+\left(\frac{37}{60}\Delta\nu +\frac{3047}{1008} +\frac{37}{15}\nu^2
-\frac{1357}{840}\nu -\frac{839}{3360}\Delta\right) p_\infty^3
+O(p_\infty^5)\right]$\\
  &rr, rec & $-\pi \nu \left[\left(\frac{37}{60} -\frac{37}{15}\nu -\frac{37}{60}\Delta\right) p_\infty
+\left(-\frac{1357}{840}\nu -\frac{839}{3360}\Delta+\frac{37}{60}\Delta\nu+\frac{37}{15}\nu^2+\frac{839}{3360}\right) p_\infty^3
+O(p_\infty^5)\right] $\\
& rr, tot & 
$\left[\frac{37}{ 15 p_\infty}+\frac{1357}{840}p_\infty+\frac{27953}{10080}p_\infty^3+O(p_\infty^5)\right]\nu\pi$
\\
\end{tabular}
\end{ruledtabular}
\end{table*}


\begin{table*}  
\caption{\label{tab:table8} PN expansion of the various coefficients of $\widehat c_{u_1}^1=\mu \sum_{n=2}^\infty [\widehat c_{u_1,G^n}^{\rm cons}+\widehat c_{u_1,G^n}^{\rm rr, rel}+\widehat c_{u_1,G^n}^{\rm rr, rec}]\left( \frac{GM}{b}\right)^n$, with $n\leq3$.}
\begin{ruledtabular}
\begin{tabular}{cclll}
  &cons & $(1+\Delta)\left( \frac{1}{p_\infty}+2p_\infty \right)^2 $ \\
$G^2$  &rr, rel  & $-$\\
  &rr, rec & $-$\\
\hline
  &cons & $\pi (1 +\Delta)\left(\frac{3}{p_\infty^2}+\frac{39}{4}+\frac{15}{2}p_\infty^2  \right)$ \\
$G^3$  &rr, rel & $0$\\
  &rr, rec & $0$\\
\end{tabular}
\end{ruledtabular}
\end{table*}


\begin{table*}  
\caption{\label{tab:table_EnJnPn} Fractionally 2PN-accurate expansion of the coefficients $E_n$, $J_n$ and $P_n$ of the PM expansions of the energy, angular momentum and linear momentum radiative losses, Eqs. \eq{PMexpEJrad} and \eq{PMexpPrad}, up to $n=7$.}
\begin{ruledtabular}
\begin{tabular}{ll}
$E_{3}$&$ \pi  \left[\frac{37 }{15}p_\infty^4
+\left(\frac{1357}{840}-\frac{74 \nu }{15}\right)p_\infty^6
+\left(\frac{37 \nu ^2}{5}-\frac{839 \nu}{420}+\frac{27953}{10080}\right)p_\infty^8
+O(p_\infty^{10})\right]$\\
$E_{4}$&$ \frac{1568 }{45}p_\infty^3
+\left(\frac{18608}{525}-\frac{1424 \nu }{15}\right)p_\infty^5
+\frac{3136 }{45}p_\infty^6
+\left(172 \nu ^2-\frac{31036 \nu }{525}+\frac{220348}{11025}\right)p_\infty^7
+O(p_\infty^8)$\\
$E_{5}$&$ \pi  \left[\frac{122 }{5}p_\infty^2
+\left(\frac{13831}{280}-\frac{933 \nu }{10}\right)p_\infty^4
+\frac{297 \pi ^2 }{20}p_\infty^5
+\left(\frac{2067 \nu ^2}{10}-\frac{187559 \nu}{1680}-\frac{64579}{5040}\right)p_\infty^6
+O(p_\infty^7)\right]$\\
$E_{6}$&$ \frac{4672 }{45}p_\infty
+\left(\frac{142112}{315}-\frac{26464 \nu }{45}\right)p_\infty^3
+\left(\frac{9344}{45}+\frac{88576 \pi ^2}{675}\right)p_\infty^4
+\left(\frac{24424 \nu ^2}{15}-\frac{6732728 \nu}{4725}-\frac{293992}{1701}\right)p_\infty^5
+O(p_\infty^6)$\\
$E_{7}$&$ \pi  \left[\frac{85}{3}
+\left(\frac{2259}{8}-265 \nu\right) p_\infty^2
+\left(\frac{1579 \pi ^2}{3}-\frac{2755 \pi^4}{64}\right) p_\infty^3
+\left(\frac{7605 \nu ^2}{8}-\frac{432805 \nu}{336}+\frac{19319}{378}\right) p_\infty^4
+O(p_\infty^5)\right]$\\
\hline
$J_{2}$&$ \frac{16 }{5}p_\infty^3
+\left(\frac{176}{35}-\frac{16 \nu }{5}\right)p_\infty^5
+\left(\frac{16 \nu ^2}{5}-\frac{148 \nu }{35}-\frac{608}{315}\right)p_\infty^7
+O(p_\infty^9)$\\
$J_{3}$&$ \pi  \left[\frac{28 }{5}p_\infty^2
+\left(\frac{739}{84}-\frac{163\nu }{15}\right) p_\infty^4
+\left(\frac{50 \nu ^2}{3}-\frac{5339 \nu}{420}-\frac{5777}{2520}\right) p_\infty^6
+O(p_\infty^8)\right]$\\
$J_{4}$&$ \frac{176 }{5}p_\infty
+\left(\frac{8144}{105}-\frac{1072 \nu }{9}\right)p_\infty^3
+\frac{448 }{5}p_\infty^4
+\left(\frac{11056 \nu ^2}{45}-\frac{247724 \nu }{1575}-\frac{93664}{1575}\right)p_\infty^5
+O(p_\infty^6)$\\
$J_{5}$&$ \pi  \left[12
+\left(\frac{107}{2}-\frac{374 \nu }{5}\right)p_\infty^2
+\frac{69 \pi ^2 }{5}p_\infty^3
+\left(209 \nu ^2-\frac{15856 \nu }{105}-\frac{101219}{1512}\right)p_\infty^4
+O(p_\infty^5)\right]$\\
$J_{6}$&$ \frac{304}{15 }\frac1{p_\infty}
+\left(\frac{84368}{315}-\frac{13456 \nu }{45}\right)p_\infty
+\left(\frac{128}{15}+\frac{4352 \pi ^2}{45}\right)p_\infty^2
+\left(\frac{3568 \nu ^2}{3}-\frac{31828 \nu }{27}-\frac{552320}{1701}\right)p_\infty^3
+O(p_\infty^4)
$\\
$J_{7}$&$ \pi  \left[\frac{359}{4}-\frac{235 \nu }{3}
+\left(303 \pi^2-\frac{423 \pi ^4}{16}\right) p_\infty
+\left(\frac{1484 \nu ^2}{3}-\frac{8497 \nu}{12}-\frac{2999}{216}\right) p_\infty^2
+O(p_\infty^3)\right]$\\
\hline
$P_{3}$&$ \pi  \left[\frac{37}{30} p_\infty^5
+\left(\frac{839}{1680}-\frac{37\nu }{15}\right) p_\infty^7
+\left(\frac{37 \nu ^2}{10}-\frac{107 \nu}{280}+\frac{2699}{2016}\right) p_\infty^9
+O(p_\infty^{11})\right]$\\
$P_{4}$&$ \frac{64}{3} p_\infty^4
+\left(\frac{1664}{175}-\frac{160 \nu }{3}\right)p_\infty^6
+\frac{128}{3} p_\infty^7
+\left(\frac{280 \nu ^2}{3}-\frac{1096 \nu }{105}+\frac{227776}{33075}\right)p_\infty^8
+O(p_\infty^9)$\\
$P_{5}$&$ \pi  \left[\frac{1097 }{60}p_\infty^3
+\left(\frac{148507}{10080}-\frac{3529 \nu }{60}\right)p_\infty^5
+\frac{1509 \pi ^2 }{140}p_\infty^6
+\left(\frac{608 \nu ^2}{5}-\frac{55009 \nu}{2016}-\frac{1131443}{40320}\right)p_\infty^7
+O(p_\infty^8)\right]$\\
$P_{6}$&$ \frac{4384 }{45}p_\infty^2
+\left(\frac{813248}{4725}-\frac{18896 \nu }{45}\right)p_\infty^4
+\left(\frac{8768}{45}+\frac{521216 \pi ^2}{4725}\right)p_\infty^5
+\left(\frac{15652 \nu ^2}{15}-\frac{2269324 \nu}{4725}-\frac{115582624}{212625}\right)p_\infty^6
+O(p_\infty^7)$\\
$P_{7}$&$ \pi  \left[\frac{2841 }{80}p_\infty
+\left(\frac{5666863}{40320}-\frac{4373 \nu }{20}\right)p_\infty^3
+\left(\frac{142391 \pi ^2}{280}-\frac{36885 \pi ^4}{896}\right)p_\infty^4
+\left(\frac{26789 \nu ^2}{40}-\frac{5607509 \nu }{10080}-\frac{37 \pi^2}{80}-\frac{178354019}{362880}\right)p_\infty^5
+O(p_\infty^6)\right]$\\
\end{tabular}
\end{ruledtabular}
\end{table*}


\begin{table*}
\caption{\label{tab:table_chi_cons} Expansion coefficients of $\frac{\chi^{\rm cons}}{2}=\sum_{n\geq1}\frac{\chi^{\rm cons}_{n}}{j^n}$ up to $n=7$, evaluated at the (absolute) 2PN level of accuracy.}
\begin{ruledtabular}
\begin{tabular}{ll}
$\chi^{\rm cons}_{1}$&$ \frac{1}{p_\infty} + 2 p_\infty\eta^2$\\
$\chi^{\rm cons}_{2}$&$\pi \left[ \frac{3}{2}  \eta^2  + \left(\frac{15}{8} - \frac{3}{4} \nu \right)p_\infty^2\eta^4\right] $\\
$\chi^{\rm cons}_{3}$&$ -\frac{1}{3 p_\infty^3}+ \frac{4}{p_\infty}\eta^2 +\left(- 8\nu + 24 \right)p_\infty\eta^4 $\\
$\chi^{\rm cons}_{4}$&$\pi\left[ -\frac{15}{4}\nu  + \frac{105}{8}  \right] $\\
$\chi^{\rm cons}_{5}$&$ \frac{1}{5 p_\infty^5} - \frac{2}{p_\infty^3}\eta^2
+\left( 32  -  8\nu  \right)\frac{\eta^4}{p_\infty}$\\
$\chi^{\rm cons}_{6}$&$ 0 $\\
$\chi^{\rm cons}_{7}$&$- \frac{1}{ 7 p_\infty^7}+ \frac{8}{5p_\infty^5}\eta^2  +\left(- 16  +  \frac{16\nu}{5}\right)\frac{\eta^4}{p_\infty^3} $\\ 
\end{tabular}
\end{ruledtabular}
\end{table*}


\begin{table*}  
\caption{\label{tab:table_chi_rr} Expansion coefficients of $\frac{\delta^{\rm rr }\chi^{\rm rel}}{2} =\sum_{n\geq3} \frac{\chi^{\rm rr}_{n}}{j^n}$ up to $n=7$, evaluated at the fractional 2PN level of accuracy.}
\begin{ruledtabular}
\begin{tabular}{ll}
$\chi^{\rm rr}_{3}$&$ \frac{8}{5}p_\infty^2\nu +\left(-\frac{8}{5}\nu^2 + \frac{40}{7} \nu\right)p_\infty^4\eta^2 
+\left(\frac{256}{63}\nu - \frac{186}{35}\nu^2 + \frac{8}{5}\nu^3 \right)p_\infty^6 \eta^4 $\\
$\chi^{\rm rr}_{4}$&$ \frac{121}{30} p_\infty\nu\pi  
+\left(\frac{23111}{1680}  \nu - \frac{437}{60} \nu^2  \right)\pi p_\infty^3\eta^2
+\left(- \frac{75253}{3360} \nu^2  + \frac{511}{48} \nu^3  + \frac{44759}{2240} \nu \right)\pi p_\infty^5 \eta^4$\\
$\chi^{\rm rr}_{5}$&$\frac{1504}{45}\nu +\left(-\frac{4352}{45}\nu^2 + \frac{43184}{525}\nu + \frac{42}{5} \pi^2\nu  
\right)\eta^2 p_\infty^2
+\left(
- \frac{373}{20}\nu^2\pi^2  + \frac{267}{14}\nu\pi^2  - \frac{22888}{105}\nu^2  + \frac{544}{3}\nu^3  + \frac{1741664}{11025}\nu \right)p_\infty^4\eta^4$\\
$\chi^{\rm rr}_{6}$&$  \frac{85}{6p_\infty}\pi\nu 
+\left(- \frac{212}{3} \nu^2  + \frac{15679}{120} \nu   \right)\pi p_\infty \eta^2
+ \left(\frac{8273}{48}\nu^3  + \frac{3402881}{8640}\nu   - \frac{514431}{1120}\nu^2\right)\eta^4 \pi p_\infty^3 $\\
$\chi^{\rm rr}_{7}$&$  \frac{1288}{45 p_\infty^2}\nu  
+\left(-  \frac{2888}{9}\nu^2  +  \frac{793928}{1575}\nu  + 18\pi^2\nu \right)\eta^2
+\left(  \frac{127546256}{99225}\nu   +  \frac{48328}{45}\nu^3  -  \frac{1449}{10}\nu^2\pi^2 + 204\nu\pi^2  -  \frac{3424658}{1575}\nu^2  \right)p_\infty^2 \eta^4$\\ 
\end{tabular}
\end{ruledtabular}
\end{table*}

\end{widetext}


\end{document}